\documentclass{article}
\usepackage{PRIMEarxiv}

\usepackage{hyperref}
\usepackage{pifont}
\usepackage{fancyhdr}       
\usepackage{graphicx}     
\usepackage[numbers]{natbib}

\usepackage{ragged2e}
\usepackage{multirow}
\usepackage{multicol}
\usepackage[table]{xcolor}
\usepackage{enumitem}
\usepackage{wrapfig}
\usepackage{paralist}
\usepackage{booktabs,tabularx}
\usepackage{longtable}
\usepackage{array}
\usepackage{colortbl}

\usepackage[normalem]{ulem}
\usepackage[most]{tcolorbox}
\usepackage{tikz}
\usetikzlibrary{arrows.meta, positioning, shapes.geometric, chains, fit}
\usepackage{xcolor}
\usepackage{adjustbox}

\definecolor{lightblue}{HTML}{ADD8E6} 
\definecolor{steelblue}{HTML}{4682B4} 
\tcbset{
  colback=lightblue!5!white, 
  colframe=steelblue!95!black, 
  left=3mm, 
  right=3mm 
}
\tcbset{
  myinset/.style={
    enhanced jigsaw,
    sharp corners,
    boxrule=0.6pt,
    colback=blue!5!white,
    colframe=blue!75!black,
    fonttitle=\bfseries,
    title=Important Note,
    float*,             
    sidebyside,         
    sidebyside align=top,
    righthand width=0.4\textwidth, 
    lefthand width=0.55\textwidth, 
    before upper={\parindent15pt},
    after upper={\par\vspace{1ex}},
  }
}

\makeatletter
\g@addto@macro\@floatboxreset\centering
\makeatother

\tikzset{
    every picture/.style={
        node distance=1.2cm and 1cm,
        >=stealth
    },
    flowchartNode/.style={
        draw=steelblue,
        fill=lightblue!15!white,
        rectangle,
        rounded corners=4pt,
        inner sep=3pt,
        font=\sffamily\small\bfseries,
        thick,
        minimum width=2cm,
        minimum height=0.7cm,
        align=center,
    },
    flowArrow/.style={
        thick,
        color=steelblue!80!black, 
        >={Stealth[length=3mm, width=3mm]}, 
        ->
    }
}

\newcommand{\revision}[1]{\textcolor{black}{#1}}

\usepackage{cleveref}

\title{
Software Engineering for Self-Adaptive Robotics: A Research Agenda
}

\author{
  Hassan Sartaj \\
  Simula Research Laboratory \\
  Oslo, Norway\\
  \texttt{hassan@simula.no} \\
  \And
  Shaukat Ali \\
  Simula Research Laboratory \\
  Oslo, Norway\\
  \texttt{shaukat@simula.no} \\
  \And
  Ana Cavalcanti \\
  University of York \\
  York, UK\\
  \texttt{ana.cavalcanti@york.ac.uk} \\
  \And
  Lukas Esterle \\
  Aarhus University \\
  Aarhus, Denmark\\
  \texttt{lukas.esterle@ece.au.dk} \\
  \And
  Cláudio Ângelo Gonçalves Gomes \\
  Aarhus University \\
  Aarhus, Denmark\\
  \texttt{claudio.gomes@ece.au.dk} \\
  \And
  Peter Gorm Larsen \\
  Aarhus University \\
  Aarhus, Denmark\\
  \texttt{pgl@ece.au.dk} \\
  \And
  Anastasios Tefas \\
  Aristotle University of Thessaloniki \\
  Thessaloniki, Greece\\
  \texttt{tefas@csd.auth.gr} \\
  \And
  Jim Woodcock \\
  Southwest University Chongqing, China, \\
  Aarhus University, Aarhus, Denmark\\
  University of York, York, UK\\
  \texttt{jim.woodcock@york.ac.uk} \\
  \And
  Houxiang Zhang \\
  Norwegian University of Science and Technology \\
  Aalesund, Norway\\
  \texttt{hozh@ntnu.no} \\
}

\begin{document}
\maketitle

\begin{abstract}
Self-adaptive robotic systems operate autonomously in dynamic and uncertain environments, requiring robust real-time monitoring and adaptive behaviour. 
Unlike traditional robotic software with predefined logic, self-adaptive robots exploit artificial intelligence (AI), machine learning, and model-driven engineering to adapt continuously to changing conditions, thereby ensuring reliability, safety, and optimal performance. 
This paper presents a research agenda for software engineering in self-adaptive robotics, structured along two dimensions. 
The first concerns the software engineering lifecycle---requirements, design, development, testing, and operations---tailored to the challenges of self-adaptive robotics. 
The second focuses on enabling technologies such as digital twins and AI-driven adaptation, which support runtime monitoring, fault detection, and automated decision-making.
We identify open challenges, including verifying adaptive behaviours under uncertainty, balancing trade-offs between adaptability, performance, and safety, and integrating self-adaptation frameworks like MAPE-K/MAPLE-K. 
By consolidating these challenges into a roadmap toward 2030, this work contributes to the foundations of trustworthy and efficient self-adaptive robotic systems capable of meeting the complexities of real-world deployment. 
\end{abstract}


\section{Introduction} \label{sec:intro}

Robotic systems are increasingly expected to operate in dynamic, uncertain, and unstructured environments, making self-adaptation a crucial capability. Unlike traditional robots that follow pre-programmed behaviours, self-adaptive robots need to exploit AI and data-driven techniques to autonomously modify their behaviour in response to environmental changes, operational faults, and evolving objectives. This adaptability is crucial in autonomous driving, industrial automation, search-and-rescue, and assistive robotics. For such a level of autonomy, a fundamental characteristic these robots need to possess is self-management capability, including self-configuration, self-optimisation, self-healing, and self-protection~\cite{kephart2003vision}.

Self-adaptive robotic systems require sophisticated mechanisms to ensure reliability, safety, and robustness while balancing competing demands, including energy efficiency, computational cost, and ethical considerations. The \emph{Monitor-Analyse-Plan-Execute-Knowledge (MAPE-K) loop} serves as a fundamental framework for self-adaptation, enabling robots to continuously assess their environment, plan adaptive responses, and execute modifications while maintaining a knowledge base. Recent advances, such as \emph{MAPLE-K}~\cite{larsen2024robotic}, extend this framework to verify the legitimacy of adaptations. 
We adopt the widely used MAPE-K loop and its recent extension MAPLE-K, which adds legitimacy checks to improve trustworthiness.

By framing self-adaptation as a continuous loop throughout the entire lifecycle, this roadmap highlights adaptability challenges across all phases of software engineering. 
Furthermore, it discusses the transformative role of enabling technologies: AI for intelligent and dynamic decision-making and digital twins for real-time simulation. 
Unlike existing roadmaps for self-adaptive systems (e.g., those by~\citet{de2013software,weyns2023towards,brugali2024future}) or domain-specific robotics (e.g., those by~\citet{garcia2020robotics,goues2024software}), we focus specifically on self-adaptive robotics. We identify open challenges and research opportunities across the software engineering lifecycle, covering key enabling technologies, and highlighting the critical quality aspects required for deploying future self-adaptive robots in real-world environments that are reliable, safe, and robust.

This paper provides a research agenda for the software engineering of self-adaptive robotics, addressing challenges from two key perspectives:
\begin{inparaenum}[(1)]
\item \textbf{software engineering lifecycle}, covering requirements engineering, software design, software development, software testing, and operations tailored for self-adaptive robotics; and 
\item \textbf{key enabling technologies}, exploring the role of digital twins, and AI in supporting autonomous adaptation and decision-making in robotic software systems.
\end{inparaenum}
By structuring the discussion around these two dimensions, we present a 2030 research agenda to advance self-adaptive robotic software engineering, ensuring robotic systems can effectively navigate unpredictable environments.

Developing and maintaining self-adaptive robotic software presents several challenges from a software engineering perspective. These include
\begin{inparaenum}[(1)]
\item requirements engineering for dynamic and evolving specifications;
\item design of scalable and modular architectures that facilitate real-time adaptation;
\item integration of AI techniques, while ensuring explainability and robustness;
\item verification and validation of adaptive behaviours under uncertainty; and
\item exploitation of digital twins for runtime monitoring, fault prediction, and real-time adaptation validation.
\end{inparaenum}

The remainder of the paper is structured as follows. 
\Cref{sec:background} provides an overview of self-adaptive robotics. 
\Cref{sec:outline} presents an overview of our roadmap. 
\Cref{sec:requirements,sec:design,sec:development,sec:testing,sec:operations} cover all phases of the software engineering lifecycle. 
\Cref{sec:keytech} discusses key enabling technologies in self-adaptive robotics. 
\Cref{sec:agenda2030} presents a forward-looking research agenda for 2030. 
\Cref{sec:relatedworks} discusses related work and finally \Cref{sec:conclusion} concludes the paper.

\section{Background}\label{sec:background}

In this section, we provide an overview of software engineering practices for self-adaptive systems. 
These are systems capable of autonomously adjusting their behaviour in response to changes in their environment or internal state~\cite{de2013software}. 
Self-adaptive systems dynamically adapt to varying conditions without direct human intervention. 
Due to their broad applicability, self-adaptive systems have become a critical research focus across diverse domains, including autonomic computing, embedded systems, multi-agent systems, and robotics. 
At the core, such adaptability is enabled by software, which serves as the key driver for self-adaptive capabilities. 

\begin{figure}[htbp]
\centerline{\includegraphics[width=0.85\linewidth, keepaspectratio]{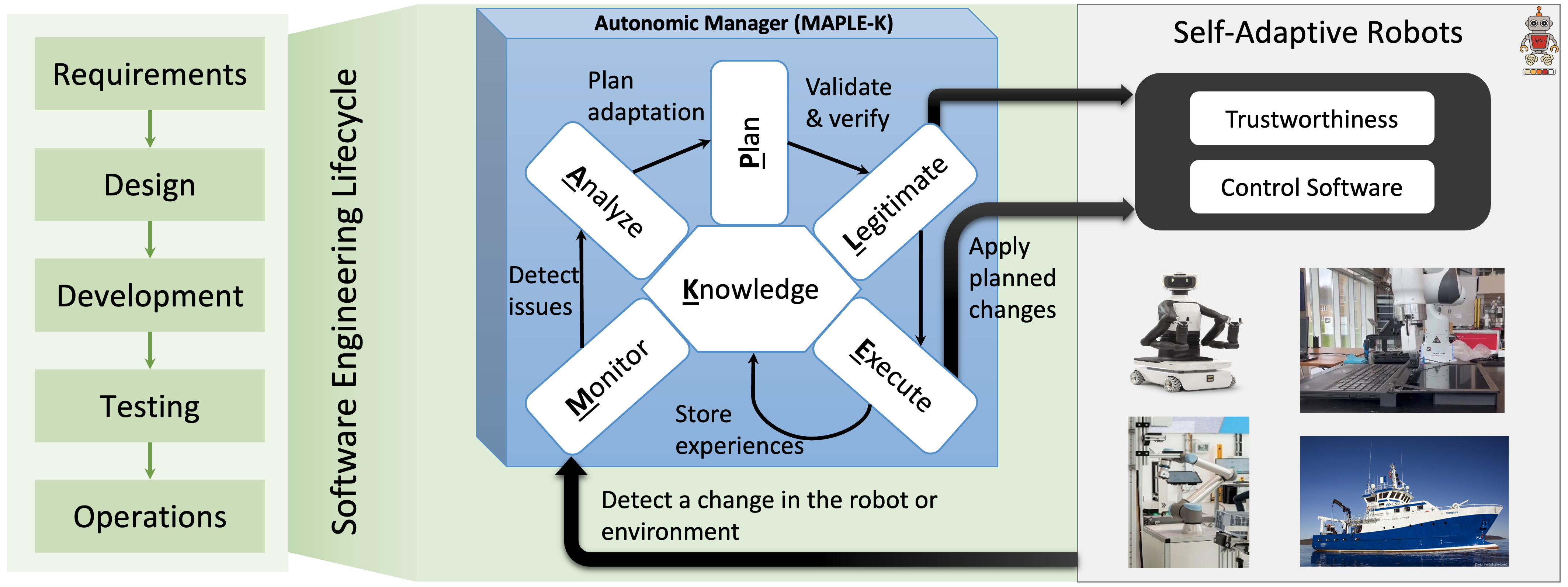}}
\caption{A conceptual view of the MAPLE-K loop for SAR and the phases of the software engineering lifecycle. }
\label{fig:se4sar}
\end{figure}

As an example, \Cref{fig:se4sar} presents an overview of an approach to developing self-adaptation in robotic systems, highlighting key components of the software architecture, such as the autonomic manager, control software, the trustworthiness analyser, and the associated phases of the software engineering lifecycle. 
The autonomic manager is central to robotic self-adaptation and operates based on the MAPLE-K loop~\cite{larsen2024robotic}---an extension of the foundational MAPE-K framework~(Monitor, Analyse, Plan, Execute, Knowledge)~\cite{kephart2003vision}---with the addition of the \textit{Legitimate} phase to address trustworthiness considerations. 

In the \textit{Monitor} phase, the robot continuously collects data from various components, including sensors and the environment, in response to changing conditions. 
The collected data is then analysed in the \textit{Analyse} phase to understand the current state and predict potential issues and future states. 
The next phase is \textit{Plan}, which involves determining appropriate actions to adapt to the changes and address potential issues. 
The subsequent phase is \textit{Legitimate}, which involves verifying and validating the safety of the suggested plan. 
Finally, in the \textit{Execution} phase, the robot adjusts its behaviour and implements the planned actions. 
At the core of the MAPLE-K loop is the \textit{Knowledge} phase, which involves maintaining and updating a knowledge base that includes information such as past experiences, environmental conditions, and the outcomes of previous actions. 

Software engineering for self-adaptive robotics (SAR) spans the entire software development lifecycle, with each phase addressing distinct adaptability considerations~\cite{de2013software}. Requirements engineering involves eliciting and specifying adaptation goals and constraints, as well as runtime evolution and uncertainty. 
Software design and modelling focus on creating architectures that support runtime monitoring, decision-making, and reconfiguration, often using modelling languages, formal methods, and simulation models (e.g., Simulink~\cite{pozzi2022modeling})  to capture variability and adaptation strategies. 
Software development involves implementing control-loop mechanisms, such as the MAPLE-K loop, encompassing monitoring, analysing, planning, legitimising, executing, and managing knowledge to enable adaptive and intelligent system behaviour, as illustrated above. 
Lastly, quality assurance includes verification and validation activities, such as testing, formal verification, and simulation, to analyse reliability under uncertain and dynamic conditions.

\revision{
While self-adaptive robots are a special case of cyber-physical systems (CPS), their key differences from other CPS types and software systems arise from robot-specific characteristics such as \textit{embodiment} and \textit{physical interaction}. 
For example, in traditional CPS and self-adaptive systems, adaptation is largely confined to the computational domain, involving reconfiguring components, reallocating resources, or tuning parameters. 
In contrast, self-adaptive robots are embodied agents whose physical form may itself vary during operation, requiring the software to remain continuously aware of and responsive to the robot’s physical configuration. 
}
Below, we list the characteristics of self-adaptive robots that raise challenges discussed in this roadmap.\\
\begin{inparaenum}[$\bullet$]
  \item \textbf{Embodiment:} Individual robots are embodied agents that have to interact with their environment. The shape of their body may vary, potentially even during the operation, requiring the software to adapt accordingly. Overall, a robot needs to be aware of its physical shape and capabilities to operate safely and in accordance with its requirements~\cite{esterle2016cyber}.
  \\
  \item \textbf{Physical Interaction:} All robots have a physical interaction with their environment, so there are inherent dynamics with which the robot has to deal. This may include interaction with other robots or even with humans. Any of these interactions, whether active or passive, may introduce uncertainty and unexpected situations that the robot must handle~\cite{Barnes2019uncertainty}. In addition, accurate system simulations are required to effectively predict and adapt to such dynamic interactions. 
  \\
   \item \textbf{Localised:} Within the physical environment, the robots need not only to be aware of their own extent but also have an understanding of how and where they are located in the environment. This might be either an absolute, global position or a relative position. This is specifically complex in self-adaptive robotics compared to pure self-adaptive software systems\revision{~\cite{Maurelli2022localisation}}.
  \\
  \item \textbf{Heterogeneity:} Robots might encounter other robots or systems, different in physical shape, capabilities, resources, goals, or behaviour. This heterogeneity can bring about benefits or additional complexities in identifying potential collaboration or competition among the other robots~\revision{\cite{Esterle2022heterogeneity}}.
  \\
  \item \textbf{\revision{Distributed Nature:}} While a single operator might operate multiple robots, they are usually distributed in the environment. This might require autonomous behaviour and self-adaptation in the absence of a centralised controller or view of the systems as a whole\revision{~\cite{Xing2023DistRobot, Rizk2019heteroRobots}}.
  \\
  \item \textbf{Modularity:} Modular robots are composed of multiple smaller units, allowing for flexible reconfiguration and adaptation to new tasks or environments. This modularity can enhance the robots' capabilities and facilitate collaboration among them. However, this also introduces additional complexity in terms of coordination, communication, and control~\cite{Yim2002modularRobots}.
\end{inparaenum}

\noindent
In our roadmap, as we discuss challenges and opportunities, we refer to the points above. 

\begin{figure*}[t]
\centerline{\includegraphics[width=1.0\linewidth, keepaspectratio]{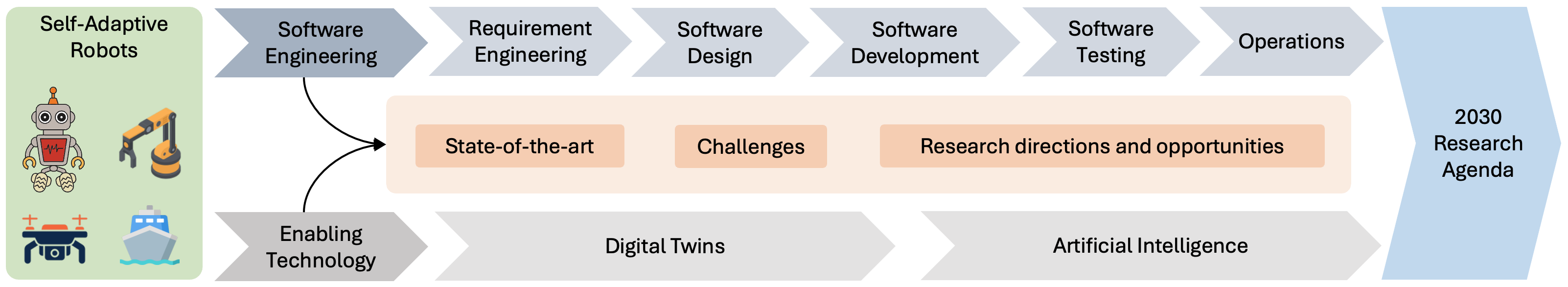}}
\caption{An overview of the roadmap, illustrating its structure and the aspects covered in the software engineering lifecycle and enabling technologies.}
\label{fig:outline}
\end{figure*}

\section{Overview of the Roadmap}\label{sec:outline} 

\Cref{fig:outline} illustrates the structure of the roadmap, organised to cover the software engineering lifecycle, key enabling technologies, and a research agenda. 
First, we examine the phases of the software engineering lifecycle, which encompass requirements engineering, software design, software development, software testing, and operations. 
Since software engineering for self-adaptive robotics spans the entire lifecycle, we focus on the phases that have received the most attention in research, while also identifying critical challenges that remain unresolved. 

Secondly, we focus on key enabling technologies: digital twins and AI. 
Technologies such as digital twins and AI have already demonstrated a significant role in advancing self-adaptive robotics by enabling real-time simulations, decision-making, and automation. We provide a detailed roadmap by analysing the current state of the art, identifying open challenges, and highlighting potential future research directions. 

Finally, we present our comprehensive research agenda for software engineering for self-adaptive robotics towards 2030.  This agenda outlines a forward-looking vision and provides a detailed roadmap for researchers in robotics interested in addressing uncertainty.

\section{Software Requirements Engineering for SAR}\label{sec:requirements} 

We begin with an overview of related research~(Section~\ref{section:jcpw}), providing context for the state of the art discussion. 
Following this, we describe key challenges that remain to be addressed~(Section~\ref{section:jcpw-challenges}). 

\subsection{Foundations and Related Research} 
\label{section:jcpw}

Software Requirements Engineering (SRE) plays a critical role in ensuring the success of self-adaptive robotic systems.
These systems operate in highly dynamic and unpredictable environments, making traditional requirements specification methods insufficient. For robotics, SRE must capture both functional requirements, such as perception, navigation, and interaction with the physical world, and non-functional requirements, such as safety, adaptability, and real-time performance.

Goal-oriented requirements engineering (GORE), formal methods, and model-driven approaches have been used for robotic SRE. Goal-oriented approaches, such as those based on KAOS and i frameworks, enable the capture and refinement of high-level goals into operational requirements~\cite{vanLamsweerde2009}. Formal methods, such as Z notation and temporal logic, are gaining traction in robotics for specifying safety-critical requirements and verifying their correctness~\cite{letier2025obstacle}. Model-driven approaches, such as SysML-based methods, facilitate the integration of requirements with system design and verification~\cite{Fitzgerald2014}. Recent efforts have focused on runtime requirements monitoring for self-adaptive systems, where requirements are treated as dynamic entities that evolve as the robot interacts with its environment \cite{Kristensen&25}.

\subsection{Challenges and Open Questions}
\label{section:jcpw-challenges}

Despite significant advances, several challenges remain in applying SRE to self-adaptive robotics.  

\subsubsection{Dynamic and Context-Aware Requirements Specification} Self-adaptive robotics requires continuous adaptation to changing environments and tasks, raising the need for context-aware and evolving requirements. Current methods struggle to handle requirements that must be updated at runtime while maintaining system consistency and safety guarantees. How can requirements be dynamically specified, validated, and adapted in real-time?  

\begin{wrapfigure}{r}{0.6\textwidth} 
  \vspace{-10pt} 
  \begin{tcolorbox}[
    colback=blue!5!white,
    colframe=blue!75!black,
    title=Normative requirements,
    fonttitle=\bfseries,
    boxrule=0.8pt,
    arc=4pt,
    left=6pt, right=6pt, top=6pt, bottom=6pt
  ]
Shortage of skilled labour across industries (healthcare, agriculture, logistics, and so on) motivates the deployment of autonomous robots in human environments. To achieve public acceptance, these robots need to satisfy a complex set of normative requirements, encompassing social, legal, ethical, and cultural expectations. There is significant progress~\cite{Bremner2019Proactive,Fjeld2020PrincipledAI} in defining normative concepts~\cite{Inverardi2022HumanDignity}, and a growing number of compliant applications~\cite{Alfieri2022HHAI}. 
The UK Trustworthy Autonomous Systems~(TAS) Resilience project has created a collaborative method~\cite{TPANCCHT22} for defining normative rules for a robot captured in SLEEC (Social, Legal, Ethical, Empathetic, and Cultural), a flexible, highly expressive language, incorporating features of defeasible logic to handle exceptions, and defining time constraints~\cite{YRCCPT25,FMYTCCC23,FMYBA24}. 
Work by others has established a logic semantics for a simplified version of SLEEC and a connection to Prolog~\cite{Troquard2024SLEEC}. Verification efforts have recently focused on the transparency, ethical, and legal aspects of high-level autonomous system designs~\cite{Bhuiyan2020Encoding,Dennis2016FormalVerification}. Separately, studies on multi-robot systems primarily focus on inter-robot trust rather than on the design of individual norm-aware robots~\cite{Frasheri2022CAESAR}. Preliminary work is just beginning to explore the realisability of normative requirements, taking into account factors like the controllability. There are two key challenges:~(\textbf{\mbox{NR-Ch-1}})~assuring that the robot meets all normative requirements upon initial deployment, and continuously satisfies norms while working in uncertain, evolving scenarios; and (\textbf{\mbox{NR-Ch-2}})~providing clear explanations for their norm-sensitive decisions.
\end{tcolorbox}
\vspace{-15pt}
\end{wrapfigure}

\subsubsection{Trade-offs between Conflicting Requirements} Self-adaptive robotic systems must balance conflicting requirements, such as energy efficiency, speed, and safety. Resolving these trade-offs at design time is complex. It becomes even more challenging during runtime adaptations~\cite{Robinson2003,hosseini2023safety}. Developing automated techniques to manage and prioritise requirements trade-offs in self-adaptive scenarios dynamically remains an open question.

\subsubsection{Requirements Traceability and Verification} It is challenging to ensure traceability from high-level requirements to implementation and maintain it across system adaptations. Verification and validation (V\&V) approaches must evolve to provide runtime assurances after adaptation. How can we ensure that traceability and V\&V processes are scalable and efficient in self-adaptive robotics? 

\subsubsection{Stakeholder Involvement and Elicitation for Complex Scenarios} The involvement of diverse stakeholders (roboticists, end-users, and domain experts) makes elicitation challenging. The unpredictable nature of robotics applications, particularly in search and rescue, healthcare, and autonomous driving, complicates requirements gathering. Collaborative approaches and user-centred design methodologies must be extended to better capture evolving requirements.  

\subsubsection{Ethical and Societal Requirements} Self-adaptive robotics brings ethical concerns, including privacy, fairness, and accountability. Defining and enforcing ethical requirements in robotic systems is an emerging research area that requires multidisciplinary collaboration. Questions about operationalising these requirements and integrating them into existing SRE processes remain open.  We refer to the inset for more details and an example.

\subsubsection{Challenges Summary}
Addressing the challenges mentioned above requires new methodologies and tools that combine formal, goal-driven, and adaptive approaches. Integrating run-time monitoring, feedback loops, and automated reasoning into requirements engineering processes will be crucial for advancing the state of the art in this domain. 
In the following, we present a summary of key SRE challenges for SAR.
\revision{The graph in~\Cref{fig:sre_graph} shows the dependencies among the challenges, indicating their priorities and timelines. Stakeholder involvement and elicitation~(SRE-Ch-4) is foundational, since the system cannot manage adaptation responsibly unless it first captures the goals, contexts, and constraints of the relevant actors. This, in turn, enables the systematic incorporation of ethical and societal requirements~(SRE-Ch-5), which should be addressed early because they constrain acceptable adaptation behaviour. Once these foundations are in place, the next priority is managing conflicting requirements~(SRE-Ch-2), followed by traceability and scalable verification and validation~(SRE-Ch-3), which together provide the assurance infrastructure needed for safe runtime requirement updates~(SRE-Ch-1). In parallel, the normative requirements line should progress from ensuring that deployed robots satisfy applicable norms in evolving situations~(NR-Ch-1) to providing explanations for norm-sensitive decisions~(NR-Ch-2). The graph should therefore be read as a flexible roadmap rather than a rigid sequence.}

\begin{tcolorbox}[title=Challenges in Software Requirements Engineering for SAR]
  \textbf{SRE-Ch-1:} Handling runtime requirement updates while ensuring system consistency and safety.\\
  \textbf{SRE-Ch-2:} Balancing conflicting requirements.\\
  \textbf{SRE-Ch-3:} Ensuring traceability of requirements and scalability of verification and validation activities.\\
  \textbf{SRE-Ch-4:} Eliciting requirements due to diverse stakeholders and the nature of robotics applications.\\
  \textbf{SRE-Ch-5:} Integrating ethical requirements into SRE processes. 
\end{tcolorbox}

\begin{figure}[htbp]
    \begin{adjustbox}{max width=\textwidth}
    \begin{tikzpicture}[
        start chain=pos1 going right,   
        start chain=pos2 going right    
    ]

        \node [flowchartNode, on chain=pos1] (s1-4) {SRE-Ch-4};
        \node [flowchartNode, on chain=pos1] (s1-5) {SRE-Ch-5};
        \node [flowchartNode, on chain=pos1] (s1-2) {SRE-Ch-2};
        \node [flowchartNode, on chain=pos1] (s1-3) {SRE-Ch-3};
        \node [flowchartNode, on chain=pos1] (s1-1) {SRE-Ch-1};

        \node [flowchartNode, below=of s1-5] (s2-1) {NR-Ch-1};
        \node [flowchartNode, right=of s2-1]  (s2-2) {NR-Ch-2};

        \draw [flowArrow] (s1-4) -- (s1-5);
        \draw [flowArrow] (s1-5) -- (s1-2);
        \draw [flowArrow] (s1-2) -- (s1-3);
        \draw [flowArrow] (s1-3) -- (s1-1);
        
        \draw [flowArrow] (s2-1) -- (s2-2);

    \end{tikzpicture}
    \end{adjustbox}
    \caption{\revision{Dependency graph for challenges in SRE and normative requirements for SAR.}}
    
    \label{fig:sre_graph}
\end{figure}

\section{Software Design for SAR}\label{sec:design}
In this section, we begin with an overview of the SAR software design and its associated challenges. 
We then discuss key aspects of SAR software design in the context of model-driven engineering and simulation by reviewing literature and highlighting open research challenges. 
Finally, we conclude with a summary of the key insights presented. 

\subsection{Overview and Principles}

Software design for SAR is a multidisciplinary endeavour combining principles from software engineering, robotics, AI, and control systems. Its goal is to provide robust, adaptable, and scalable architectures that support continuous changes in tasks, environments, and configurations while ensuring safety and efficiency.
Several principles underpin design for SAR.

\paragraph{\textbf{Modularity and Scalability}} \revision{Component-based and service-oriented architectures promote modular and scalable development. Middleware such as the Robot Operating System (ROS) and architecture-centric approaches to robotic software provide essential building blocks for integration and evolution across heterogeneous platforms~\cite{Brugali2010,medvidovic2010engineering,alberts2025software}.}

\paragraph{\textbf{Abstraction and Automation}} \revision{Model-driven development (e.g., UML and SysML) facilitates high-level abstraction, analysis, and automated code generation, enabling systematic transitions from design models to implementation~\cite{Fitzgerald2014,Dhouib2012,RRRW15}.}

\paragraph{\textbf{Cross-cutting Concerns}} \revision{Architectural and aspect-oriented techniques help to manage concerns such as fault tolerance, resource management, and assurance obligations without scattering them across multiple modules~\cite{ZH11,medvidovic2010engineering,baxter2025formal}.}

\paragraph{\textbf{Self-Adaptive Architectures}} \revision{Feedback control loops such as MAPE-K provide a principled basis for runtime monitoring, analysis, planning, and reconfiguration, and recent work has begun to adapt these ideas explicitly to trustworthy robotic software~\cite{kephart2003vision,de2013software,larsen2024robotic,nezhad2025towards}. AI-based techniques are increasingly integrated to enhance decision-making and adaptation strategies~\cite{de2013software,alberts2025software}.}

\subsection{Core Challenges in Software Design}

Despite advances in methodologies, SAR software faces several unresolved research questions:
\begin{description}
\item \emph{Scalable and modular architectures.} How can architectures balance flexibility and efficiency while supporting continuous adaptation across diverse robotic platforms?
\item \emph{Trade-offs between adaptability, safety, and performance.} How can systems remain reliable without sacrificing adaptability or degrading performance?
\item \emph{Runtime evolution and reconfiguration.} How can software be safely updated during operation while maintaining consistency and fault tolerance?
\item \emph{Handling uncertainty in AI-based adaptation.} How can design practices mitigate unpredictability when AI is used for adaptation?
\item \emph{Model-driven engineering integration.} How can runtime models remain consistent with design-time models to support lifecycle adaptation?
\item \emph{Human-in-the-loop adaptation.} What frameworks and interfaces best support human-system collaboration while minimising cognitive load?
\item \emph{Verification and validation.} What novel V\&V methods can assure the correctness of highly dynamic and safety-critical self-adaptive software?
\end{description}

In the following, we present a concise description of challenges corresponding to the research questions outlined above. 
\revision{These challenges arise directly from the design principles introduced above. SD-Ch-1 follows from the need for modularity and scalability; SD-Ch-5 from abstraction and automation in model-driven engineering workflows; SD-Ch-3 and SD-Ch-6 from self-adaptive architectures that must support safe runtime reconfiguration and meaningful human oversight; SD-Ch-4 from the growing use of AI within adaptation mechanisms; SD-Ch-2 cuts across all of these principles because architectural decisions must balance adaptability, safety, and performance; and SD-Ch-7 is a cross-cutting assurance challenge that spans them all. The graph in~\Cref{fig:design_graph} shows the dependencies among the challenges, indicating their priorities and timelines. From a timing perspective, scalable and modular architectures~(SD-Ch-1) and the management of trade-offs between adaptability, safety, and performance~(SD-Ch-2) provide the architectural foundation and should be addressed first, and largely in parallel. Once that foundation is in place, three strands can advance: maintaining consistency between runtime and design-time models~(SD-Ch-5), handling uncertainty in AI-based adaptation~(SD-Ch-4), and strengthening verification and validation strategies~(SD-Ch-7). Progress on runtime-model consistency then enables safer runtime evolution and reconfiguration~(SD-Ch-3). Safe reconfiguration and principled handling of AI uncertainty together support effective human-in-the-loop adaptation~(SD-Ch-6). The graph should therefore be read as a flexible roadmap rather than a rigid sequence: early work should focus on architectural structure and design trade-offs, mid-term work on model consistency and AI uncertainty, and later integrated work on safe reconfiguration, human oversight, and system-level assurance.}
\begin{tcolorbox}[title=Challenges in Software Design for SAR]
  \textbf{SD-Ch-1:} Scalable and modular architectures that balance flexibility and efficiency while supporting continuous adaptation across diverse platforms.\\
  \textbf{SD-Ch-2:} Managing trade-offs between adaptability, safety, and performance so that reliability is preserved without degrading system behaviour.\\
  \textbf{SD-Ch-3:} Safe runtime evolution and reconfiguration with consistency and fault tolerance during software updates in operation.\\
  \textbf{SD-Ch-4:} Handling uncertainty introduced by AI-driven adaptation, including mitigating unpredictability and ensuring trustworthiness.\\
  \textbf{SD-Ch-5:} Keeping runtime models consistent with design-time models to support lifecycle adaptation in MDE workflows. \\
  \textbf{SD-Ch-6:} Human-in-the-loop adaptation with interfaces and processes that minimise cognitive load and enable effective collaboration.\\
  \textbf{SD-Ch-7:} Verification and validation techniques suited to highly dynamic and safety-critical self-adaptive software. 
\end{tcolorbox}

\begin{figure}[htbp]
    \begin{adjustbox}{max width=\textwidth}
    \begin{tikzpicture}[
        start chain=going right 
    ]
        
        \node[flowchartNode] (n1) {SD-Ch-1};
        \node[flowchartNode] (n2) [right=of n1] {SD-Ch-2};
        
        \node[flowchartNode] (n4) [below=of n1] {SD-Ch-4}; 
        \node[flowchartNode] (n5) [left=of n4] {SD-Ch-5};
        \node[flowchartNode] (n7) [right=of n4] {SD-Ch-7};
        
        \node[flowchartNode] (n3) [below=of n5] {SD-Ch-3};
        \node[flowchartNode] (n6) [below=of n4] {SD-Ch-6};

        \draw[flowArrow, <->] (n1) -- (n2);

        \draw[flowArrow] (n1) -- (n5);
        \draw[flowArrow] (n1) -- (n4);
        \draw[flowArrow] (n1) -- (n7);

        \draw[flowArrow] (n5) -- (n3);
        \draw[flowArrow] (n3) -- (n6);
        \draw[flowArrow] (n4) -- (n6);

        \draw[flowArrow, <->] (n6) -- (n7);

    \end{tikzpicture}
    \end{adjustbox}
    \caption{\revision{Dependency graph for challenges in software design for SAR.}}
    
    \label{fig:design_graph}
\end{figure}

\subsection{Model-Driven Engineering for SAR}
\subsubsection{Foundations and Related Research}
Model-Driven Engineering (MDE) has been successfully applied in domains such as aerospace and automotive, and offers strong potential for SAR. By treating abstract models as central artefacts, MDE supports requirements clarification, early validation, and automated code generation.

\paragraph{Domain-Specific Languages (DSLs).} Numerous DSLs have been proposed for robotics, many with formal semantics enabling reasoning. Examples include UML derivatives such as mUML~\cite{GS13}, UML profiles for human--robot collaboration~\cite{ALLIRV21}, SysML with CSP-like semantics~\cite{LMCCISHLL15}, AADL with rewriting logic~\cite{OBM10}, RobotML~\cite{Dhouib2012}, SafeRobots~\cite{RMT14}, and RoboChart~\cite{hierons2021mutation}. Frameworks such as MontiArcAutomaton~\cite{RRRW15} integrate multiple modelling languages and generators for heterogeneous platforms.

\paragraph{Verification Support.} Several approaches incorporate formal verification, including contract languages and temporal logic (Mauve~\cite{GLD14}, Orccad~\cite{kortenkamp2016robotic}, SPECTRA~\cite{SPECTRA}), theorem proving (AutoFocus and Isabelle/HOL transformations~\cite{SHT2012}), and model checking (GenoM with Petri Nets~\cite{Foughali2016}, JavaPathFinder~\cite{miyazawa2025diagrammatic}). RoboStar~\cite{cavalcanti2021robostar} provides domain-specific notations and transformations to support modelling, simulation, testing, and verification.

\subsubsection{Challenges and Open Questions}
There are, however, many open challenges in the effective use of these notations for MDE. That requires automation based on sound techniques that can ensure the values of the derived artefacts and compensate for the effort required to develop models. 
Some open problems are as follows.

    \paragraph{\textbf{Formalisation of Architectures for Adaptation}} There is no standard or formal understanding of what it means for a self-adaptive system to adopt the MAPE-K architecture. This impacts the traceability between the code and the MAPE-K design of an adaptive system. It also makes it difficult to exploit the features of the MAPE-K architecture at the code level.
    
    \paragraph{\textbf{Support for Code Generation}} The lack of MAPE-K formalisation leads to the lack of customised facilities to model and generate code that adopts that architecture. While general modelling languages and MDE tools can be employed, none provides domain-specific support, i.e., no native support to ensure the communication patterns and control flow of MAPE-K, for example, are enforced at either the model or coding levels. 
    
    \paragraph{\textbf{Hybrid Reasoning}} Adaptation is often a response to robotic hardware and environment changes, which are often better described via hybrid models. While MBE is often focused on the software component, we need to employ hybrid models to generate simulations, identify meaningful tests, and reason about adaptation in self-adaptive robotics. There is a rich literature on hybrid model checkers and theorem provers. However, a practical approach capable of handling the scale and complexity of robotic models remains missing.
    
    \paragraph{\textbf{Integration with AI}} MBE techniques are often component-based to a certain level, with at least some structure in the models to reflect commonly used architectures or computational resources. Extensions of the notations and techniques for dealing with AI components are crucial to enabling the system-level approach necessary for adaptation. 
    
    \paragraph{\textbf{Human Factors}} Again, with the observation that adaptation is a system-level concern, we have to consider the impact of human factors. There are two perspectives on human behaviour, which lead to a requirement for adaptation and to an assessment of the impact of adaptation on human stakeholders. Much work on ethical factors is being undertaken, but we require full operationalisation to support the deployment of suitable adaptations.

\paragraph{\textbf{Challenges Summary}}
In the following, we outline key challenges in MDE for SAR.
\revision{The graph in~\Cref{fig:mde_graph} shows the dependencies among the challenges, indicating their priorities and timelines. A standard architecture~(MDE-Ch-1) is a key enabling result. It is more useful if it can support code generation~(MDE-Ch-2), hybrid reasoning~(MDE-Ch-3), support for AI~(MDE-Ch-4), and consideration of human factors and ethics~(MDE-Ch-5). Conversely, the existence of a standard architecture can enable (compositional) reasoning approaches~(MDE-Ch-3).  In addition, code generation is more~(MDE-Ch-2) if it can cater for AI techniques~(MDE-Ch-4). 
}
\begin{tcolorbox}[title=Challenges in MDE for SAR]
  \textbf{MDE-Ch-1:} Lack of a formal standard for self-adaptive architectures (e.g., MAPE-K), hindering traceability between models and code. \\
  \textbf{MDE-Ch-2:} Limited support for automated code generation that reflects adaptation-specific semantics.\\
  \textbf{MDE-Ch-3:} Need for hybrid reasoning that combines software with models of hardware and environment.\\
  \textbf{MDE-Ch-4:} Insufficient integration of AI components into MDE frameworks.\\
  \textbf{MDE-Ch-5:} Limited treatment of human and ethical factors in adaptation decisions.
\end{tcolorbox}

\begin{figure}[htbp]
    \begin{adjustbox}{max width=\textwidth}
    \begin{tikzpicture}
        \node[flowchartNode] (n1) at (0,0) {MDE-Ch-1};
        
        \node[flowchartNode, above=of n1] (n4) {MDE-Ch-4};
        
        \node[flowchartNode, left=of n4] (n2) {MDE-Ch-2};
        
        \node[flowchartNode, right=of n4] (n5) {MDE-Ch-5};
        
        \node[flowchartNode, right=of n1] (n3) {MDE-Ch-3};

        \draw[flowArrow] (n4) -- (n2);
        
        \draw[flowArrow] (n2.south) -- (n1.west);
        \draw[flowArrow] (n4) -- (n1);
        \draw[flowArrow] (n5.south) -- (n1.north east);
        
        \draw[flowArrow, <->] (n1) -- (n3);

    \end{tikzpicture}
    \end{adjustbox}
    \caption{\revision{Dependency graph for challenges in MDE for SAR.}}
    
    \label{fig:mde_graph}
\end{figure}

\subsection{Simulation Approaches for SAR}
\subsubsection{Foundations and Related Research}
Simulation offers a risk-free and cost-effective means of validating designs and exploring adaptation strategies before deployment.  

\paragraph{Foundations.} A model is a representation of a system capturing relevant properties. Simulations produce traces of behaviour under varying conditions, with fidelity, substitutability, and verifiability determining their usefulness~\cite{Kuhne2005a}.

\paragraph{Formal Simulation Approaches.} ActivFORMS~\cite{Weyns2023} employs timed automata to model MAPE-K loops, supporting runtime verification and execution of adaptation logic. Digital twin frameworks integrate simulations into monitoring and planning~\cite{Feng2022a}, enabling predictive what-if analyses and reachability-based verification~\cite{Wright2022}. Other approaches include agent-based models~\cite{Loo2021} and adaptive Abstract State Machines (ASMs)~\cite{Arcaini2017}, which support distributed adaptation and emergent behaviours.

\paragraph{Co-simulation.} Heterogeneous subsystem models can be coupled using frameworks such as FMI~\cite{Junghanns2021}, HLA~\cite{Nagele2017}, and DEVS~\cite{blas2022devs}. Orchestration algorithms synchronise simulators, but challenges remain in accurately capturing reconfiguration dynamics and ensuring intellectual property protection.

\subsubsection{Challenges and Open Questions}
Despite the tremendous work by the community, challenges and open research questions remain in simulating self-adaptive systems. 

\paragraph{\textbf{Adapting Simulations}} Simulation of self-adaptive systems requires models that have a broad range of scenarios where the model is relevant. Typically, this is achieved by selecting the most appropriate model for any given scenario \cite{Biglari2022}. This highlights the challenge of representing the range of valid scenarios a model can be used in, in a compact format, that is \emph{shipped} alongside the model. The representation of this range must be machine-readable and highly expressive, as it can vary from parameter and initial condition ranges to more dynamic descriptions, such as the maximum frequency of model inputs. So far, no universal description has emerged in the state of the art. 

\paragraph{\textbf{Coupling Simulations}} In modern supply chains, a self-adaptive system is formed by subsystems provided by external suppliers \cite{feng2021developing}. This leads to difficulties in simulating the entire system using a single formalism, such as an ASM. An existing solution to couple heterogeneous system models is co-simulation \cite{Gomes2018}. In this technique, subsystem simulators are heterogeneous, but they implement the same interface. This allows the coupling of these simulators to form a composite simulation, but it has not been widely used in the context of self-adaptive system simulation.

\paragraph{\textbf{Orchestration Challenge in Co-simulation}} There are multiple interfaces that sub-simulations can implement, such as the Functional Mock-up Interface (FMI~\cite{Junghanns2021}), High-Level Architecture (HLA~\cite{Nagele2017}), DEVS~\cite{blas2022devs}, and others. When the subsystems implement a standard interface, co-simulation is achieved by intentionally stepping each simulator forward in simulated time while exchanging data between each step with the other simulators~\cite{Kubler2000}. The algorithm that manages this process is called the orchestration algorithm. Orchestration algorithms proposed in the state of the art range from simple and general~\cite{Bastian2011} to complex and highly specialised~\cite{Inci2021}, exhibiting self-adaptive properties~\cite {Inci2023}. 
Co-simulation of self-adaptive robotics requires orchestration algorithms that differ from traditional ones to ensure that system reconfigurations are accurately captured in the simulation dynamics, including dynamically changing simulator dependencies~\cite{ejersbo2023dynamic,barros2024pi}. This is an open area of research.

\paragraph{\textbf{Limitations of Co-simulation Interfaces}} Co-simulation interfaces may not be rich enough to enable the orchestrator to capture dynamics correctly. For instance, traditional co-simulation of multi-rate systems is challenging in FMI version 2.0 \cite{Blochwitz2012}. Recently, FMI 3.0 has mitigated this issue by introducing the notions of clocks and clocked variables~\cite{Gomes2021a}. Conversely, error-free simulation of continuous dynamics in discrete-event simulation interfaces (such as DEVS and HLA) is impossible to achieve due to quantisation \cite{blas2022devs}.

\paragraph{\textbf{Intellectual Property Protection in Simulation}} In modern supply chains, there is a need to protect the intellectual property (IP) of subsystems, as external companies often provide these. To avoid expensive contracts, the co-simulation interface must promote IP protection. The question remains whether traditional IP protection mechanisms in existing co-simulation interfaces are sufficient for simulating self-adaptive systems.

\paragraph{\textbf{Challenges Summary}}
A summarised list of key challenges related to simulations in SAR is given below.
\revision{The graph in~\Cref{fig:simulations_graph} highlights a tightly coupled core formed by heterogeneous coupling~(Sim-Ch-2), orchestration~(Sim-Ch-3), and interface limitations~(Sim-Ch-4). Progress in any one of these areas depends on and reinforces progress in the others: richer interfaces enable better orchestration, while the needs of orchestration and coupling expose limitations in current interfaces. Addressing this core also enables more practical support for compact machine-readable scenario ranges~(Sim-Ch-1) and stronger IP protection mechanisms~(Sim-Ch-5) for supplier-provided subsystem models.}
\begin{tcolorbox}[title=Challenges in Simulations for SAR]
  \textbf{Sim-Ch-1:} Compact representation of scenario ranges for adaptive simulations. \\
  \textbf{Sim-Ch-2:} Coupling heterogeneous subsystem simulations to form a composite simulation.\\
  \textbf{Sim-Ch-3:} Orchestration algorithms that can dynamically adapt to system reconfiguration.\\
  \textbf{Sim-Ch-4:} Co-simulation interfaces often lack the richness required to capture multi-rate and hybrid dynamics.\\
  \textbf{Sim-Ch-5:} Protection of intellectual property when coupling externally provided subsystem models.
\end{tcolorbox}

\begin{figure}[htbp]

    \begin{adjustbox}{max width=\textwidth}
    \begin{tikzpicture}
        \node[flowchartNode] (n2) {Sim-Ch-2};
        \node[flowchartNode] (n3) [right=of n2, xshift=1cm] {Sim-Ch-3};
        \node[flowchartNode] (n4) [below=of n2, xshift=2.0cm] {Sim-Ch-4};

        \node[draw=steelblue, thick, rounded corners=20pt, 
              inner sep=15pt, fit=(n2) (n3) (n4)] (cluster) {};

        \node[flowchartNode, right=of cluster, yshift=0.8cm] (n5) {Sim-Ch-5};
        \node[flowchartNode, right=of cluster, yshift=-0.8cm] (n1) {Sim-Ch-1};

        \draw[flowArrow, <->] (n2) -- (n3);
        \draw[flowArrow, <->] (n2) -- (n4);
        \draw[flowArrow, <->] (n3) -- (n4);

        \draw[flowArrow] (cluster.east |- n5) -- (n5.west);
        \draw[flowArrow] (cluster.east |- n1) -- (n1.west);

    \end{tikzpicture}
    \end{adjustbox}
    \caption{\revision{Dependency graph for challenges in simulations for SAR.}}
    
    \label{fig:simulations_graph}
\end{figure}

\subsection{Summary}

Software design for SAR must integrate architectural principles, MDE, and simulation to achieve adaptability without compromising safety or performance. The open challenges (SD-Ch-1 to SD-Ch-7) highlight the need for scalable architectures, robust integration of AI and human input, and advanced V\&V methods. Bridging MDE and simulation approaches---while extending them to handle hybrid models, AI, and ethical concerns---remains a critical research direction.

\section{Software Development of SAR}\label{sec:development} 
In this section, we begin with an overview of research on the development of SAR software. 
Following this, we outline open challenges that need to be addressed.

\subsection{Foundations and Related Research}
Software development for SAR focuses on implementing advanced self-adaptation mechanisms, such as MAPE-K/MAPLE-K loops, to enable robots to autonomously adjust their behaviour, configurations, and decision-making processes in response to dynamic and uncertain environments~\cite{baxter2025formal}.
Given the inherent complexity of SAR systems, modularity, scalability, and maintainability are crucial design principles to ensure seamless integration across subsystems and with evolving requirements. 
Therefore, the development of SAR software is closely aligned with design models and architectures such as AADL or RoboArch~\cite{baxter2025formal}, which provide systematic abstractions from high-level designs to implementation. 
Additional considerations include integrating advanced frameworks and technologies, such as machine learning and data analysis, to support real-time decision-making, interactions in uncertain environments, and collaboration with humans or other robots.

A common method for developing SAR software is to utilise the models and architectures created in the design phase~\cite{alberts2025software}. 
In this regard, several techniques have been proposed in the literature targeting different aspects of self-adaptation. 
\citet{edwards2009architecture} employed high-level architectural components to develop self-management capabilities, including monitoring and self-adaptation. 
To improve runtime reconfigurations, \citet{bao2015architecture} proposed a model-based approach that employs meta-models of a robotic system to generate online configurations.
Moreover, \citet{rivera2021software} presented an initial effort on developing self-recovering and self-healing capabilities in robotic software.

To support automation in SAR software development, several studies have focused on using design models, such as meta-models and runtime models, for automatic code generation~\cite{ingles2010using,ingles2011towards}.
Similarly, \citet{medvidovic2010engineering} presented RoboPrism, a framework for architecture-driven development of robotic software. 
Focusing on uncertainty as a key concern, \citet{alberts2023development} introduced a reasoning-based approach for implementing and integrating uncertainty-aware adaptation strategies. 
\citet{silva2023suave} presented SUAVE, a framework for developing self-adaptation in autonomous underwater vehicles (UUVs) considering uncertainties.  
In a recent work, \citet{nezhad2025towards} presented a framework for model-based development of the MAPE-K loop that accounts for the trustworthiness of robotic software. 

\subsection{Challenges and Open Questions}
Despite significant advances in SAR software development, several challenges remain to be addressed. 

\subsubsection{Interfacing with Heterogeneous Components} 
Self-adaptive robots often rely on a wide range of hardware components (e.g., sensors, actuators, cameras, and LIDAR) and software platforms (e.g., control systems, communication protocols, and AI models). 
These components are typically developed by different vendors using varying standards, making seamless integration a complex task. 
For example, a robotic system may have an LIDAR sensor that requires a proprietary driver and actuators that rely on custom firmware. 
Developing software that effectively interfaces with heterogeneous components and platforms, ensuring seamless communication, compatibility, and adaptability, remains a critical challenge. 

\subsubsection{Rapidly Changing Hardware and Software} 
Both the hardware and software of robots continuously evolve at different paces, with robotic software upgrades occurring more frequently than hardware upgrades. 
To keep up with adaptations to different environments, AI components need to evolve, which requires retraining or fine-tuning of AI models. 
Similarly, whenever hardware components are upgraded or replaced, the corresponding software must be changed to accommodate hardware advancements. 
Thus, supporting the continuous integration of new/updated hardware and software into self-adaptive robotic systems without requiring a complete overhaul of the existing software architecture is a major challenge.

\subsubsection{Sensor Fusion for Adaptation} 
The software components for data analysis rely on data from multiple sensors, such as cameras, LIDAR, and GPS, to perceive the robots' operating environment and decide actions. 
However, these sensors often provide data in different formats, sampling rates, accuracies, and noise levels. 
Misaligned or inconsistent sensor data can lead to incorrect decisions, such as failing to detect obstacles, which can in turn result in unsafe actions. 
Developing software capable of real-time sensor fusion to integrate data from various sensors while ensuring reliability and accuracy remains an open challenge. 
Given the recent rise of large language models with advanced analytical capabilities, one potential direction is to explore their applicability to real-time data fusion and analysis. 

\subsubsection{Energy-Efficient Software Development} 
Developing intelligent software components commonly involves developing AI models for various types of decision-making. 
For example, the perception module requires training AI models on image/video data. 
As the robot operating environment changes, AI models need to be retrained on the new imagery data. 
Furthermore, robotic software involves several computationally expensive algorithms, such as path planning and adaptive navigation, to run during the robot's operations in real environments. 
Developing energy-efficient robotic software that minimises computational overhead while maintaining the robot’s ability to adapt in real-time is a critical challenge.

\subsubsection{Ethical and Security Concerns} 
As self-adaptive robots operate autonomously and interact with humans, they raise significant ethical and security concerns. 
For example, a caregiver robot that assists older people must ensure privacy while handling their personal data, such as health records or behavioural patterns. 
Such concerns, although specified in software requirements, are critical from an implementation perspective. 
For example, developing stronger security and privacy mechanisms to make software more resilient to cyberattacks, hijacking, or other malicious attacks. 
Therefore, developing robotic software capable of handling unknown and dynamic security and privacy concerns at runtime remains an open challenge.

\subsubsection{Challenges Summary}
In the following sections, we outline the challenges associated with SAR software development. 
\revision{
The graph in~\Cref{fig:development_graph} illustrates the dependencies among the challenges, indicating their relative priorities and timelines. 
At the foundation, sensor fusion (Dev-Ch-3), sustainable robotic software (Dev-Ch-4), and resilience against cyber attacks (Dev-Ch-5) are interdependent and should be addressed concurrently, with Dev-Ch-5 having the highest priority among the three. 
Once these foundational challenges are resolved, the next priority is to address hardware/software integration (Dev-Ch-2). 
Finally, implementing adaptations and evolution (Dev-Ch-1) has comparatively lower criticality and can be addressed in the later stages of development.  
}
\begin{tcolorbox}[title=Challenges in Software Development of SAR]
  \textbf{Dev-Ch-1:} Ensuring effective integration and compatibility between various hardware and software components from different vendors.\\
  \textbf{Dev-Ch-2:} Adapting to continuously evolving hardware and software without rebuilding existing architectures.\\
  \textbf{Dev-Ch-3:} Developing real-time sensor fusion to integrate diverse sensor data.\\
  \textbf{Dev-Ch-4:} Developing robotic software in an energy-efficient manner and providing sustainable operations in real environments. \\
  \textbf{Dev-Ch-5:} Ensuring privacy, data security, and robustness against cyberattacks at runtime. 
\end{tcolorbox}

\begin{figure}[htbp]
    \begin{adjustbox}{max width=\textwidth}
    \begin{tikzpicture}
        \node[flowchartNode] (n5) {Dev-Ch-5};
        
        \node[flowchartNode, above right=of n5, yshift=-0.3cm] (n3) {Dev-Ch-3};
        \node[flowchartNode, below right=of n5, yshift=0.3cm] (n4) {Dev-Ch-4};
        
        \node[flowchartNode, below right=of n3, yshift=0.3cm] (n1) {Dev-Ch-1};
        
        \node[flowchartNode, right=of n1] (n2) {Dev-Ch-2};

        \draw[flowArrow, <->] (n5) -- (n3);
        \draw[flowArrow, <->] (n5) -- (n4);
        
        \draw[flowArrow] (n3) -- (n1);
        \draw[flowArrow] (n4) -- (n1);
        
        \draw[flowArrow] (n1) -- (n2);

    \end{tikzpicture}
    \end{adjustbox}
    \caption{\revision{Dependency graph for challenges in software development of SAR.}}
    
    \label{fig:development_graph}
\end{figure}

\section{Software Testing of SAR}\label{sec:testing} 
This section begins with an overview of research on software testing in SAR. Then it highlights a set of open challenges that remain to be addressed.

\subsection{Foundations and Related Research}
Testing plays a key role in ensuring that self-adaptive robots function as intended while meeting all safety and dependability requirements. 
However, the dynamic and autonomous nature of these systems makes testing significantly challenging. 
Therefore, SAR system testing is conducted in different setups: software-in-the-loop (SiL), which evaluates robot software components within a simulated environment; model-in-the-loop (MiL), which verifies robot control logic through simulation models; and hardware-in-the-loop (HiL), which tests interaction between software and hardware by incorporating hardware components, such as sensors and actuators, into the testing process. 
Moreover, operational testing with physical robots in a controlled environment is often performed, typically supported by advanced technologies such as digital twins, which allow real-time monitoring and validation of the robot's performance.

Many existing works focus on testing various aspects of robots, with model-driven approaches being prevalent. 
One such framework, RoboStar, provides a model-based solution for automatic test generation for autonomous robots~\cite{cavalcanti2021robostar}. 
Another approach utilises mutation testing to automatically generate tests by mutating RoboChart models~\cite{hierons2021mutation}. 
For uncrewed aerial vehicles (UAVs), a model-based approach has been proposed to automate the testing of interactive remote control systems~\cite{sartaj2021testing}. 
Similarly, a system-level testing framework, AITester, combines model-driven techniques with reinforcement learning to enable automatic and dynamic testing of unmanned aerial systems~\cite{sartaj2021automated,sartaj2024automated}. 
In addition to model-driven approaches, search-based test generation has been applied to test autonomous systems using requirements and constraint specifications~\cite{adigun2023risk,sartaj2025search}. Furthermore, field-based testing techniques for robots, such as in-house, online, and offline field testing, are also common~\cite{caldas2024runtime}.

When it comes to testing the self-adaptive behaviour of robots, only a limited number of works are available in the literature. In this context, a SiL testing framework powered by machine-learning techniques has been proposed to test the behaviour of autonomous systems, such as UUVs~\cite{mullins2018adaptive}. 
To test self-healing behaviour, a model-based framework was introduced to automate testing of autonomous CPS under uncertainty~\cite {ma2019modeling}. 
Subsequently, reinforcement learning algorithms were evaluated to test the self-healing behaviour of various autonomous vehicles, including UAVs and mobile robots~\cite{ma2021testing}.
In a similar line of work, a model-based approach was proposed to test the self-adaptive behaviour of UAVs~\cite{javed2024automated}. 
Building on this, a hybrid approach was introduced that combines model-based and search-based methods to test self-adaptive UAVs~\cite{javed2025hybrid}. 
Recently, an approach has been developed for testing autonomous mobile robots that uses robot requirements and vision language models to generate test scenarios~\cite{wu2025vision}. 
Another direction of research focuses on the use of digital twins for verifying robot behaviour at runtime. 
For instance, digital twins have been employed to monitor and verify the behaviour of mobile robots under uncertainty~\cite{betzer2024digital}. 
Similarly, a digital twin-based approach has been introduced for real-time out-of-distribution detection in self-adaptive robots, including autonomous vessels and service robots~\cite{isaku2025digital,isaku2025oodisar}. 
Furthermore, an initial AI-based digital twin framework has been introduced to enable task verification for medical robots~\cite{mania2025towards}.  

\subsection{Challenges and Open Questions} 
Although considerable progress has been made in SAR software testing, several challenges remain to be addressed.

\subsubsection{Simulation-based Testing in Varied Reality Gap}  A key challenge in simulation-based testing (e.g., in a SiL setup) is the reality gap—the discrepancy between simulation and real-world conditions. This makes realistic testing in simulation difficult, as test scenarios may not accurately reflect real-world situations (e.g., weather changes). Highly realistic simulations can be costly, yet full realism may not always be necessary. This raises a key question: What level of realism strikes the balance between effectiveness and cost? 
\revision{
In practice, this question is often addressed through a structured progression of simulation and testing environments, such as MiL simulation for rapid development in early stages, SiL simulation for integration testing, and HiL simulation for testing with real hardware components~\cite{cheng2024survey}. 
Although such progressions are effective, transferring behaviours, scenarios, and test artefacts across these stages remains challenging due to differing assumptions, sensor characteristics, and timing behaviour~\cite{tobin2017domain,cai2021modular}. 
This presents an opportunity to exploit AI-based techniques to relate and adapt test scenarios, models, and observed behaviours across multiple stages of the testing pipeline. Specifically, AI foundation models with multimodal capabilities (e.g., vision-language models) offer promising ways to enhance simulation realism and improve testing effectiveness. 
}

\subsubsection{Testing AI and non-AI Components} 
AI is increasingly integrated into self-adaptive robotic software (including MAPE-K) for tasks such as object identification and autonomous decision-making. However, this introduces testing challenges for both AI and non-AI components independently and due to their interactions, including uncertainty, lack of explainability, and ethical concerns. Effective testing methodologies must address these issues.

\subsubsection{Testing under Uncertainty} Uncertainty is inherent in self-adaptive software. Thus, testing methods must treat uncertainty as a first-class entity to identify faults effectively. Holistic approaches are needed to quantify uncertainty across different sources, including AI components. Furthermore, continuous assessment and management of uncertainty in real-world operations are crucial, along with strategies for handling cases where uncertainty exceeds acceptable limits, to ensure system reliability and safety.

\subsubsection{Testing Self-Adaptive MAPE-K Loop} Self-adaptation, often via the MAPE-K loop, introduces testing challenges. New methods are needed to test each component, such as whether Analyse triggers adaptations, Plan generates appropriate strategies, and adaptations execute correctly. Testing must assess both individual components in MAPE-K and the entire loop to ensure overall correctness.

\subsubsection{Real-Time Testing through Digital Twins} Digital twins are becoming increasingly prevalent in robotics. Our focus here is on testing challenges; more details on digital twins in the context of self-adaptive robotics are provided in~\Cref{subsec:dt}. Whether built using model-driven approaches, AI, or a combination of both, digital twins provide a continuous means of assessing and testing self-adaptive robots, helping prevent failures in real-world operations. However, there are several testing challenges. One key challenge is enabling effective real-time testing while accounting for potential synchronisation delays between the physical system and its digital twin. Such delays are quite common due to communication issues. Within digital twins, there is also a need for validation methods (e.g., AI-based) that can predict or forecast potential failures, enabling the digital twin to prevent them before they occur.  

\subsubsection{Testing under Continuous Evolution} Self-adaptive robotic software evolves continuously, requiring cost-effective testing of modified parts while ensuring existing functionality remains intact. This requires new regression testing methods for the MAPE-K loop and AI components to support continuous evolution.

\subsubsection{Testing Robotic Foundation Models}
Recent advancements in foundation models, such as large language models (LLMs), have led to the development of robotic foundation models. 
These models include LLMs that are specifically fine-tuned for robotic tasks, as well as vision-language-action (VLA) models designed for various types of robots. 
However, this paradigm shift in robotics also introduces several critical challenges that need to be addressed. 
A key challenge in robotic foundation models is hallucinations, which pose a significant safety risk in robotics as they can lead to incorrect or unsafe behaviours in real-world environments~\cite{xiao2025robot}. 
Developing testing techniques to assess the robustness of robotic foundation models remains an open challenge. 
Another challenge is inherent nondeterminism and output variability in robotic foundation models, which lead to inconsistent actions and decision-making~\cite{schreiter2025evaluating}. 
Moreover, these models lack grounding, i.e., a limited understanding of real-world and dynamic contexts, which is crucial for effectively controlling robots in complex environments~\cite{greengard2025can}.  
Therefore, the development of reliability testing techniques for robotic foundation models, especially when integrated with physical robots, remains a critical and open research problem.

\subsubsection{Security Testing of Self-adaptive Robotics}
With the increasing integration of AI, especially LLMs in robotics, several new types of vulnerabilities and security threats are emerging~\cite{wu2024safety}. 
These include prompt-based attacks on LLMs and perception-based attacks on vision-language models (VLMs), intended to manipulate the robot's decision-making. 
Although recent efforts have focused on addressing these attacks and mitigating associated risks~\cite{wang2025gsce,zhang2025enhancing}, developing effective security testing techniques to ensure robots' resilience against adversarial threats remains an open research challenge. 

\subsubsection{Challenges Summary}
In the following, we present a summary of key challenges in testing SAR software.
\revision{The graph in~\Cref{fig:testing_graph} illustrates the dependencies among the testing challenges, indicating their relative priorities and timelines. 
A fundamental challenge is ensuring the correctness of MAPE-K (ST-Ch-4), which must be addressed first. 
Subsequently, resilience against adversarial threats (ST-Ch-8), testing AI and non-AI component interactions (ST-Ch-2), and testing under uncertainty (ST-Ch-3) can then be tackled in parallel, with ST-Ch-2 and ST-Ch-3 being interdependent and ST-Ch-2 relying on ST-Ch-8. 
Once ST-Ch-2 and ST-Ch-3 are addressed, they collectively form the basis for addressing robustness and reliability (ST-Ch-7), which is the next priority. 
Following this, bridging the reality gap (ST-Ch-1) and testing digital twins (ST-Ch-5) can be pursued concurrently, as the two are interdependent. 
Finally, regression testing (ST-Ch-6) represents the last challenge to be resolved once all preceding priorities have been addressed. 
}
\begin{tcolorbox}[title=Challenges in Software Testing for SAR]
  \textbf{ST-Ch-1:} Bridging the reality gap in simulation-based testing to ensure realistic yet cost-effective testing.\\
  \textbf{ST-Ch-2:} Testing the interactions between AI and non-AI components, while considering uncertainty and ethical concerns.\\
  \textbf{ST-Ch-3:} Testing SAR software in the presence of uncertainty. \\
  \textbf{ST-Ch-4:} Ensuring the correctness of individual MAPE-K components and the overall loop during testing.\\
  \textbf{ST-Ch-5:} Real-time testing of digital twins, while accounting for synchronisation delays with physical systems.\\
  \textbf{ST-Ch-6:} Developing cost-effective regression testing methods to support the continuous evolution of SAR. \\
  \textbf{ST-Ch-7:} Testing robustness and reliability of robotic foundation models. \\
  \textbf{ST-Ch-8:} Ensuring resilience against adversarial threats, such as prompt-based and perception-based attacks. 
\end{tcolorbox}

\begin{figure}[htbp]
    \begin{adjustbox}{max width=\textwidth}
    \begin{tikzpicture}[
        node distance=1.2cm and 1.5cm 
    ]
        \node[flowchartNode] (n4) {ST-Ch-4};

        \node[flowchartNode, right=of n4] (n2) {ST-Ch-2};
        \node[flowchartNode, above=of n2] (n8) {ST-Ch-8};
        \node[flowchartNode, below=of n2] (n3) {ST-Ch-3};

        \node[flowchartNode, right=of n2] (n7) {ST-Ch-7};

        \node[flowchartNode, right=of n7, yshift=0.8cm] (n1) {ST-Ch-1};
        \node[flowchartNode, right=of n7, yshift=-0.8cm] (n5) {ST-Ch-5};

        \node[flowchartNode, right=of n5, xshift=0.5cm, yshift=0.8cm] (n6) {ST-Ch-6};

        \draw[flowArrow] (n4.east) -- (n2.west);
        \draw[flowArrow] (n4.east) -- (n3.west);

        \draw[flowArrow] (n8) -- (n2);
        \draw[flowArrow, <->] (n2) -- (n3);

        \draw[flowArrow] (n2.east) -- (n7.west);
        \draw[flowArrow] (n3.east) -- (n7.west);

        \draw[flowArrow] (n7.east) -- (n1.west);
        \draw[flowArrow] (n7.east) -- (n5.west);

        \draw[flowArrow, <->] (n1) -- (n5);
        \draw[flowArrow] (n1.east) -- (n6.west);
        \draw[flowArrow] (n5.east) -- (n6.west);

    \end{tikzpicture}
    \end{adjustbox}
    \caption{\revision{Dependency graph for challenges in software testing for SAR.}}
    
    \label{fig:testing_graph}
\end{figure}

\section{Operations for SAR}\label{sec:operations} 
This section starts by reviewing research contributions in the field of operations for SAR. 
Next, it presents key open challenges that require further exploration. 

\subsection{Foundations and Related Research}
Many recent works focus on the operation of SARs. These works can be classified by the main artefacts used to trigger and plan adaptations: Models/Data, Non-functional Requirements, and Assurance cases.

\paragraph{Models/Data.} Many approaches rely on models and data to monitor the system and its environment and to trigger and plan adaptations. \Citet{Bencomo2019} provides an extensive survey of 275 papers, providing a taxonomy to classify work along dimensions such as modelled artefacts, runtime model types, purposes (e.g., self-adaptation, assurance), techniques (e.g., model transformation, reflection, reasoning), application domains, and intersecting research areas.

Concrete examples of contributions in this category include \citet{Aldrich2019} and \citet{Feng2022a}. 
In \citet{Aldrich2019}, the authors present an approach to making robotic software long-lived and more adaptable by leveraging formal models of software and its environment to enable automated adaptation. They argue that today's robotic systems require extensive manual effort to handle hardware upgrades, software changes, and environmental variations, which limits their long-term deployment. To address this, they propose intelligent model-based adaptation, where models capture architecture, behaviour, utility, and environment, and are used at runtime to detect when adaptation is needed, explore alternative configurations, and automatically select and enact suitable adaptations. Their work encompasses techniques such as sensitivity analysis for performance/resource trade-offs, state-transition repair using SMT-based solving with human feedback, task-switching policies for dynamic environments, architectural adaptation that integrates ROS-based components, multi-model integration with probabilistic planning, and automatic code repair using tools like Houston.

Similarly, \citet{Feng2022a} demonstrates how the MAPE-K feedback loop can be integrated into a Digital Twin (DT) to enable self-adaptation in CPS. Using a thermal incubator case study, they designed a DT with services for monitoring, anomaly detection (via a Kalman filter), recalibration, controller optimisation, and reconfiguration. When unexpected conditions occurred (e.g., opening the incubator lid), the DT automatically recalibrated its model and optimised the controller to maintain stable operation, without human intervention.

\paragraph{Non-functional Requirements.} \Citet{Lotz2013} and \citet{Brugali2018} represent example works in this category. \Citet{Lotz2013} proposes a modelling approach to handle dynamic variability in service robots by separating concerns between operational robustness and QoS (quality of service) optimisation. At design-time, they introduce two domain-specific languages: SmartTCL, which models operational variability by defining flexible task coordination strategies that allow robots to react to contingencies at runtime, and VML (Variability Modelling Language), which models variability in QoS by specifying how non-functional properties such as safety, energy consumption, and performance should be optimised under changing conditions. At runtime, these two mechanisms are integrated, enabling SmartTCL to handle functional variability. At the same time, VML resolves trade-offs in non-functional requirements, with different strategies proposed for their safe and consistent coordination. 

\Citet{Brugali2018} proposes a model-driven methodology for designing robotic systems that can dynamically reconfigure themselves at runtime while ensuring QoS guarantees. Their approach builds on three pillars: (1) explicitly modelling runtime variabilities in environment, system resources, and tasks, (2) annotating system variants with QoS-related properties (e.g., execution times, safety constraints), and (3) leveraging UML MARTE and Queuing Network models to evaluate performance and dependability trade-offs.

\paragraph{Assurance cases.} These works make use of assurance cases in the SAR adaptation. \Citet{Cheng2020} introduces AC-ROS, a framework that integrates assurance cases into the MAPE-K feedback loop to manage runtime adaptations in ROS-based autonomous systems. Assurance cases, expressed in Goal Structuring Notation (GSN), provide structured arguments that system requirements are met and are supported by evidence such as tests or monitored data. AC-ROS uses these models at runtime to guide adaptation decisions, ensuring that self-adaptive behaviours continue to satisfy safety and functional requirements.

\subsection{Challenges and Open Questions} 
Due to their self-adaptive nature, SAR systems require continuous monitoring. As argued in \cite{larsen2024robotic}, legitimation of new plans and system configurations is crucial for continued safe and performant operation.

\subsubsection{State Estimation and Model Synchronisation}
To achieve this monitoring, several models need to be kept up to date: the environment model, the system model, and models of its subsystems. State estimation is the preferred technique for maintaining the model states and parameters up to date as the system operates. However, due to the overlap between models (e.g., system models share parameters and states with their subsystem models), it is challenging to ensure consistency across them. This is also an opportunity to develop coupled state estimation techniques that operate on an overlapping ensemble of models.

\subsubsection{Managing Uncertainty in SAR Operations}
A key challenge in SAR operations is to characterise and manage the uncertainty in the system. This goes beyond maintaining uncertainty measures on model parameters and states (which state estimators already do) to quantify uncertainty at every phase of the MAPLE-K loop. For instance, a planner's optimal solution needs to be accompanied by a measure of confidence in the solution, indicating how robust its optimality is to uncertainties in the system and the models used to quantify it.

\subsubsection{Experimentation in Real Environment}
During operations, there might not be enough information to accurately identify the relationships between the system and its environment for planning purposes. For instance, a robot that never changes speed will not have a sense of its own inertia. This means that SAR planners may need to actively instruct the managed robot to conduct experiments to gather crucial data about itself or its environment, enabling them to plan effectively. This presents a challenge, as the experiments must be designed to avoid compromising the safety of the system or its environment.

\subsubsection{Adaptive Software Updates}
Since both the MAPLE-K and SAR require software updates throughout their operations, it is essential to plan and execute these updates without disrupting system operation. Ideally, techniques from dynamic software updating~\cite{ahmed2020dynamic} should be applied here, except these need to be adapted from cloud systems to robotic systems. Here, the challenge is that software components may not be replicable or their state may not be so easily transferable to new versions, as these techniques commonly require.

\subsubsection{\revision{Dev(Sec)Ops Pipelines for SAR Fleets}}
\revision{
Current operational practice for SAR often adopts selected DevOps activities (e.g., automated builds, containerised deployment, and remote updates), but typically in a fragmented manner and without a unified Dev(Sec)Ops methodology tailored to heterogeneous robot fleets. 
Compared to cloud-native systems, SAR deployment pipelines must account for embodiment, intermittent connectivity, heterogeneous hardware/software stacks, and mission/safety constraints during rollout. 
An open challenge is therefore to define end-to-end pipelines that continuously build, qualify, and secure updates, and then roll them out incrementally (e.g., canary/staged deployment) with robust rollback strategies when runtime evidence indicates elevated risk. 
This challenge directly intersects with software testing: qualification must combine staged MiL/SiL/HiL evidence, regression testing under continuous evolution, and runtime validation to decide whether a change can be promoted from digital and controlled environments to fleet-wide operation.
}

\subsubsection{Challenges Summary}
In the following, we present key challenges related to runtime operations in SAR. 
\revision{The graph in~\Cref{fig:operations_graph} suggests that uncertainty-aware operation~(Op-Ch-2) is foundational, since confidence estimates are needed to decide when active uncertainty reduction~(Op-Ch-3) or software changes should be triggered. Dynamic software updating~(Op-Ch-4) is another prerequisite for continuous Dev(Sec)Ops pipelines~(Op-Ch-5), which must safely qualify, roll out, and roll back changes across heterogeneous fleets. In turn, both active uncertainty reduction and pipeline evidence support more dependable overlapping model state estimation~(Op-Ch-1) during operation.}
\begin{tcolorbox}[title=Challenges in Operations for SAR]
  \textbf{Op-Ch-1:} Overlapping model state estimation.\\
  \textbf{Op-Ch-2:} Uncertainty as first-class citizen in MAPLE-K Loops.\\
  \textbf{Op-Ch-3:} Active uncertainty reduction.\\
  \textbf{Op-Ch-4:} Dynamic software updating for SARs.\\
  \revision{\textbf{Op-Ch-5:} Continuous Dev(Sec)Ops pipelines for heterogeneous SAR fleets.}
\end{tcolorbox}

\begin{figure}[htbp]
    \begin{adjustbox}{max width=\textwidth}
    \begin{tikzpicture}
        \node[flowchartNode] (n2) {Op-Ch-2};
        \node[flowchartNode, below=of n2] (n4) {Op-Ch-4};
        
        \node[flowchartNode, right=of n2, xshift=1cm, yshift=-1.0cm] (n3) {Op-Ch-3};
        \node[flowchartNode, right=of n3] (n5) {Op-Ch-5};
        \node[flowchartNode, right=of n5] (n1) {Op-Ch-1};

        \draw[flowArrow] (n2.east) -- (n3.west);
        
        \draw[flowArrow, rounded corners=10pt] (n2.east) -| ([xshift=-0.5cm]n5.west) -- (n5.west);
        \draw[flowArrow, rounded corners=10pt] (n4.east) -| ([xshift=-0.5cm]n5.west) -- (n5.west);
        
        \draw[flowArrow] (n3) -- (n5);
        \draw[flowArrow] (n5) -- (n1);

    \end{tikzpicture}
    \end{adjustbox}
    \caption{\revision{Dependency graph for challenges in operations for SAR.}}
    
    \label{fig:operations_graph}
\end{figure}

\section{Key Enabling Technologies}\label{sec:keytech}
In this section, we provide an overview of \revision{two} key enabling technologies shaping the future of self-adaptive robotics, \revision{namely DTs (\Cref{subsec:dt}) and AI (\Cref{subsec:ai}), along with their associated challenges and opportunities.}
\revision{
Several other technologies can play an important role in specific SAR contexts, such as decentralised network and communication systems in multi-robot and swarm robotics; however, these fall outside the scope of this paper. 
The focus on DTs and AI is motivated by their prominence, growing importance, and broad applicability across robotics in the literature~\cite{Mazumder2023,billard2025roadmap}. 
Although neither is a mandatory prerequisite for SAR, both serve as powerful enablers that substantially enhance how SAR systems are engineered and operated. 
Their increasing influence on state-of-the-art practices in modelling, validation, optimisation, and runtime adaptation makes them a timely focus for this roadmap. 
}

\subsection{Digital Twins}\label{subsec:dt} 
This section starts with a literature overview and then outlines key open challenges.

\subsubsection{Foundations and Related Research}
Many definitions of Digital Twins (DTs) exist~\cite{Barricelli&19}. However, we use the following from \cite{Fitzgerald&24}.
\begin{inparaenum}[(1)] 
\item A Digital Twin (DT) is a digital representation of a real-world entity called the Physical Twin (PT). 
\item The DT and PT are connected by a communications infrastructure that allows the DT to maintain a known level of fidelity to the PT it represents. 
\item A DT offers its stakeholders a range of services that add value to the PT without unduly compromising the PT's operation. 
\end{inparaenum}
\revision{This definition on purpose does not distinguish between DTs and Digital Shadows (DSs) (without a direct bi-directional communication) \cite{Kritzinger2018a} because in reality DTs initially start as DSs where a human is involved. When a satisfactory level of trust in the DT's bidirectional communication is established, autonomy can be enabled. We imagine that this autonomy will be introduced gradually, as is the case with autonomous cars.}

DTs are a generic technology that can be applied to any CPS, and naturally, for robots as well~\cite{malik2021digital,Gil&24b}. DTs can be used in various domains, offering distinct services. At present, DT is utilised across multiple industries, including manufacturing~\cite{Tao2017}, automotive~\cite{Bhatti2021}, aviation~\cite{Xiong2022}, maritime~\cite{Zhang2022}, construction~\cite{Xie2023}, and healthcare~\cite{Sun2023,sartaj2025medet,sartaj2024modelbased}, serving diverse functions like simulation, integration, testing, monitoring, and maintenance.

There is growing interest in using DTs in robotics, for example, to improve the design, performance, and maintenance. DTs are applied in various subfields, such as robotics design~\cite{Kousi2021}, motion planning and control~\cite{Yang2024}, human-robot interaction~\cite{malik2021digital,Li2021}, autonomous robots~\cite{Hoebert2019}, smart manufacturing~\cite{Li2022}, and prognostics and health management (PHM)~\cite{han2021fault,Song2023}. 
Mazumder et al.~\cite{Mazumder2023} explored trends in DT-integrated robotics to identify gaps, potential scope, challenges, and future perspectives.
Yang et al.~\cite{Yang2024} proposed a DT-based autonomous navigation and control method for omnidirectional mobile robots (OMRs), achieving a physical-virtual synchronisation tracking error of 0.061 m and effectively handling various tasks.
Song et al. \cite{Song2023} introduced a DT-assisted fault diagnosis system for robot joints, using a CycleGAN-based model to map virtual entity data to physical data. Their method increased fault diagnosis accuracy from 32.5\% to 98.86\% using only 400 measured data sets.
The purpose of a DT is essential, as it determines the kind of predictive power its internal models need. One DT can have multiple models (aiming to constantly monitor the state of the PT and make future predictions) inside, and there can be different ways to determine which of these to use in other circumstances.

\subsubsection{Challenges and Open Questions}
Various publications address the most significant research challenges in DTs~\cite{Fuller&20,Botín-Sanabria&22,Larsen&23}. Challenges for DTs and future research directions can be described in different dimensions. 
However, from a robotics perspective, we think that the most important ones are the following, which deserve attention from the research community.

\paragraph{\textbf{Data Quality from PT}} Data from a physical robot is mainly derived from sensors. Still, it is delivered with different kinds of uncertainties (noise affecting the values and time delay in delivery). Depending on the purpose of a DT, such uncertainties may make it impossible to provide the predictions in a trustworthy manner.

\paragraph{\textbf{Providing Predictions Promptly}} When data from a PT is fed into one of the models inside a DT, it is essential that it is possible to provide predictions to indicate that some action is required promptly. A human user can make an actual decision autonomously if one trusts the predictions. The challenge here is ensuring that the underlying computing infrastructure is enough for the model to make predictions.

\paragraph{\textbf{State Estimation for the PT}} When debugging an application on a computer, one has complete insight into what state the system is in at all times. However, when dealing with physical processes, it is much more challenging to determine the exact moment when a state transition occurs. 

\paragraph{\textbf{Composition of DTs}} If we have established DTs for different collaborating robots, the composition of the DTs is far from trivial. One root cause is that the IP embedded in models with predictive power can be essential to the organisations that produce them. Thus, it is paramount to ensure that such IP is protected.

\paragraph{\textbf{Flexibility and Multi-Purpose Demands}} DTs have significant potential to improve lifecycle engineering. For DTs to be effective, their models must adapt across all lifecycle stages with fidelity adjusted to each phase---lower fidelity for design and higher fidelity for control, verification, and operations. However, most current research focuses on specific lifecycle stages, rather than examining the broader value chain~\cite{zhang2024digital}. Enhancing the adaptability and flexibility of DT models remains a key challenge.

\paragraph{\textbf{Limited Complexity and Extensibility}} Current DT study in virtual prototyping (VP)~\cite{major2021use} often focuses on small-scale modules or isolated systems. However, a typical robot involves a complex interplay of mechanical, thermal, hydraulic, and control subsystems. The existing literature offers limited insights into the development of large-scale, fully integrated DT systems. The key challenge lies in the comprehensive integration of these subsystems, including sensors, control mechanisms, and mechanical systems.

\paragraph{\textbf{Dynamic Interaction between DTs and PTs}} DTs differ from traditional models through their real-time, dynamic interaction with PTs. Unlike conventional models used in isolated software for tasks like design and control, DTs evolve alongside the systems they represent, integrating live sensor data, simulation updates, and software upgrades throughout the system’s life cycle~\cite{mao2025survey}. The complexity of modern robotic systems, with their multidisciplinary nature (e.g., mechanical, electrical, and software), makes it challenging to model all aspects on a single platform. As a result, achieving a fully integrated and responsive representation of real-world systems remains a significant challenge for DTs.

\paragraph{\textbf{Challenges Summary}}
In the following, we present a summarised list of challenges in the use of digital twins for SAR.
\revision{The graph in~\Cref{fig:dt_graph} shows the dependencies among the challenges, indicating their priorities and timelines. From a timing perspective, determining state transitions of PTs is the most important challenge (DT-Ch-3). This is necessary before trustworthy predictions (DT-Ch1), timely predictions (DT-Ch-2), and adaptation (DT-Ch-5) can depend on it. Afterwards, composition (DT-Ch-4) and complex subsystems (DT-Ch-6) will enable more complex DT combinations. Finally, a fully integrated, dynamic and responsive DT (DT-Ch-7) is the ultimate target from a DT research perspective.} 
\begin{tcolorbox}[title=Challenges in Digital Twins for SAR]
  \textbf{DT-Ch-1:} Ensuring trustworthy predictions due to data quality and uncertainties from physical systems.\\
  \textbf{DT-Ch-2:} Requires sufficient computing infrastructure to support timely predictions for decision-making.\\
  \textbf{DT-Ch-3:} Determining the exact moment of state transitions in physical systems.\\
  \textbf{DT-Ch-4:} Composing DTs for collaborating robots.\\
  \textbf{DT-Ch-5:} Adapting DT models across all software engineering lifecycle phases with sufficient fidelity.\\
  \textbf{DT-Ch-6:} Integrating complex subsystems in large-scale robotic DTs.\\
  \textbf{DT-Ch-7:} Achieving a fully integrated, dynamic, and responsive DT representation of multidisciplinary robotic systems.
\end{tcolorbox}

\begin{figure}[htbp]

    \begin{adjustbox}{max width=\textwidth}
    \begin{tikzpicture}

        \node[flowchartNode] (n3) at (-1.2, 0) {DT-Ch-3};

        \node[flowchartNode] (n1) at (2, 1.2)  {DT-Ch-1};
        \node[flowchartNode] (n2) at (2, 0)    {DT-Ch-2};
        \node[flowchartNode] (n5) at (2, -1.2) {DT-Ch-5};

        \node[flowchartNode,draw=steelblue, thick, rounded corners=10pt, inner sep=10pt, fit=(n1) (n2) (n5), fill=none] (group) {};
 
        \node[flowchartNode,right of=n2, xshift=2.5cm, yshift=.55cm] (n4) {DT-Ch-4};

        \node[flowchartNode,right of=n2, xshift=2.5cm,  yshift=-.55cm] (n6) {DT-Ch-6};

        \node[flowchartNode,draw=steelblue, thick, rounded corners=15pt, inner sep=10pt, fit=(n4) (n6), fill=none] (group2) {};

        \node[flowchartNode,right of=group2, xshift=2cm] (n7) {DT-Ch-7};

        \draw[flowArrow] (n3.east) -- (group.west);
 
        \draw[flowArrow] (group.east) -- (group2.west);
        \draw[flowArrow] (group2.east) -- (n7.west);
 
    \end{tikzpicture}
    \end{adjustbox}
    \caption{\revision{Dependency graph for challenges in digital twins for SAR.}}
    
    \label{fig:dt_graph}
\end{figure}

\subsection{Artificial Intelligence} \label{subsec:ai} 
This section first presents an overview of the research and then highlights open challenges.

\subsubsection{Foundations and Related Research}
AI has become a cornerstone in the evolution of robotics, enabling robots to perform complex tasks with increasing autonomy and adaptability. 
AI techniques, including machine learning (ML), deep learning (DL), and reinforcement learning (RL), have facilitated advancements in perception, planning, control, and human-robot interaction. 
AI in this context is characterised by advances in several areas, including computer vision, natural language processing, and reinforcement learning (RL). These technologies have enabled robots to perceive, reason, and act in complex environments with minimal human intervention.

\paragraph{Perception and Sensing}
Modern robotic systems use deep learning-based computer vision techniques to interpret visual data. Convolutional Neural Networks (CNNs) and transformer-based architectures have revolutionised object detection, segmentation, and scene understanding. For example, models such as YOLO~\cite{redmon2016you} and Mask R-CNN~\cite{he2017mask} are widely used for real-time object detection and instance segmentation in robotics applications. 
Additionally, multimodal sensing, which combines visual, LiDAR, and radar data, has improved the robustness of perception systems in autonomous vehicles and drones. 
Furthermore, modern autonomous vehicles employ occupancy networks to reconstruct 3D environments~\cite{roddick2020predicting} and incorporate flow information to predict object motion~\cite{agro2023implicit}, further improving situational awareness and navigation.

\paragraph{Decision-Making and Planning} 
AI has enhanced robotic decision-making through techniques such as RL and LLMs as autonomous agents. 
RL algorithms, such as Deep Q-Networks (DQN) and Proximal Policy Optimization (PPO), have been widely applied in robotic manipulation~\cite{nguyen2019review}, autonomous navigation~\cite{wang2019autonomous}, and multi-agent coordination. Moreover, hierarchical RL structures~\cite{pateria2021hierarchical} allow robots to decompose complex tasks into manageable subtasks, improving long-term planning and adaptability. LLMs are increasingly integrated into robotics to enable high-level reasoning and task planning. For instance, LLMs can be used as the brain for generating a robot behaviour tree~\cite{lykov2024llm}. 

\paragraph{Control and Actuation} 
AI has become crucial in robotic control and actuation, enabling precise, adaptive, and efficient movement. One line of research focuses on using AI to learn unmodelled dynamics, including complex aerodynamic effects, vehicle friction models~\cite{spielberg2019neural}, and hydrodynamics~\cite{wang2021incorporating}, resulting in more accurate dynamic models and improved control performance. Another line of research exploits RL to learn controllers directly through simulation, followed by sim-to-real transfer techniques to deploy them on physical robots. This approach has demonstrated the ability to develop robust controllers that enable robots to adapt to varying terrains and dynamic environments.

\paragraph{Human-Robot Interaction}
NLP and affective computing have enabled more natural and intuitive interactions between humans and robots. Chatbots and voice assistants powered by transformer models, such as GPT and BERT, are being integrated into service robots for healthcare, education, and customer service. Additionally, emotion recognition systems allow robots to adapt their behaviour based on human emotional states.

\subsubsection{Challenges and Open Questions}
Despite considerable efforts, AI for software engineering of self-adaptive robots still faces several challenges. 
\paragraph{\textbf{Inherent Challenges Related to Data}} Self-adaptation poses specific challenges in AI-powered robotics systems. The most popular AI systems are trained on vast amounts of data. Thus, self-adaptation can usually be considered as gathering more data that represents the new state of the world and the task to which the systems should adapt. This is very expensive or infeasible in many situations. 

\paragraph{\textbf{Overcoming Catastrophic Forgetting in Continual Learning}} Another challenge is retaining previously acquired knowledge when adapting to new data, thereby avoiding catastrophic forgetting and promoting smooth, continual learning. Model fine-tuning, transfer learning, knowledge distillation, and RL-based continual training are methods to consider for systems that adapt to new data.

\paragraph{\textbf{Overfitting}} AI often overfits training data, leading to poor generalisation and causing operational anomalies. These should be detected and analysed to provide adaptation plans. More research is needed to address AI overfitting by implementing self-adaptive robots.  

\paragraph{\textbf{AI-specific Challenges}} AI in self-adaptive robots presents new challenges, including hallucinations that can lead to unsafe decisions and inherent uncertainty that requires quantification. Real-time handling of uncertainty remains underexplored and needs further study. Regardless of the adaptation method, trustworthiness is key. AI-powered robots must adapt safely, ensuring legitimacy through explainability, fairness, transparency, and robustness. Assuring trustworthiness in self-adaptive robots is an understudied area and requires novel solutions. \revision{There are several additional challenges related to human-robot interaction that focus more on human aspects, e.g., developer/user experience, operator trust, organisational policy, transparency, accountability, etc., which are indeed important but are not in the main scope of this paper and will not be discussed in detail.}

\paragraph{\textbf{Challenges Summary}}
In the following, we present a summary of key challenges in applying AI to SAR. 
\revision{The graph in~\Cref{fig:ai_graph} shows the dependencies among the challenges, indicating their priorities and timelines. From a timing perspective, gathering sufficient data (AI-Ch-1) relevant to the self-adaptation to be performed is the first challenge to be addressed. Retraining or fine-tuning the AI algorithms for adaptation, ensuring that previous knowledge is not severely altered (AI-Ch-2) and at the same time avoiding overfitting (AI-Ch-3), is challenging but necessary for reliable self-adaptation. Finally, the adapted AI models should be trustworthy (AI-Ch-4) without hallucinations or raised ethical concerns.}

\begin{tcolorbox}[title=Challenges in AI for SAR]
  \textbf{AI-Ch-1:} Gathering sufficient data for self-adaptation in AI-powered robotics is often expensive or infeasible. \\ 
  \textbf{AI-Ch-2:} Retaining previously acquired knowledge while adapting to new data and avoiding catastrophic forgetting.\\
  \textbf{AI-Ch-3:} Addressing AI overfitting to improve generalisation and prevent operational anomalies in self-adaptive robots.\\
  \textbf{AI-Ch-4:} Ensuring trustworthiness in self-adaptive robots, including handling hallucinations, real-time uncertainty, and ethical concerns. 
\end{tcolorbox}

\begin{figure}[htbp]

    \begin{adjustbox}{max width=\textwidth}
    \begin{tikzpicture}
        \node[flowchartNode] (n1) {AI-Ch-1};
        \node[flowchartNode, right=of n1] (n2) {AI-Ch-2};
        \node[flowchartNode, below=of n2, yshift=0.5cm] (n3) {AI-Ch-3};

        \node[flowchartNode, right=of n2, xshift=2cm] (n4) {AI-Ch-4};

        \node[draw=steelblue, dashed, rounded corners, thick, 
              inner sep=10pt, fit=(n1) (n2) (n3)] (container) {};

        \draw[flowArrow] (n1) -- (n2);
        \draw[flowArrow] (n1) -- (n3);

        \draw[flowArrow] (container.east) -- (n4.west);

    \end{tikzpicture}
    \end{adjustbox}
    \caption{\revision{Dependency graph for challenges in AI for SAR.}}
    
    \label{fig:ai_graph}
\end{figure}

\section{2030 Research Agenda}\label{sec:agenda2030} 
In this section, we begin by analysing the challenges outlined in the previous sections with respect to quality aspects (\Cref{subsec:chqas}), aiming to identify key areas that require future attention. 
\revision{
We then examine how SAR characteristics are relevant to the challenges outlined in this roadmap (\Cref{subsec:characteristics}). 
Next, we analyse the relationships between each enabling technology and the five phases of the software engineering lifecycle (\Cref{subsec:relation}). 
Finally,
}
we explore potential research opportunities (\Cref{subsec:opportunities}) to drive advancements in the field of SAR. 

\subsection{Challenges and Quality Aspects}\label{subsec:chqas} 
For all challenges identified in the software engineering lifecycle and enabling technologies, we analysed the quality aspects that they highlight. 
Specifically, we analysed: \begin{inparaenum}[(1)] 
\item the relationships between challenges and quality aspects, 
\item the relative criticality of challenges based on their impact across multiple quality aspects (i.e., challenges influencing several aspects are considered more critical), and 
\item the overall importance of each quality aspect (i.e., quality aspects requiring further attention). 
\end{inparaenum} 
The results of this analysis are presented in~\Cref{tab:analysis},
\revision{
and a detailed interactive version of the table is available at \url{https://simula-complex.github.io/robomap/}. 
}

\revision{
The analysis was conducted as follows. 
First, we identified the set of quality aspects from the SAR literature~\cite{alberts2025software} and the ISO/IEC 25010:2023 standard~\cite{international2011systems}. 
Next, we examined the description of each challenge to identify which quality aspects were highlighted. 
For example, the description of \textit{\mbox{SRE-Ch-1}} highlights safety, reliability, and adaptability; these three aspects are therefore marked for that challenge in the table. 
We then computed the criticality of each challenge by counting the number of quality aspects associated with it (i.e., the number of rows in the table). 
In the case of \textit{\mbox{SRE-Ch-1}}, this yields a criticality score of 3. 
A higher criticality count indicates that the corresponding challenge requires greater attention. 
Finally, to assess the overall importance of each quality aspect, we counted how many challenges reference it (i.e., the number of vertical lines in the table). 
A higher value indicates that the quality aspect is widely emphasised across multiple challenges, reflecting its overall significance in the context of SAR. 
}

When examining the criticality of challenges across various quality aspects, it is evident that challenges in MDE and software testing span most of them. 
This indicates that these challenges require significant attention in future research. 
Similarly, challenges related to SAR simulations and operations appear to be the second most critical, underscoring their importance for advancing SAR research. 
Among enabling technologies, the challenges in AI and QC are closely related to multiple quality aspects, indicating that applying these technologies to SAR will require more attention in the future.

\setlength{\tabcolsep}{4pt}
\newcommand{\groupmultirow}[2]{\multirow[t]{#1}{=}{\centering\textbf{#2}}}
\newcolumntype{C}[1]{>{\centering\arraybackslash}p{#1}}

\begin{longtable}
{|C{2.5cm}|p{0.15\textwidth}|>{\centering\arraybackslash}p{0.05\textwidth}|>{\centering\arraybackslash}p{0.05\textwidth}|>{\centering\arraybackslash}p{0.05\textwidth}|>{\centering\arraybackslash}p{0.05\textwidth}|>{\centering\arraybackslash}p{0.05\textwidth}|>{\centering\arraybackslash}p{0.05\textwidth}|>{\centering\arraybackslash}p{0.05\textwidth}||>{\centering\arraybackslash}p{0.05\textwidth}|}
\caption{Mapping of challenges in self-adaptive robotics to cross-cutting quality aspects. Rows with more “+” symbols indicate higher criticality of the corresponding challenge, while columns with more “+” symbols show higher importance of the associated quality aspect.} 
\label{tab:analysis}
\\ 
\hline
\rowcolor{teal!5!white}
\textbf{Topic} & \textbf{Challenge} & \rotatebox{90}{\textbf{Safety}} & \rotatebox{90}{\textbf{Performance}} & \rotatebox{90}{\textbf{Reliability}} & \rotatebox{90}{\textbf{Explainability}}  & \rotatebox{90}{\textbf{Privacy}} & \rotatebox{90}{\textbf{Adaptability}} & \rotatebox{90}{\textbf{Security}} & \rotatebox{90}{\textbf{Ch. Criticality}} \\ 
\hline
\endfirsthead
\hline
\rowcolor{teal!5!white}
\textbf{Topic} & \textbf{Challenge} & \rotatebox{90}{\textbf{Safety}} & \rotatebox{90}{\textbf{Performance}} & \rotatebox{90}{\textbf{Reliability}} & \rotatebox{90}{\textbf{Explainability}} & \rotatebox{90}{\textbf{Privacy}} & \rotatebox{90}{\textbf{Adaptability}} & \rotatebox{90}{\textbf{Security}} & \rotatebox{90}{\textbf{Ch. Criticality}} \\ 
\hline
\endhead
\hline
\endfoot

\multirow{7}{=}{\centering \textbf{Requirements Engineering}}
& SRE-Ch-1  & + &   & + &  &  & + &  & \textbf{3} \\ 
& SRE-Ch-2  &   &  & + &  &  & + &  & \textbf{2} \\
& SRE-Ch-3  &  + &  & + &   &  & + &  & \textbf{3} \\
& SRE-Ch-4  &  + &   &   & + &  & + &  & \textbf{3} \\
& SRE-Ch-5  &   &   &   & + & + &  &  & \textbf{2} \\

\arrayrulecolor{lightgray}\cline{2-10}
& NR-Ch-1  &  + &   & +  &  &  & + &  & \textbf{3} \\
& NR-Ch-2  &   &   &   & + &  &  &  & \textbf{1} \\
\hline

\multirow{17}{=}{\centering \textbf{Software Design}} 
& SD-Ch-1   &   & + &   &   & & + &  & \textbf{2} \\ 
& SD-Ch-2   & + & + & + &   &  & + &  & \textbf{4} \\
& SD-Ch-3   & + &   & + &   & & + &  & \textbf{3} \\
& SD-Ch-4   &   &   & + &   &  & + &  & \textbf{2} \\
& SD-Ch-5   &   &   & +  &   & & + &  & \textbf{2} \\
& SD-Ch-6   & + &   &   & + &  & + &  & \textbf{3} \\
& SD-Ch-7   & + &   & + &   & & + &  & \textbf{3} \\

\arrayrulecolor{lightgray}\cline{2-10} 
& MDE-Ch-1  &   &  &  + & + &  & + &  & \textbf{3} \\
& MDE-Ch-2  &   &  + &  + &   & & + &  & \textbf{3} \\
& MDE-Ch-3  &  + &  + &  + &  & & + &  & \textbf{4} \\
& MDE-Ch-4  &   &  &  + &  + & & + &  & \textbf{3} \\
& MDE-Ch-5  &  + &   &   &  + & + & + &  & \textbf{4} \\

\arrayrulecolor{lightgray}\cline{2-10} 
& Sim-Ch-1  &   &   &  + &  & & + &  & \textbf{2} \\ 
& Sim-Ch-2  &   &  &  + &  & & + &  & \textbf{2} \\
& Sim-Ch-3  &   &  + &  + &   & & + &  & \textbf{3} \\
& Sim-Ch-4  &   &  + &  + &   &  &  &  & \textbf{2} \\
& Sim-Ch-5  &   &   &   &   & + & & + & \textbf{2} \\
\hline

\multirow{5}{=}{\centering \textbf{Software Development}} 
& Dev-Ch-1  &   &  + &  + &   & &+ &  & \textbf{3} \\ 
& Dev-Ch-2  &   &  + &  + &   & & +&  & \textbf{3} \\
& Dev-Ch-3  & +  &  + & +  &   &  & + &  & \textbf{4} \\
& Dev-Ch-4  &   &  + &  + &   &  & + &  & \textbf{3} \\
& Dev-Ch-5  &  + &   &   &   & + & & + & \textbf{3} \\
\hline

\multirow{8}{=}{\centering \textbf{Software Testing}}
& ST-Ch-1  & + &  + & + &   &   &  + &   & \textbf{4} \\ 
& ST-Ch-2  & + &   & + & + &   &   &   & \textbf{3} \\
& ST-Ch-3  & + &   & + &   &   &  + &   & \textbf{3} \\
& ST-Ch-4  & + &  + & + &   &   &  + &   & \textbf{4} \\
& ST-Ch-5  &  & + & + &   &   & + &   & \textbf{3} \\
& ST-Ch-6  &  + &  & + &  + &   & + &   & \textbf{4} \\
& ST-Ch-7  & + &   & + & + &   &   &   & \textbf{3} \\
& ST-Ch-8  & + &   & + & + &  &   & + & \textbf{4} \\
\hline

\multirow{4}{=}{\centering \textbf{Operations}} 
& Op-Ch-1   &   &   &  + &   &   &  + &   & \textbf{2} \\ 
& Op-Ch-2   &  + &   &  + &  + &   & +  &   & \textbf{4} \\
& Op-Ch-3   &  + &   &  + &   &   &  + &   & \textbf{3} \\
& Op-Ch-4   &  + &   &  + &   &   &  + &   & \textbf{3} \\
&\revision{Op-Ch-5}   &  \revision{+} &   & \revision{+}  &   &   & \revision{+}  &  \revision{+} & \revision{\textbf{4}} \\
\hline

\multirow{7}{=}{\centering \textbf{Digital Twins}} 
& DT-Ch-1   & + &   & + & + &   &   &   & \textbf{3} \\ 
& DT-Ch-2   &  + & + &  + &   &  &  &   & \textbf{3} \\
& DT-Ch-3   &   &   & + & + &   &  + &   & \textbf{3} \\
& DT-Ch-4   &   &   &   &   & + &  &  + & \textbf{2} \\
& DT-Ch-5   &   & + & + &   &   & + &   & \textbf{3} \\
& DT-Ch-6   &   & + & + &   &   & + &   & \textbf{3} \\
& DT-Ch-7   &   & + & + &   &   &  + &   & \textbf{3} \\
\hline

\multirow{4}{=}{\centering \textbf{Artificial Intelligence}} 
& AI-Ch-1   &   & + &  + &   &   & + &   & \textbf{3} \\ 
& AI-Ch-2   &  + &   & + &   &   & + &   & \textbf{3} \\
& AI-Ch-3   & + &   & + &  &   &  + &   & \textbf{3} \\
& AI-Ch-4   & + &   & + &  + & + &   & + & \textbf{5} \\
\hline

\arrayrulecolor{black}
\hline\hline
\multicolumn{2}{|c|}{\textbf{Quality Aspect Focus}} 
& \revision{\textbf{28}} & \revision{\textbf{18}} & \revision{\textbf{44}} & \revision{\textbf{15}} 
& \revision{\textbf{6}} & \revision{\textbf{41}} & \revision{\textbf{6}} & \textbf{-} \\
\hline 
\end{longtable}

In terms of the importance of quality aspects highlighted across the challenges, reliability is the most frequently cited concern, underscoring the need for further research to ensure the dependability of SAR systems throughout the software engineering lifecycle. 
Adaptability follows as the second most emphasised aspect, suggesting a growing demand for SAR systems that can seamlessly adapt to evolving environments and requirements. 
Other most highlighted quality aspects are safety and performance, especially in challenges involving AI-driven components, indicating the need for further research. 
However, privacy and security are less highlighted across challenges, appearing mainly in challenges related to simulations, software development, software testing, AI, and digital twins. 
Given the growing integration of AI in robotics, the increasing importance of software testing to ensure reliability, and the key role of digital twins, both privacy and security require future attention, especially as future robots are expected to operate in real-world, human-centric environments.

The various types of robotic systems, such as soft robotics, modular robotics, and swarm robotics, may lead to greater complexity.
Adaptive and autonomous robots can adapt to changing goals, requirements, and environmental conditions. This can range from simple parameter adjustments to more complex behavioural changes, such as switching between modes of operation, adapting to new tasks, or modifying their current goals to operate safely and in accordance with their requirements.

Soft robotics encompasses a field of robotics where robots are built using highly flexible materials, allowing them to adapt to their environment and perform tasks that require a high degree of flexibility~\cite{Oncay2023softrobotics}. This allows them to interact with their environment in a more varied way, whether this regards their movement, their way of grasping or interacting with objects, or their response to changing environmental conditions. This high flexibility, however, leads to a state-space explosion in the robot's potential behaviour. Programming and controlling such robots can therefore become challenging~\cite{Rafsanjani2019programmingSoftRobots}. 

Modular robots can interact with each other and adapt to new complexities and tasks through self-composition. Specifically, robots are designed to interlock and create new, larger physical structures~\cite{Yim2002modularRobots}. While the individual robots are usually very simple shapes, e.g., cubes, they can, through their ability to compose with other robots, take on more complex shapes. This, in turn, may allow them to overcome additional changes in their environment, e.g., by forming a bridge to cross a gap. The ability to self-compose, however, also leads to a combinatorial explosion in the number of possible configurations and shapes that robots can take on, which in turn leads to higher complexity in the programming and control of such robots~\cite{Gilpin2010modularRobots}. This poses significant challenges for the distributed planning and verification of both individual and composed robotic systems~\cite{Esterle2021verification}. 

Swarm robotics emphasises the individuality of robots and their ability to operate autonomously while also interacting with their peers and the environment~\cite{Brambilla2013swarmRobotics}. The core idea in swarm robotics is that collective behaviour emerges from local interactions among robots. As they are tasked with achieving their common goals, they work towards their individual objectives. For this, they coordinate locally and adapt individually to changes in their local environment. Through this emergent behaviour, the swarm can achieve complex tasks that would be difficult or impossible for individual robots to accomplish on their own. While swarm robotics offers higher robustness compared to individual robots due to their redundancy, designing or identifying local rules, guaranteeing their effectiveness and efficiency, and ensuring the overall system's performance remain challenging tasks~\cite{Casadei2025collectives,Pianini2022coordinatemultirobot, Kephart2017collectiveSelfAware}.

\revision{
\subsection{Challenges and SAR Characteristics}\label{subsec:characteristics}
\Cref{tab:analysis2} highlights the relevance of the challenges with the SAR characteristics outlined in~\Cref{sec:background}. A detailed interactive version of this table is available at \url{https://simula-complex.github.io/robomap/}. 
Each “+” indicates that the respective characteristic plays a significant role as part of this challenge. 
In the following, we provide illustrative examples of how SAR characteristics are relevant to the challenges highlighted in this paper.\\
\begin{inparaenum}[$\bullet$]
    \item \textbf{Embodiment:} Robots handling uncertainty during their adaptation are required to have an understanding of their physical shape to adapt in a predictable and trustworthy manner (SD-Ch-4). Another example is real-time sensor fusion, which requires the robot to understand where the diverse sensors are located and how their fusion needs to be adapted based on physical changes of the robot (Dev-Ch-3).\\
    \item \textbf{Physical Interaction:} As robots interact with their environment, there is a need for combining software with models of hardware and the immediate environment (MDE-Ch-3). When interacting physically with the environment, and specifically with other robots, robots may adapt using AI and online learning to improve their performance. With this, we need to address AI overfitting to improve generalisation and prevent operational anomalies (AI-Ch-3).\\
    \item \textbf{Localised:} To provide energy-efficient robotic software and sustainable operations, robots need to have an understanding of their local environment and their own embedding within to optimise their activities and adaptations (Dev-Ch-4). As ethical requirements may vary by location, robots need to understand their environments and localise themselves within the relevant ethical environment (SRE-Ch-5). Similarly, this holds for norm-sensitive decisions (NR-Ch-2).\\
    \item \textbf{Heterogeneity:} In dynamic, open environments, robotic systems may encounter other robots; coupling these systems in a composite simulation requires them to deal with their respective heterogeneity (Sim-Ch-2). 
    The heterogeneity of SAR fleets is also relevant when defining the underlying Dev(Sec)Ops pipelines (Op-Ch-5). \\
    \item \textbf{Distributed Nature:} As self-adaptive multi-robot systems and fleets are inherently distributed in the physical environment, this distributed nature needs to be explicitly considered in the algorithms that dynamically adapt to system reconfiguration (Sim-Ch-3). Similarly, when operating fleets of DTs for collaborative robots, the distributed nature needs to be taken into account (DT-Ch-4). \\
    \item \textbf{Modularity:} As robots are composed of various hardware that might change during runtime (e.g., changing the gripper on a robotic arm), this modularity should be a fundamental aspect to avoid rebuilding existing architectures (Dev-Ch-2). With modular SAR, MAPE-K components might be available in different modules. With this, we have to ensure the correctness of individual MAPE-K components and the overall loop during testing (ST-Ch-4).
\end{inparaenum}
}

\begingroup
\color{black}
\begin{longtable}
{|C{2.5cm}|p{0.25\textwidth}|>{\centering\arraybackslash}p{0.05\textwidth}|>{\centering\arraybackslash}p{0.05\textwidth}|>{\centering\arraybackslash}p{0.05\textwidth}|>{\centering\arraybackslash}p{0.05\textwidth}|>{\centering\arraybackslash}p{0.05\textwidth}|>{\centering\arraybackslash}p{0.05\textwidth}|}
\caption{\revision{Mapping of challenges in self-adaptive robotics to SAR characteristics. Rows with more “+” symbols indicate the relevance of the characteristic to the challenge.}} 
\label{tab:analysis2}
\\ 
\hline
\rowcolor{teal!5!white}
\textbf{Topic} & \textbf{Challenge} & \rotatebox{90}{\textbf{Embodiment}} & \rotatebox{90}{\textbf{Physical Interaction}}  & \rotatebox{90}{\textbf{Localised}} & \rotatebox{90}{\textbf{Heterogeneity}} & \rotatebox{90}{\textbf{Distributed Nature}} & \rotatebox{90}{\textbf{Modularity}} \\ 
\hline
\endfirsthead
\hline
\rowcolor{teal!5!white}
\textbf{Topic} & \textbf{Challenge} & \rotatebox{90}{\textbf{Embodiment}} & \rotatebox{90}{\textbf{Physical Interaction}}  & \rotatebox{90}{\textbf{Localised}} & \rotatebox{90}{\textbf{Heterogeneity}} & \rotatebox{90}{\textbf{Distributed Nature}} & \rotatebox{90}{\textbf{Modularity}} \\ 
\hline
\endhead
\hline
\endfoot

\multirow{7}{=}{\centering \textbf{Requirements Engineering}}
& SRE-Ch-1  &   & + &   & + & + & +     \\ 
& SRE-Ch-2  & + &   &   & + & + &       \\
& SRE-Ch-3  &   &   &   &   & + &     \\
& SRE-Ch-4  & + & + &   &   &   &    \\
& SRE-Ch-5  & + &   & + &   &   &    \\

\arrayrulecolor{lightgray}\cline{2-8}
& NR-Ch-1  & + & + &   & + &   & + \\
& NR-Ch-2  &   &   & + & + & + & + \\
\hline

\multirow{17}{=}{\centering \textbf{Software Design}} 
& SD-Ch-1   &   &   &   &   & + & +  \\ 
& SD-Ch-2   & + & + &   & + &   &     \\
& SD-Ch-3   & + & + &   & + &   &    \\
& SD-Ch-4   & + & + & + &   &   &   \\
& SD-Ch-5   &   &   &   & + &   & +  \\
& SD-Ch-6   & + & + & + &   &   &    \\
& SD-Ch-7   &   & + & + & + &   & + \\

\arrayrulecolor{lightgray}\cline{2-8} 
& MDE-Ch-1  & + &   & + & + & + & +  \\
& MDE-Ch-2  & + &   & + & + &   & +  \\
& MDE-Ch-3  &   & + & + & + &   &     \\
& MDE-Ch-4  & + &   &   & + &   & + \\
& MDE-Ch-5  & + & + & + &   &   &    \\

\arrayrulecolor{lightgray}\cline{2-8} 
& Sim-Ch-1  & + & + & + &   &   &   \\ 
& Sim-Ch-2  &   & + &   & + & + & + \\
& Sim-Ch-3  &   &   & + & + & + & \\
& Sim-Ch-4  &   &   &   & + &   & +  \\
& Sim-Ch-5  & + &   &   &   &   &  \\
\hline

\multirow{5}{=}{\centering \textbf{Software Development}} 
& Dev-Ch-1  & + &   &   &   &   & + \\ 
& Dev-Ch-2  & + &   &   &   &   & +  \\
& Dev-Ch-3  & + & + & + &   &   &   \\
& Dev-Ch-4  & + &   & + & + & + &  \\
& Dev-Ch-5  & + & + & + &   &   &     \\
\hline

\multirow{8}{=}{\centering \textbf{Software Testing}}
& ST-Ch-1  & + & + &   &   &   &    \\ 
& ST-Ch-2  &   &   &   & + &   & +   \\
& ST-Ch-3  & + & + &   &   &   &    \\
& ST-Ch-4  & + &   & + & + &   & +    \\
& ST-Ch-5  &   & + &   & + & + &    \\
& ST-Ch-6  & + &   &   & + &   &     \\
& ST-Ch-7  & + & + &   & + &   & +    \\
& ST-Ch-8  & + &   &   &   &   &    \\
\hline

\multirow{4}{=}{\centering \textbf{Operations}} 
& Op-Ch-1   & + &   &   & + &   & +  \\ 
& Op-Ch-2   & + & + & + &   &   &    \\
& Op-Ch-3   & + & + & + & + & + & +  \\
& Op-Ch-4   & + &   &   & + &   & +  \\
& Op-Ch-5   &  + &   &   & +  & +  &    \\
\hline

\multirow{7}{=}{\centering \textbf{Digital Twins}} 
& DT-Ch-1   & + & + & + & + &   &     \\ 
& DT-Ch-2   &   & + & + &   & + &     \\
& DT-Ch-3   &   & + &   &   &   &      \\
& DT-Ch-4   &   &   &   & + & + &    \\
& DT-Ch-5   & + &   &   & + &   &      \\
& DT-Ch-6   &   &   &   & + & + & +    \\
& DT-Ch-7   &   &   &   & + & + & +     \\
\hline

\multirow{4}{=}{\centering \textbf{Artificial Intelligence}} 
& AI-Ch-1   & + & + & + &   &   &       \\ 
& AI-Ch-2   & + & + &   &   &   &       \\
& AI-Ch-3   & + & + &   &   &   &      \\
& AI-Ch-4   & + & + &   &   &   &       \\
\hline

\end{longtable}
\endgroup

\revision{
\subsection{Enabling Technologies across SAR Software Engineering Lifecycle}\label{subsec:relation}
\Cref{tab:phase_tech_matrix} presents the relationships between each enabling technology and the five phases of the software engineering lifecycle for SAR. 
As shown in the table, DTs and AI play key roles across these phases. 
DTs, in particular, provide high-fidelity, executable models of the robotic system and its environment that can be used at multiple phases of the lifecycle. 
In \emph{Requirements Engineering}, DTs enable engineers to validate requirements across various operational contexts (e.g., different layouts, environmental conditions, or human–robot interaction patterns) and verify that the requirements remain coherent under these conditions. 
During \emph{Software Design}, they support virtual prototyping and architecture validation through high-fidelity twin simulations, allowing engineers to explore and evaluate different design configurations and refine system architectures. 
As development progresses into Software Development, Testing, and Operations, DTs continue to provide substantial value. 
In \emph{Software Development}, they provide real-time feedback that guides implementation decisions, validates new features against the intended system behaviour, and helps developers iteratively refine adaptive logic. 
For \emph{Software Testing}, DTs enable safe and cost-effective testing in hardware- and software-in-the-loop setups, allowing engineers to expose the system to a wide range of scenarios, including edge cases and hazardous conditions, that would be difficult or dangerous to replicate in the physical world. 
Finally, in \emph{Operations}, DTs support runtime monitoring, predictive maintenance, and what-if analysis of adaptation strategies by maintaining an up-to-date virtual model of the system.  This allows operators to anticipate failures and trigger adaptive responses proactively, a capability recently demonstrated in~\cite{isaku2025oodisar}. 
}
\begin{table}[htbp]
\centering
\caption{\revision{Relation between enabling technologies and the software engineering lifecycle phases in the context of SAR}}
\label{tab:phase_tech_matrix}
\setlength{\tabcolsep}{4pt} 
\revision{
\begin{tabular}{|
  >{\columncolor{teal!5!white}\raggedright\arraybackslash}p{0.16\linewidth}|
  p{0.38\linewidth}|p{0.38\linewidth}|}
\hline
\rowcolor{teal!5!white}
\textbf{} &
\textbf{Digital Twins (DTs)} &
\textbf{Artificial Intelligence (AI)} \\
\hline
\textbf{Requirements Engineering} & Validate requirements in virtual environments before physical deployment. & Support automated requirements elicitation, conflict detection, and adaptive requirement prioritisation. \\
\hline
\textbf{Software Design} & Enable virtual prototyping and architecture validation through high-fidelity twins. & Support design space exploration, trade-off analysis, and synthesis of adaptive control and decision policies. \\
\hline
\textbf{Software Development} & Provide real-time feedback to validate new features in the twin before deployment. & Automate code generation, self-adaptive logic, and static-analysis support. \\
\hline
\textbf{Software Testing} & Enable safe and cost-effective hardware-/software-in-the-loop testing in realistic twin scenarios. & Support automated test generation and prioritisation, fault localisation, adaptive test execution, and test oracle inference. \\
\hline
\textbf{Operations} & Enable continuous runtime monitoring, predictive maintenance, and what-if analysis of adaptation strategies in the twin. & Facilitate runtime adaptation, predictive analytics, online anomaly and fault detection. \\
\hline
\end{tabular}
}
\end{table}

\revision{Artificial intelligence contributes more pervasively across all phases of the software engineering lifecycle for SAR. 
In \emph{Requirements Engineering}, AI techniques (e.g., natural language processing and LLMs) can support automated requirements elicitation, the detection of conflicts and inconsistencies in requirements, and the prioritisation of requirements based on risk or stakeholder impact. 
In \emph{Software Design}, AI enables intelligent design space exploration, trade-off analysis, and the synthesis and evaluation of adaptive control and decision policies, helping engineers navigate the complexity of designing systems that must balance quality aspects such as safety, reliability, and performance. 
During \emph{Software Development}, AI techniques assist with automated code generation and completion, the implementation of self-adaptive logic, and support for static analysis (e.g., detecting potential defects or security weaknesses). 
In \emph{Software Testing}, AI is increasingly used to generate and prioritise test cases, to perform intelligent fault detection and localisation, to develop adaptive test execution strategies, and to infer test oracles from specifications or historical traces. 
During \emph{Operations}, AI serves as the cognitive backbone of self-adaptive robotic systems, enabling autonomous decision-making, real-time adaptation, predictive analytics, and anomaly detection in dynamic, unpredictable environments. 
This enables robotic systems to respond intelligently to unforeseen changes without human intervention. 
}

\revision{
Looking ahead to 2030, we expect both DTs and AI to become even more deeply integrated into the software engineering lifecycle for SAR. 
DTs are likely to evolve into more intelligent, semantically rich, and real-time synchronised models, capable not only of reflecting physical systems but also of predicting future states and autonomously triggering adaptive responses. 
AI, and in particular LLMs and foundation models specialised for robotics and software engineering, will increasingly support end-to-end lifecycle activities, from interactive requirements elicitation and automated design-space exploration, through AI-assisted implementation and test generation, to autonomous runtime adaptation and fleet-level optimisation in operations. 
Together, these technologies are likely to give rise to a new generation of self-learning robotic systems that can continuously refine their behaviour, anticipate environmental changes, and self-optimise across the entire lifecycle, ultimately bringing SAR closer to full autonomy in safety-critical real-world applications. 
}

\subsection{Research Opportunities} \label{subsec:opportunities} 
Given the numerous challenges currently being addressed across various fields of self-adaptive robotics, we aim to identify and outline future research directions and long-term challenges. An overview of these opportunities is given in the green box below.

Within \textbf{Requirements Engineering} (O1--O2), there is a need for incorporating insights from social sciences and humanities to improve the interaction of humans and robots. Specifically, we need to include social and ethical norms and enable robots to identify, understand, and reason on them. For example, robots might be trusted more easily in some societies than in others. In response, SARs need to adjust their behaviour according to their social environment. With this, not only will the safety of interactions with robots improve, but it can also lead to better performance of the robot in its required tasks.

In \textbf{Software Design} (O3--O4), we need to develop scalable, modular software architectures that enable safe runtime reconfiguration. This includes the ability to dynamically compose and recompose software modules based on the current context and requirements. This, in itself, creates several challenges, such as ensuring safe reconfiguration, consistency, and minimal disruption, while also providing semantic information for each software module to enable greater autonomy in the reconfiguration process.
Furthermore, we need to advance hybrid MDE and simulation frameworks. These frameworks need to provide simple approaches to test various types of AI, software components, and hardware models. They further need to provide adaptive co-simulation to enable more accurate and efficient design and testing of SARs.

In \textbf{Software Development} (O5--O7), we need to efficiently integrate heterogeneous hardware and software components and enable the evolution of these components. This includes the ability to seamlessly integrate new and replace existing sensors, actuators, and software modules into existing robotic systems. This requires standardised interfaces, protocols, and middleware capable of handling the complexity and diversity of robotic systems.
Moreover, foundation models are required for robotic applications. This needs to capture all aspects of robotics, including real-time sensor fusion, which enables self-adaptation. These models should be capable of processing and integrating data from multiple sensors in real time, thereby reducing development effort. This, in turn, will provide the robots with a better understanding of the context and environment, advancing their self-adaptive capabilities. Additionally, we need to focus on energy-efficient AI-driven robotic software development. This includes developing algorithms and techniques that run efficiently on resource-constrained robotic platforms while maintaining high performance and adaptability.

In \textbf{Software Testing} (O8--O10), future research over the coming years needs to address the reality gap in SAR simulation. This includes developing techniques to bridge the gap between simulated environments and real-world scenarios, ensuring that tests conducted in simulation accurately reflect robot behaviour under real-world conditions. This can be achieved through techniques such as domain randomisation, transfer learning, and sim-to-real adaptation.
Furthermore, we need to advance testing techniques to address uncertainty, nondeterminism, and safety risks in AI components embodied in SARs. This includes developing methods to systematically and rigorously test and validate AI algorithms under various conditions. While exhaustive testing may not be possible, providing guarantees and verification techniques can help ensure the reliability and safety of SARs in dynamic and uncertain environments. Additionally, we need to advance security testing to strengthen SARs' resilience against emerging threats. This includes developing techniques to identify and mitigate vulnerabilities in robotic systems, ensuring their integrity and availability in the face of cyber-attacks.

\begin{tcolorbox}[colback=green!2!white, colframe=teal!70, title=Research Opportunities]
\textbf{\textit{Requirements Engineering}}\\[3pt]
    \textbf{O1:} Interdisciplinary research combining safety and human behaviour.\\
    \textbf{O2:} Incorporating normative requirements in SARs.\\
    
\textbf{\textit{Software Design}}\\[3pt]
    \textbf{O3:} Develop scalable and modular software architectures that ensure safe runtime reconfiguration.\\
    \textbf{O4:} Advance hybrid MDE and simulation frameworks incorporating AI, hardware models, and adaptive co-simulation.\\

\textbf{\textit{Software Development}}\\[3pt]
    \textbf{O5:} Efficient integration of heterogeneous and evolving hardware and software components. \\
    \textbf{O6:} Large language models for real-time sensor fusion to advance self-adaptation. \\
    \textbf{O7:} Energy-efficient AI-driven robotic software development. \\

\textbf{\textit{Software Testing}}\\[3pt]
    \textbf{O8:} Address reality gap in simulation-based testing of SARs. \\
    \textbf{O9:} Advance testing techniques to address uncertainty, nondeterminism, and safety risks in AI components embodied in SARs. \\
    \textbf{O10:} Advance security testing to strengthen SARs' resilience against emerging threats.\\

\textbf{\textit{Operations}}\\[3pt]
    \textbf{O11:} Uncertainty management across all phases of the MAPLE-K loop.\\
    \textbf{O12:} Dynamic deployment and maintenance \revision{through continuous Dev(Sec)Ops pipelines.}\\
    
\textbf{\textit{Digital Twins}}\\[3pt]
    \textbf{O13:} Enabling IP to be protected in DTs supporting multiple legal entities.\\
    \textbf{O14:} Develop scalable and adaptable DT architectures to support integration of complex robotic subsystems.\\
    
\textbf{\textit{Artificial Intelligence}}\\[3pt]
    \textbf{O15:} Foundational models for physics and general behaviour.\\
    \textbf{O16:} Autonomous dynamic training and learning.\\

\end{tcolorbox}

When it comes to \textbf{Operations} (O11--O12) of SARs, we will need to manage uncertainty within all stages of the MAPE-K loop. This includes developing techniques to quantify, model, and mitigate uncertainty in the perception, decision-making, and adaptation processes of SARs. This will enable robots to make more informed decisions and adapt more effectively to changing environments and requirements. Additionally, we need to focus on dynamic deployment and maintenance through continuous Dev(Sec)Ops pipelines for heterogeneous fleets. This includes developing techniques to continuously build, qualify, secure, and deploy updates in real-time while preserving safety and availability. Such pipelines should support staged and canary rollouts, runtime monitoring, and dependable rollback strategies, and they should explicitly integrate evidence from simulation/testing stages (e.g., MiL/SiL/HiL and regression testing under continuous evolution) before fleet-wide promotion.

In the area of \textbf{Digital Twins} (O13--O14), we need to enable IP protection in DTs, particularly when collaborating across multiple legal entities. This includes developing techniques to ensure the confidentiality, integrity, and availability of data and models used in DTs, while also allowing collaboration and sharing among different stakeholders. Additionally, we need to develop scalable and adaptable DT architectures to support the integration of complex robotic subsystems. This includes developing modular, flexible architectures that accommodate the diverse and evolving needs of SARs, enabling seamless integration and interaction among components and systems. Additionally, we must consider the different fidelities of DTs for various use cases and situations, e.g., high-fidelity DTs for highly critical situations and low-fidelity DTs for safe states of the SAR operation. This will require work to generate these DTs, potentially on the fly, as well as using and switching between them during SAR operation.

\revision{Finally, in the field of \textbf{Artificial Intelligence} (O15--O16), we need to develop foundational models that capture the underlying physical principles and dynamics of robotic systems, enabling more accurate and efficient perception, decision-making, and control.} These models should be capable of generalising across different tasks and environments, reducing the need for task-specific training and adaptation.
Furthermore, we need to improve autonomous dynamic training and online learning. This includes developing techniques that enable robots to autonomously explore their environment and acquire new skills and knowledge through interaction with the same. Generating the scenario and executing it safely, without damaging or endangering the SAR and others, is crucial.

\section{Related Works}\label{sec:relatedworks}
Several efforts have been made to outline future roadmaps for robotics software engineering. 
\citet{garcia2020robotics} presented research challenges and future directions based on an empirical evaluation with industry practitioners from the service robotics domain. 
The main differences are: (i) our work is more comprehensive without targeting any specific robotics domain, and (ii) our focus is on self-adaptive robotics. 
\citet{goues2024software} presented a roadmap for robotics software engineering based on a discussion with experts from industry and academia. 
Our work differs in that it focuses specifically on self-adaptive robotics and presents a roadmap based on a detailed review of state-of-the-art research across key phases of the software engineering lifecycle.

For self-adaptive systems, \citet{de2013software} outlined a roadmap for their software development lifecycle. 
\citet{weyns2023towards} presented challenges and future directions for uncertainty in self-adaptive systems. 
Recently, \citet{li2024exploring} surveyed the research potential and challenges of large language models for self-adaptive systems.
Although these works generally cover self-adaptive systems, we mainly focus on self-adaptive robots.  
\citet{brugali2024future} outlined open challenges and future directions across four aspects of robotic software engineering: domain expertise and architecture, verification and validation, prototype-to-product transition, and trustworthiness. 
Our work differs in several key aspects: (i) it focuses on current challenges and future opportunities specific to self-adaptive robotics, (ii) structures the roadmap across software engineering phases, spanning from requirements to operations, (iii) provides a detailed analysis of enabling technologies like DTs, and (iv) examines critical quality aspects essential for self-adaptive robotics.

Some related systematic reviews also exist.
\citet{bozhinoski2019safety} presented a review of available software engineering solutions for mobile robot safety.
Similarly, \citet{casalaro2022model} presented a systematic review of MDE for mobile robots. 
These works primarily focus on literature reviews on mobile robots or on specific aspects, such as MDE. 
In comparison, we present a forward-looking roadmap for self-adaptive robotics, identifying open challenges and future research opportunities through a comprehensive review of the existing literature, and covering key phases of the software engineering lifecycle and enabling technologies such as AI.

\section{Conclusion}\label{sec:conclusion}
Self-adaptive robotics poses fundamental software engineering challenges across the lifecycle, from evolving requirements to runtime operations. 
Addressing these challenges requires rigorous methods that integrate runtime verification with continuous learning, and development frameworks that combine model-driven engineering, AI-based decision-making, and formal verification. Promising directions include co-simulation and digital twins, which can close the reality gap by enabling safe and efficient testing and validation of adaptive behaviours before deployment. 
A central open problem remains how to ensure trustworthiness by capturing safety, reliability, explainability, and ethical accountability when data and uncertainty drive adaptation. 
Looking toward 2030, progress will depend on combining advances in adaptive architectures, assurance techniques, and enabling technologies such as AI and digital twins. 
By consolidating these perspectives into a unified roadmap, this paper establishes a structured foundation for the next generation of self-adaptive robotic systems, enabling them to achieve trustworthy autonomy in critical domains such as healthcare, manufacturing, and autonomous transportation.

\section*{Acknowledgments}
This research is supported by the RoboSAPIENS project, funded through the European Commission Horizon Europe program under Grant Agreement No. 101133807.

\bibliographystyle{plainnat}  
\bibliography{refs,espfor,publications,robochart,jim}

@ARTICLE{ALLIRV21,
  author = "M.~Askarpour and L.~Lestingi and S.~Longoni and N.~Iannacci and M.~Rossi and F.~Vicentini",
  title = "Formally-based Model-Driven Development of Collaborative Robotic Applications",
  journal = "Journal of Intelligent {\&} Robotic Systems",
  year = "2021",
  volume = "102",
  number = "3",
  pages = "59",
}

@INPROCEEDINGS{GLD14,
  author = "N.~Gobillot and C.~Lesire and D.~Doose",
  editor = "D.~Brugali and J.~F.~Broenink and T.~Kroeger and B.~A.~MacDonald",
  title = "A Modeling Framework for Software Architecture Specification and Validation",
  booktitle = "Simulation, Modeling, and Programming for Autonomous Robots",
  year = "2014",
  publisher = "Springer International Publishing",
  address = "Cham",
  pages = "303-314",
}

@INCOLLECTION{GS13,
  author = {H.~Giese and W.~Sch{\"{a}}fer},
  editor = {J.~C{\'{a}}mara and R.~Lemos and C.~Ghezzi and A.~Lopes},
  title = {{Model-Driven Development of Safe Self-optimizing Mechatronic Systems with MechatronicUML}},
  booktitle = {Assurances for Self-Adaptive Systems - Principles, Models, and Techniques},
  series = {Lecture Notes in Computer Science},
  volume = {7740},
  pages = {152-186},
  publisher = {Springer},
  address = {Berlin Heidelberg},
  year      = {2013},
}

@INPROCEEDINGS{OBM10,
  author = "P.~C.~{\"O}lveczky and A.~Boronat and J.~Meseguer",
  editor = "J.~Hatcliff and E.~Zucca",
  title = "Formal Semantics and Analysis of Behavioral AADL Models in Real-Time Maude",
  booktitle="Formal Techniques for Distributed Systems",
  year="2010",
  publisher="Springer",
  address="Berlin Heidelberg",
  pages="47-62",
}

@INPROCEEDINGS{RMT14,
  author = {A. Ramaswamy and B. Monsuez and A. Tapus},
  booktitle = {2014 IEEE/RSJ International Conference on Intelligent Robots and Systems},
  title = {SafeRobots: A model-driven Framework for developing Robotic Systems},
  year = {2014},
  pages = {1517-1524},
  publisher={IEEE},
  address = {New York, NY, USA},
}

@ARTICLE{RRRW15,
  author = {J.~O.~Ringert and A.~Roth and B.~Rumpe and A.~Wortmann},
  title = {Code Generator Composition for Model-Driven Engineering of Robotics Component {\&} Connector Systems},
  journal = {Journal of Software Engineering for Robotics},
  volume = {6},
  number = {1},
  year = {2015},
  pages = {33-57}
}

@INPROCEEDINGS{ZH11,
  author = {L.~Zhang and He~Jifeng},
  title = {A Formal Framework for Aspect-Oriented Specification of Cyber Physical Systems},
  booktitle = {Convergence and Hybrid Information Technology},
  pages = {391-398},
  editor = {G.~Lee and D.~Howard and D.~Slezak},
  publisher = {Springer},
  address = {Berlin, Heidelberg},
  series = {Communications in Computer and Information Science},
  volume = {206},
  year = {2011},
}

@MANUAL{RoboChart,
  title = {{RoboChart Reference Manual}},
  organization = {University of York},
  note = {{\url{www.cs.york.ac.uk/circus/RoboCalc/robotool/}}},
}

@book{vanLamsweerde2009,
  author    = {Axel van Lamsweerde},
  title     = {Requirements Engineering: From System Goals to UML Models to Software Specifications},
  year      = {2009},
  publisher = {Wiley},
  address = {NJ, USA}
}

@book{Fitzgerald2014,
  author    = {John Fitzgerald and Peter G. Larsen and Marcel Verhoef},
  title     = {Collaborative Design for Embedded Systems: Co-modelling and Co-simulation},
  year      = {2014},
  publisher = {Springer},
  address = {Berlin, Heidelberg},
  doi       = {10.1007/978-3-642-54118-6}
}

@article{Kritzinger2018a,
author = {Kritzinger, Werner and Karner, Matthias and Traar, Georg and Henjes, Jan and Sihn, Wilfried},
year = {2018},
month = {01},
pages = {1016-1022},
doi = {10.1016/j.ifacol.2018.08.474},
title = {Digital Twin In Manufacturing: A Categorical Literature Review And Classification},
volume = {51},
journal = {IFAC-PapersOnLine},
}

@article{Robinson2003,
  author    = {William N. Robinson and Suzanne D. Pawlowski and Vladimir Volkov},
  title     = {Requirements interaction management},
  journal   = {ACM Computing Surveys},
  volume    = {35},
  number    = {2},
  pages     = {132--190},
  year      = {2003},
  doi       = {10.1145/857076.857078}
}

@article{Brugali2010,
  author    = {Davide Brugali and Azamat Shakhimardanov},
  title     = {Component-based robotics},
  journal   = {IEEE Robotics \& Automation Magazine},
  volume    = {17},
  number    = {1},
  pages     = {99--111},
  year      = {2010},
  doi       = {10.1109/MRA.2010.936952}
}

@ARTICLE{LMCCISHLL15,
  year = "2017",
  volume = "16",
  number = "3",
  journal = {Software \& Systems Modeling},
  title = {{An integrated semantics for reasoning about SysML design models using refinement}},
  publisher = {Springer},
  author = {L.~Lima and A.~Miyazawa and A.~L.~C.~Cavalcanti and M.~Corn\'elio and J.~Iyoda and A.~C.~A.~Sampaio and R.~Hains and A.~Larkham and V.~Lewis},
  pages = {1-28},
  url = {papers/LMCCISHLL15.pdf},
  doi = {10.1007/s10270-015-0492-y}
}

@ARTICLE{TPANCCHT22,
  title = "{From Pluralistic Normative Principles to Autonomous-Agent Rules}",
  journal = "Minds and Machines",
  year = "2022",
  author = "B.~Townsend and C.~Paterson and T.~T.~Arvind and G.~Nemirovsky and R.~Calinescu and A.~L.~C.~Cavalcanti and I.~Habli and A.~Thomas",
  doi = {https://doi.org/10.1007/s11023-022-09614-w},
  publisher = {Kluwer Academic Publishers},
  address = {USA},
  volume = {32},
  number = {4},
  pages = {683–715},
  numpages = {33},
}

@INPROCEEDINGS{FMYTCCC23,
  author = "N.~Feng and L.~Marsso and S.~G.~Yaman and B.~Townsend and A.~L.~C.~Cavalcanti and R.~Calinescu and M.~Chechik",
  title = "{Towards a Formal Framework for Normative Requirements Elicitation}",
  booktitle = "Automated Software Engineering",
  year = "2023",
  series = "Conference Publishing Services",
  pages = "1776-1780",
  publisher={IEEE},
  address = {New York, NY, USA},
}

@INPROCEEDINGS{FMYBA24,
  author = {N.~Feng and L.~Marsso and S.~G.~Yaman and Y.~Baatartogtokh and R.~Ayad and V.~O.~D.~Mello and B.~Townsend and I.~Standen and I.~Stefanakos and C.~Imrie and G.~N.~Rodrigues and A.~L.~C.~Cavalcanti and R.~Calinescu and M.~Chechik},
  title = {{Analyzing and Debugging Normative Requirements via Satisfiability Checking}},
  year = {2024},
  publisher = {Association for Computing Machinery},
  address = {New York, NY, USA},
  doi = {10.1145/3597503.3639093},
  booktitle = {IEEE/ACM 46th International Conference on Software Engineering},
  articleno = {214},
  numpages = {12},
}

@InProceedings{Biglari2022,
  author     = {Biglari, Raheleh and Denil, Joachim},
  booktitle  = {Proceedings of the 25th International Conference on Model Driven Engineering Languages and Systems: Companion Proceedings},
  title      = {Model validity and tolerance quantification for real-time adaptive approximation},
  year       = {2022},
  month      = oct,
  pages      = {668--676},
  publisher  = {ACM},
  address = {New York, NY, USA},
  series     = {MODELS ’22},
  collection = {MODELS ’22},
  doi        = {10.1145/3550356.3561604},
}

@article{YRCCPT25,
  title = {Specification, validation and verification of social, legal, ethical, empathetic and cultural requirements for autonomous agents},
  journal = {Journal of Systems and Software},
  volume = {220},
  pages = {112229},
  year = {2025},
  doi = {10.1016/j.jss.2024.112229},
  url = {https://www.sciencedirect.com/science/article/pii/S0164121224002735},
  author = {S.~G.~Yaman and P.~Ribeiro and A.~L.~C.~Cavalcanti and R.~Calinescu and C.~Paterson and B.~Townsend}
}

@article{Frasheri2022CAESAR,
  author  = {Frasheri, Mirgita and Struhar, Vaclav and Papadopoulos, Alessandro Vittorio and Causevic, Aida},
  title   = {Ethics of Autonomous Collective Decision-Making: The CAESAR Framework},
  journal = {Science and Engineering Ethics},
  volume  = {28},
  number  = {6},
  pages   = {61},
  year    = {2022},
  doi     = {10.1007/s11948-022-00414-0}
}

@article{Dennis2016FormalVerification,
  author  = {Dennis, Louise and Fisher, Michael and Slavkovik, Marija and Webster, Matt},
  title   = {Formal verification of ethical choices in autonomous systems},
  journal = {Robotics and Autonomous Systems},
  volume  = {77},
  pages   = {1--14},
  year    = {2016},
  doi={10.1016/j.robot.2015.11.012}
}

@inproceedings{Bhuiyan2020Encoding,
  author    = {Bhuiyan, Hridoy and Olivieri, Franco and Governatori, Guido and Islam, Md. Belal and Bond, Alan and Rakotonirainy, Andry},
  title     = {A methodology for encoding regulatory rules},
  booktitle={Proceedings of the 4th International Workshop on Mining and Reasoning with Legal texts: co-located with the 32nd International Conference on Legal Knowledge and Information Systems (JURIX 2019)},
  pages={1--13},
  year={2019},
  publisher={Sun SITE Central Europe (CEUR)},
  address = {Aachen},
  url = {https://eprints.qut.edu.au/236378/}
}

@inproceedings{Troquard2024SLEEC,
  author    = {Troquard, Nicolas and De Sanctis, Martina and Inverardi, Paola and Pelliccione, Patrizio and Scoccia, Gian Luca},
  title     = {Social, Legal, Ethical, Empathetic, and Cultural Rules: Compilation and Reasoning},
  booktitle = {Proceedings of the AAAI Conference on Artificial Intelligence (AAAI)},
  volume    = {38},
  pages     = {22385--22392},
  year      = {2024},
  doi       = {10.1609/aaai.v38i20.30245},
  publisher = {AAAI Press},
  address = {Washington, DC, USA}
}

@inproceedings{Alfieri2022HHAI,
  author    = {Alfieri, Constanza and Inverardi, Paola and Migliarini, Patrizio and Palmiero, Massimiliano},
  title     = {Exosoul: Ethical Profiling in the Digital World}, 
  booktitle = {HHAI 2022: Augmenting Human Intellect},
  editor    = {Schlobach, Stefan and P{\'e}rez-Ortiz, Mar{\'\i}a and Tielman, Mel}, 
  series    = {Frontiers in Artificial Intelligence and Applications},
  volume    = {354},
  pages     = {128--142},
  publisher = {IOS Press},
  year      = {2022},
  address   = {Amsterdam, Netherlands},
  doi={10.3233/FAIA220194}
}

@inbook{Inverardi2022HumanDignity,
  author    = {Inverardi, Paola},
  title     = {The Challenge of Human Dignity in the Era of Autonomous Systems},
  publisher = {Springer International Publishing},
  year      = {2022},
  address   = {Cham},
  chapter   = {4},
  pages     = {25--29},
  doi = {10.1007/978-3-030-86144-5_4}
}

@techreport{Fjeld2020PrincipledAI,
  author      = {Fjeld, Jessica and Achten, Nele and Hilligoss, Hannah and Nagy, Adam and Srikumar, Madhulika},
  title       = {Principled Artificial Intelligence: Mapping Consensus in Ethical and Rights-based Approaches to Principles for AI},
  institution = {Berkman Klein Center for Internet \& Society},
  year        = {2020},
  number      = {2020-1},
  address     = {Cambridge, MA},
  url         = {https://dash.harvard.edu/handle/1/42160420}
}

@article{Bremner2019Proactive,
  author    = {Paul A. Bremner and Louise A. Dennis and Michael Fisher and Alan F. T. Winfield},
  title     = {On Proactive, Transparent, and Verifiable Ethical Reasoning for Robots},
  journal   = {Proceedings of the IEEE},
  volume    = {107},
  number    = {3},
  pages     = {541--561},
  year      = {2019},
  doi       = {10.1109/JPROC.2019.2898267}
}

@InProceedings{Kristensen&25,
author="Kristensen, Morten Haahr
and Wright, Thomas
and Gomes, Cl{\'a}udio
and Esterle, Lukas
and Larsen, Peter Gorm",
editor="K{\"o}nighofer, Bettina
and Torfah, Hazem",
title="DynSRV: Dynamically Updated Properties for Stream Runtime Verification",
booktitle="Runtime Verification",
year="2026",
publisher="Springer Nature Switzerland",
address="Cham",
pages="101--119",
isbn="978-3-032-05435-7",
doi="10.1007/978-3-032-05435-7_7"
}

@article{pozzi2022modeling,
  title={Modeling and Simulation of Robotic Grasping in Simulink Through Simscape Multibody},
  author={Pozzi, Maria and Achilli, Gabriele Maria and Valigi, Maria Cristina and Malvezzi, Monica},
  journal={Frontiers in Robotics and AI},
  volume={9},
  pages={873558},
  year={2022},
  publisher={Frontiers Media SA},
  doi={10.3389/frobt.2022.873558}
}

@inproceedings{de2013software,
  title={Software engineering for self-adaptive systems: A second research roadmap},
  author={De Lemos, Rog{\'e}rio and Giese, Holger and M{\"u}ller, Hausi A and Shaw, Mary and Andersson, Jesper and Litoiu, Marin and Schmerl, Bradley and Tamura, Gabriel and Villegas, Norha M and Vogel, Thomas and others},
  booktitle={Software Engineering for Self-Adaptive Systems II: International Seminar, Dagstuhl Castle, Germany, October 24-29, 2010 Revised Selected and Invited Papers},
  pages={1--32},
  volume={7475},
  year={2013},
  publisher={Springer},
  address={Berlin, Heidelberg},
  doi={10.1007/978-3-642-35813-5_1}
}

@inproceedings{garcia2020robotics,
  title={Robotics software engineering: A perspective from the service robotics domain},
  author={Garc{\'\i}a, Sergio and Str{\"u}ber, Daniel and Brugali, Davide and Berger, Thorsten and Pelliccione, Patrizio},
  booktitle={Proceedings of the 28th ACM Joint Meeting on European Software Engineering Conference and Symposium on the Foundations of Software Engineering},
  pages={593--604},
  year={2020},
  publisher = {Association for Computing Machinery},
  address = {New York, NY, USA},
  url = {https://doi.org/10.1145/3368089.3409743},
  doi = {10.1145/3368089.3409743}
}

@misc{goues2024software,
      title={Software Engineering for Robotics: Future Research Directions; Report from the 2023 Workshop on Software Engineering for Robotics}, 
      author={Claire Le Goues and Sebastian Elbaum and David Anthony and Z. Berkay Celik and Mauricio Castillo-Effen and Nikolaus Correll and Pooyan Jamshidi and Morgan Quigley and Trenton Tabor and Qi Zhu},
      year={2024},
      eprint={2401.12317},
      archivePrefix={arXiv},
      primaryClass={cs.RO},
      url={https://arxiv.org/abs/2401.12317}, 
}

@article{bozhinoski2019safety,
  title={Safety for mobile robotic systems: A systematic mapping study from a software engineering perspective},
  author={Bozhinoski, Darko and Di Ruscio, Davide and Malavolta, Ivano and Pelliccione, Patrizio and Crnkovic, Ivica},
  journal={Journal of Systems and Software},
  volume={151},
  pages={150--179},
  year={2019},
  publisher={Elsevier},
  doi={10.1016/j.jss.2019.02.021}
}

@article{casalaro2022model,
  title={Model-driven engineering for mobile robotic systems: a systematic mapping study},
  author={Casalaro, Giuseppina Lucia and Cattivera, Giulio and Ciccozzi, Federico and Malavolta, Ivano and Wortmann, Andreas and Pelliccione, Patrizio},
  journal={Software and Systems Modeling},
  volume={21},
  number={1},
  pages={19--49},
  year={2022},
  publisher={Springer},
  doi={10.1007/s10270-021-00908-8}
}

@inproceedings{li2024exploring,
  title={Exploring the Potential of Large Language Models in Self-adaptive Systems},
  author={Li, Jialong and Zhang, Mingyue and Li, Nianyu and Weyns, Danny and Jin, Zhi and Tei, Kenji},
  booktitle={Proceedings of the 19th International Symposium on Software Engineering for Adaptive and Self-Managing Systems},
  pages={77--83},
  year={2024},
  publisher = {Association for Computing Machinery},
  address = {New York, NY, USA},
  url = {https://doi.org/10.1145/3643915.3644088},
  doi = {10.1145/3643915.3644088}
}

@article{brugali2024future,
  title={Future Directions in Software Engineering for Autonomous Robots: An Agenda for Trustworthiness [Opinion]},
  author={Brugali, Davide and Cavalcanti, Ana and Hochgeschwender, Nico and Pelliccione, Patrizio and Rebelo, Luciana},
  journal={IEEE Robotics \& Automation Magazine},
  volume={31},
  number={3},
  pages={186--204},
  year={2024},
  publisher={IEEE},
  address = {New York, NY, USA},
  doi={10.1109/MRA.2024.3417089}
}

@article{weyns2023towards,
  title={Towards a research agenda for understanding and managing uncertainty in self-adaptive systems},
  author={Weyns, Danny and Calinescu, Radu and Mirandola, Raffaela and Tei, Kenji and Acosta, Maribel and Bencomo, Nelly and Bennaceur, Amel and Boltz, Nicolas and Bures, Tomas and Camara, Javier and others},
  journal={ACM SIGSOFT Software Engineering Notes},
  volume={48},
  number={4},
  pages={20--36},
  year={2023},
  publisher={ACM},
  address = {New York, NY, USA},
  url = {https://doi.org/10.1145/3617946.3617951},
  doi = {10.1145/3617946.3617951}
}

@article{kephart2003vision,
  title={The vision of autonomic computing},
  author={Kephart, Jeffrey O and Chess, David M},
  journal={Computer},
  volume={36},
  number={1},
  pages={41--50},
  year={2003},
  publisher={IEEE},
  address = {New York, NY, USA},
  doi={10.1109/MC.2003.1160055}
}

@article{larsen2024robotic,
  title={Robotic safe adaptation in unprecedented situations: the RoboSAPIENS project},
  author={Larsen, Peter G and Ali, Shaukat and Behrens, Roland and Cavalcanti, Ana and Gomes, Claudio and Li, Guoyuan and De Meulenaere, Paul and Olsen, Mikkel L and Passalis, Nikolaos and Peyrucain, Thomas and others},
  journal={Research Directions: Cyber-Physical Systems},
  volume={2},
  pages={e4},
  year={2024},
  publisher={Cambridge University Press},
  doi={10.1017/cbp.2024.4}
}

@article{cavalcanti2021robostar,
  title={{RoboStar} technology: a roboticist’s toolbox for combined proof, simulation, and testing},
  author={Cavalcanti, Ana and Barnett, Will and Baxter, James and Carvalho, Gustavo and Filho, Madiel Conserva and Miyazawa, Alvaro and Ribeiro, Pedro and Sampaio, Augusto},
  journal={Software Engineering for Robotics},
  pages={249--293},
  year={2021},
  publisher={Springer},
  doi={10.1007/978-3-030-66494-7_9}
}

@article{hierons2021mutation,
  title={Mutation Testing for {RoboChart}},
  author={Hierons, Robert M and Gazda, Maciej and G{\'o}mez-Abajo, Pablo and Lefticaru, Raluca and Merayo, Mercedes G},
  journal={Software Engineering for Robotics},
  pages={345--375},
  year={2021},
  publisher={Springer},
  doi={10.1007/978-3-030-66494-7_11}
}

@article{sartaj2021testing,
  author={Sartaj, Hassan and Iqbal, Muhammad Zohaib and Khan, Muhammad Uzair},
  title={Testing cockpit display systems of aircraft using a model-based approach},
  journal={Software and Systems Modeling},
  volume={20},
  number={6},
  pages={1977--2002},
  year={2021},
  publisher={Springer Berlin Heidelberg Berlin/Heidelberg},
  doi={10.1007/s10270-020-00844-z}
}

@inproceedings{sartaj2021automated,
  title={Automated approach for system-level testing of unmanned aerial systems},
  author={Sartaj, Hassan},
  booktitle={2021 36th IEEE/ACM International Conference on Automated Software Engineering (ASE)},
  pages={1069--1073},
  year={2021},
  publisher={IEEE},
  address = {New York, NY, USA},
  doi={10.1109/ASE51524.2021.9678902},
}

@article{sartaj2024automated,
  author={Sartaj, Hassan and Muqeet, Asmar and Iqbal, Muhammad Zohaib and Khan, Muhammad Uzair},
  title={Automated system-level testing of unmanned aerial systems},
  journal={Automated Software Engineering},
  volume={31},
  number={64},
  pages={1--48},
  year={2024},
  publisher={Springer},
  doi={10.1007/s10515-024-00462-9}
}

@inproceedings{mania2025towards,
  title={Towards autonomous verification: Integrating cognitive AI and semantic digital twins in medical robotics},
  author={Mania, Patrick and Neumann, Michael and Kenfack, Franklin Kenghagho and Beetz, Michael},
  booktitle={2025 IEEE International Conference on Robotics and Automation (ICRA)},
  pages={1736--1742},
  year={2025},
  publisher={IEEE},
  address = {New York, NY, USA},
  doi={10.1109/ICRA55743.2025.11127803}
}

@article{xiao2025robot,
  title={Robot learning in the era of foundation models: A survey},
  author={Xiao, Xuan and Liu, Jiahang and Wang, Zhipeng and Zhou, Yanmin and Qi, Yong and Jiang, Shuo and He, Bin and Cheng, Qian},
  journal={Neurocomputing},
  volume = {638},
  number = {14},
  pages={129963},
  year={2025},
  publisher={Elsevier},
  doi={10.1016/j.neucom.2025.129963}
}

@inproceedings{schreiter2025evaluating,
  title={Evaluating efficiency and engagement in scripted and {LLM}-enhanced human-robot interactions},
  author={Schreiter, Tim and R{\"u}ppel, Jens V and Hazra, Rishi and Rudenko, Andrey and Magnusson, Martin and Lilienthal, Achim J},
  booktitle={2025 20th ACM/IEEE International Conference on Human-Robot Interaction (HRI)},
  pages={1608--1612},
  year={2025},
  publisher={IEEE},
  address = {New York, NY, USA},
  doi={10.1109/HRI61500.2025.10974124}
}

@article{greengard2025can,
  title={Can LLMs Make Robots Smarter?},
  author={Greengard, Samuel},
  year={2025},
  publisher={ACM},
  address = {New York, NY, USA},
  volume = {68},
  number = {2}, 
  doi = {10.1145/3701227},
  journal={Communications of the ACM},
  pages = {11–13}
}

@inproceedings{wu2024safety,
  title={On the safety concerns of deploying {LLMs}/{VLMs} in robotics: Highlighting the risks and vulnerabilities},
  author={Wu, Xiyang and Xian, Ruiqi and Guan, Tianrui and Liang, Jing and Chakraborty, Souradip and Liu, Fuxiao and Sadler, Brian M and Manocha, Dinesh and Bedi, Amrit},
  booktitle={First Vision and Language for Autonomous Driving and Robotics Workshop},
  pages = {1–10},
  year={2024},
}

@inproceedings{wang2025gsce,
  title={{GSCE}: A prompt framework with enhanced reasoning for reliable {LLM}-driven drone control},
  author={Wang, Wenhao and Li, Yanyan and Jiao, Long and Yuan, Jiawei},
  booktitle={2025 International Conference on Unmanned Aircraft Systems (ICUAS)},
  pages={441--448},
  year={2025},
  publisher={IEEE},
  address = {New York, NY, USA},
  doi={10.1109/ICUAS65942.2025.11007864}
}

@misc{zhang2025enhancing,
      title={Enhancing Reliability in {LLM}-Integrated Robotic Systems: A Unified Approach to Security and Safety}, 
      author={Wenxiao Zhang and Xiangrui Kong and Conan Dewitt and Thomas Bräunl and Jin B. Hong},
      year={2025},
      eprint={2509.02163},
      archivePrefix={arXiv},
      primaryClass={cs.RO},
      url={https://arxiv.org/abs/2509.02163}, 
}

@article{sartaj2025search,
  author={Sartaj, Hassan and Iqbal, Muhammad Zohaib and Jilani, Atif Aftab Ahmed and Khan, Muhammad Uzair},
  title={{Search-Based MC/DC Test Data Generation With OCL Constraints}},
  journal={Software Testing, Verification And Reliability},
  volume = {35},
  number = {1},
  pages = {e1906},
  year={2025},
  publisher={Wiley Online Library},
  doi={10.1002/stvr.1906},
  url = {https://onlinelibrary.wiley.com/doi/abs/10.1002/stvr.1906}
}

@article{sartaj2025medet,
    title={{MeDeT: Medical Device Digital Twins Creation with Few-shot Meta-learning}}, 
    author = {Sartaj, Hassan and Ali, Shaukat and Gjøby, Julie Marie},
    journal = {ACM Transactions on Software Engineering and Methodology},
    volume = {34},
    number = {6},
    pages = {1-36},
    doi = {10.1145/3708534},
    url = {https://dl.acm.org/doi/10.1145/3708534},
    year={2025},
    publisher = {Association for Computing Machinery},
    address = {New York, NY, USA}
}

@article{sartaj2024modelbased,
    author = {Sartaj, Hassan and Ali, Shaukat and Yue, Tao and Moberg, Kjetil},
    title = {Model-based digital twins of medicine dispensers for healthcare {IoT} applications},
    journal = {Software: Practice and Experience},
    volume = {54},
    number = {6},
    pages = {1172-1192},
    doi = {10.1002/spe.3311},
    url = {https://onlinelibrary.wiley.com/doi/abs/10.1002/spe.3311},
    year={2024},
    publisher={Wiley Online Library}
}

@misc{wu2025vision,
      title={Vision Language Model-based Testing of Industrial Autonomous Mobile Robots}, 
      author={Jiahui Wu and Chengjie Lu and Aitor Arrieta and Shaukat Ali and Thomas Peyrucain},
      year={2025},
      eprint={2508.02338},
      archivePrefix={arXiv},
      primaryClass={cs.SE},
      url={https://arxiv.org/abs/2508.02338}, 
}

@misc{isaku2025digital,
      title={Digital Twin-based Out-of-Distribution Detection in Autonomous Vessels}, 
      author={Erblin Isaku and Hassan Sartaj and Shaukat Ali},
      year={2025},
      eprint={2504.19816},
      archivePrefix={arXiv},
      primaryClass={eess.SY},
      url={https://arxiv.org/abs/2504.19816}, 
}

@inproceedings{betzer2024digital,
  title={Digital Twin Enabled Runtime Verification for Autonomous Mobile Robots under Uncertainty},
  author={Betzer, Joakim Schack and Boudjadar, Jalil and Frasheri, Mirgita and Talasila, Prasad},
  booktitle={2024 28th International Symposium on Distributed Simulation and Real Time Applications (DS-RT)},
  pages={10--17},
  year={2024},
  publisher={IEEE},
  address = {New York, NY, USA},
  doi={10.1109/DS-RT62209.2024.00012}
}

@article{caldas2024runtime,
  author={Caldas, Ricardo and García, Juan Antonio Piñera and Schiopu, Matei and Pelliccione, Patrizio and Rodrigues, Genaína and Berger, Thorsten},
  journal={IEEE Transactions on Software Engineering}, 
  title={Runtime Verification and Field-Based Testing for ROS-Based Robotic Systems}, 
  year={2024},
  volume={50},
  number={10},
  pages={2544-2567},
  doi={10.1109/TSE.2024.3444697}
}

@article{ma2019modeling,
  title={Modeling foundations for executable model-based testing of self-healing cyber-physical systems},
  author={Ma, Tao and Ali, Shaukat and Yue, Tao},
  journal={Software \& Systems Modeling},
  volume={18},
  pages={2843--2873},
  year={2019},
  publisher={Springer},
  doi={10.1007/s10270-018-00703-y}
}

@article{ma2021testing,
  title={Testing self-healing cyber-physical systems under uncertainty with reinforcement learning: an empirical study},
  author={Ma, Tao and Ali, Shaukat and Yue, Tao},
  journal={Empirical Software Engineering},
  volume={26},
  pages={1--54},
  year={2021},
  publisher={Springer},
  doi={10.1007/s10664-021-09941-z}
}

@article{mullins2018adaptive,
  title={Adaptive generation of challenging scenarios for testing and evaluation of autonomous vehicles},
  author={Mullins, Galen E and Stankiewicz, Paul G and Hawthorne, R Chad and Gupta, Satyandra K},
  journal={Journal of Systems and Software},
  volume={137},
  pages={197--215},
  year={2018},
  publisher={Elsevier},
  doi={10.1016/j.jss.2017.10.031}
}

@inproceedings{adigun2023risk,
  title={Risk-driven Online Testing and Test Case Diversity Analysis for ML-enabled Critical Systems},
  author={Adigun, Jubril Gbolahan and Huck, Tom Philip and Camilli, Matteo and Felderer, Michael},
  booktitle={2023 IEEE 34th International Symposium on Software Reliability Engineering (ISSRE)},
  pages={344--354},
  year={2023},
  publisher={IEEE},
  address = {New York, NY, USA},
  doi={10.1109/ISSRE59848.2023.00017}
}

@article{javed2024automated,
  title={An automated model-based testing approach for the self-adaptive behavior of the unmanned aircraft system application software},
  author={Javed, Zainab and Iqbal, Muhammad Zohaib and Khan, Muhammad Uzair and Usman, Muhammad and Jilani, Atif Aftab Ahmed},
  journal={Software: Practice and Experience},
  volume={54},
  number={12},
  pages={2375--2427},
  year={2024},
  publisher={Wiley Online Library},
  doi={10.1002/spe.3358}
}

@article{javed2025hybrid,
  title={A hybrid search and model-based approach for testing the self-adaptive unmanned aircraft system software},
  author={Javed, Zainab and Iqbal, Muhammad Zohaib and Khan, Muhammad Uzair and Usman, Muhammad and Jilani, Atif Aftab Ahmed},
  journal={Computer Standards \& Interfaces},
  volume={93},
  pages={103959},
  year={2025},
  publisher={Elsevier},
  doi={10.1016/j.csi.2024.103959}
}

@ARTICLE{Barricelli&19,
  author={Barricelli, Barbara Rita and Casiraghi, Elena and Fogli, Daniela},
  journal={IEEE Access}, 
  title="{A Survey on Digital Twin: Definitions, Characteristics, Applications, and Design Implications}", 
  year={2019},
  volume={7},
  number={},
  pages={167653-167671},
  doi={10.1109/ACCESS.2019.2953499}}

@BOOK{Fitzgerald&24,
  EDITOR        = "John Fitzgerald and Cláudio Gomes and Peter Gorm Larsen",
  TITLE         = "{The Engineering of Digital Twins}",
  PUBLISHER     = "Springer",
  YEAR          = "2024",
  ISBN = "978-3-031-66718-3",
  COMMENT       = "BIB PGL",
  address  = "Switzerland"
}

@article{Gil&24b,
author = {Santiago Gil and Bentley J Oakes and Cláudio Gomes and Mirgita Frasheri and Peter G Larsen},
title ={Toward a systematic reporting framework for Digital Twins: a cooperative robotics case study},
journal = {SIMULATION},
pages = {313-339},
volume = {101},
number = {3},
doi = {10.1177/00375497241261406},
publisher = {SAGE Publications},
address={Thousand Oaks, CA, USA},
URL = {https://doi.org/10.1177/00375497241261406},
year = {2025}
}

@article{malik2021digital,
  title={Digital twins for collaborative robots: A case study in human-robot interaction},
  author={Malik, Ali Ahmad and Brem, Alexander},
  journal={Robotics and Computer-Integrated Manufacturing},
  volume={68},
  pages={102092},
  year={2021},
  publisher={Elsevier},
  doi={10.1016/j.rcim.2020.102092}
}

@article{Larsen&23,
title={How do we Engineer Trustworthy Digital Twins?},
DOI={10.1017/cbp.2023.3},
journal={Research Directions: Cyber-Physical Systems},
publisher={Cambridge University Press},
author={Larsen, Peter Gorm and Fitzgerald, John and Woodcock, Jim},
year={2023},
pages={1–6},
volume={1}, 
}

@ARTICLE{Fuller&20,
  author={A. {Fuller} and Z. {Fan} and C. {Day} and C. {Barlow}},
  journal={IEEE Access},
  title={Digital Twin: Enabling Technologies, Challenges and Open Research},
  year={2020},
  volume={8},
  number={},
  pages={108952-108971},
  doi={10.1109/ACCESS.2020.2998358}}

@Article{Botín-Sanabria&22,
AUTHOR = {Botín-Sanabria, Diego M. and Mihaita, Adriana-Simona and Peimbert-García, Rodrigo E. and Ramírez-Moreno, Mauricio A. and Ramírez-Mendoza, Ricardo A. and Lozoya-Santos, Jorge de J.},
TITLE = {Digital Twin Technology Challenges and Applications: A Comprehensive Review},
JOURNAL = {Remote Sensing},
VOLUME = {14},
YEAR = {2022},
NUMBER = {6},
ARTICLE-NUMBER = {1335},
URL = {https://www.mdpi.com/2072-4292/14/6/1335},
ISSN = {2072-4292},
pages={1335},
DOI = {10.3390/rs14061335}
}

@Article{Weyns2023,
  author    = {Weyns, Danny and Iftikhar, Usman M.},
  journal   = {ACM Transactions on Software Engineering and Methodology},
  title     = {ActivFORMS: A Formally Founded Model-based Approach to Engineer Self-adaptive Systems},
  year      = {2023},
  issn      = {1557-7392},
  month     = jan,
  number    = {1},
  pages     = {1--48},
  volume    = {32},
  doi       = {10.1145/3522585},
  publisher = {Association for Computing Machinery (ACM)},
}

@InProceedings{Kuhne2005a,
  author =	{K\"{u}hne, Thomas},
  title =	{{What is a Model?}},
  booktitle =	{Language Engineering for Model-Driven Software Development},
  pages =	{1--10},
  series =	{Dagstuhl Seminar Proceedings (DagSemProc)},
  ISSN =	{1862-4405},
  year =	{2005},
  volume =	{4101},
  editor =	{Jean Bezivin and Reiko Heckel},
  publisher =	{Schloss Dagstuhl -- Leibniz-Zentrum f{\"u}r Informatik},
  address =	{Dagstuhl, Germany},
  URL =		{https://drops.dagstuhl.de/entities/document/10.4230/DagSemProc.04101.15},
  doi =		{10.4230/DagSemProc.04101.15}
}

@InProceedings{Feng2022a,
  author    = {Feng, Hao and Gomes, Claudio and Gil, Santiago and Mikkelsen, Peter H. and Tola, Daniella and Larsen, Peter Gorm and Sandberg, Michael},
  booktitle = {2022 Annual Modeling and Simulation Conference (ANNSIM)},
  title     = {Integration Of The Mape-K Loop In Digital Twins},
  year      = {2022},
  month     = jul,
  pages     = {102--113},
  publisher = {IEEE},
  address = {New York, NY, USA},
  doi       = {10.23919/annsim55834.2022.9859489},
}

@InBook{Wright2022,
  author    = {Wright, Thomas and Gomes, Cláudio and Woodcock, Jim},
  pages     = {89--109},
  publisher = {Springer Nature},
  address = {Switzerland},
  title     = {Formally Verified Self-adaptation of an Incubator Digital Twin},
  year      = {2022},
  isbn      = {9783031197628},
  booktitle = {Leveraging Applications of Formal Methods, Verification and Validation. Practice},
  doi       = {10.1007/978-3-031-19762-8_7},
  issn      = {1611-3349},
}

@InProceedings{Loo2021,
  author    = {Loo, Yim Ling and Ahmad, Azhana and Mustapha, Aida and Mostafa, Salama A.},
  booktitle = {2021 4th International Symposium on Agents, Multi-Agent Systems and Robotics (ISAMSR)},
  title     = {A Self-Adaptive Agent-Based Dynamic Processes Simulation Modelling Framework},
  year      = {2021},
  month     = sep,
  pages     = {124--129},
  publisher = {IEEE},
  address = {New York, NY, USA},
  doi       = {10.1109/isamsr53229.2021.9567854},
}

@Article{Arcaini2017,
  author    = {Arcaini, Paolo and Riccobene, Elvinia and Scandurra, Patrizia},
  journal   = {ACM Transactions on Autonomous and Adaptive Systems},
  title     = {Formal Design and Verification of Self-Adaptive Systems with Decentralized Control},
  year      = {2017},
  issn      = {1556-4703},
  month     = jan,
  number    = {4},
  pages     = {1--35},
  volume    = {11},
  doi       = {10.1145/3019598},
  publisher = {Association for Computing Machinery (ACM)},
}

@Article{Gomes2018,
  author    = {Gomes, Cláudio and Thule, Casper and Broman, David and Larsen, Peter Gorm and Vangheluwe, Hans},
  journal   = {ACM Computing Surveys},
  title     = {Co-Simulation: A Survey},
  year      = {2018},
  issn      = {1557-7341},
  month     = may,
  number    = {3},
  pages     = {1--33},
  volume    = {51},
  doi       = {10.1145/3179993},
  publisher = {Association for Computing Machinery (ACM)},
}

@inproceedings{feng2021developing,
  title={Developing a physical and digital twin: an example process model},
  author={Feng, Hao and Gomes, Cl{\'a}udio and Sandberg, Michael and Thule, Casper and Lausdahl, Kenneth and Larsen, Peter Gorm},
  booktitle={2021 ACM/IEEE International Conference on Model Driven Engineering Languages and Systems Companion (MODELS-C)},
  pages={286--295},
  year={2021},
  publisher={IEEE},
  address = {New York, NY, USA},
  doi={10.1109/MODELS-C53483.2021.00050}
}

@article{letier2025obstacle,
  title={Obstacle Analysis in Requirements Engineering: Retrospective and Emerging Challenges},
  author={Letier, Emmanuel and Van Lamsweerde, Axel},
  journal={IEEE Transactions on Software Engineering},
  year={2025},
  publisher={IEEE},
  volume={51},
  number={3},
  pages={795-801},
  address = {New York, NY, USA},
  doi={10.1109/TSE.2025.3534318}
}

@InProceedings{Junghanns2021,
  author     = {Andreas Junghanns and Cláudio Gomes and Christian Schulze and Klaus Schuch and Pierre R. and Matthias Blaesken and Irina Zacharias and Andreas Pillekeit and Karl Wernersson and Torsten Sommer and Christian Bertsch and Torsten Blochwitz and Masoud Najafi},
  booktitle  = {Proceedings of 14th Modelica Conference 2021, Linköping, Sweden, September 20-24, 2021},
  title      = {The Functional Mock-up Interface 3.0 - New Features Enabling New Applications},
  year       = {2021},
  month      = sep,
  pages      = {17--26},
  publisher  = {Linköping University Electronic Press},
  address    = {Linköping, Sweden},
  series     = {Modelica 2021},
  volume     = {181},
  collection = {Modelica 2021},
  doi        = {10.3384/ecp2118117},
  issn       = {1650-3686},
}

@InProceedings{Nagele2017,
  author    = {Nagele, Thomas and Hooman, Jozef},
  booktitle = {2017 IEEE 7th Annual Computing and Communication Workshop and Conference (CCWC)},
  title     = {Co-simulation of cyber-physical systems using HLA},
  year      = {2017},
  month     = jan,
  pages     = {1--6},
  publisher = {IEEE},
  address = {New York, NY, USA},
  doi       = {10.1109/ccwc.2017.7868401},
}

@Article{Kubler2000,
  author    = {Kubler, R. and Schiehlen, W.},
  journal   = {Mathematical and Computer Modelling of Dynamical Systems},
  title     = {Two Methods of Simulator Coupling},
  year      = {2000},
  issn      = {1387-3954},
  month     = jun,
  number    = {2},
  pages     = {93--113},
  volume    = {6},
  doi       = {10.1076/1387-3954(200006)6:2;1-m;ft093},
  publisher = {Informa UK Limited},
}

@InProceedings{Inci2021,
  author    = {Inci, Emin Oguz and Croes, Jan and Desmet, Wim and Gomes, Claudio and Thule, Casper and Lausdahl, Kenneth and Larsen, Peter Gorm},
  booktitle = {2021 Annual Modeling and Simulation Conference (ANNSIM)},
  title     = {The Effect and Selection of Solution Sequence in Co-Simulation},
  year      = {2021},
  month     = jul,
  pages     = {1--12},
  publisher = {IEEE},
  address = {New York, NY, USA},
  doi       = {10.23919/annsim52504.2021.9552130},
}

@InProceedings{Inci2023,
  author    = {Inci, Emin Oguz and Desmet, Wim and Gomes, Cláudio and Croes, Jan},
  booktitle = {2023 Annual Modeling and Simulation Conference (ANNSIM)},
  title     = {Error Estimators for Adaptive Scheduling Algorithm for Serial Co-Simulation},
  year      = {2023},
  pages     = {73-83},
  publisher={IEEE},
  address = {New York, NY, USA},
}

@inproceedings{ejersbo2023dynamic,
  title={Dynamic runtime integration of new models in digital twins},
  author={Ejersbo, Henrik and Lausdahl, Kenneth and Frasheri, Mirgita and Esterle, Lukas},
  booktitle={2023 IEEE/ACM 18th Symposium on Software Engineering for Adaptive and Self-Managing Systems (SEAMS)},
  pages={44--55},
  year={2023},
  publisher={IEEE},
  address = {New York, NY, USA},
  doi={10.1109/SEAMS59076.2023.00016}
}

@article{barros2024pi,
  title={$\pi$ HyFlow: formalism, semantics, and applications},
  author={Barros, Fernando},
  journal={Discrete Event Dynamic Systems},
  volume={34},
  number={1},
  pages={95--124},
  year={2024},
  publisher={Springer},
  doi={10.1007/s10626-023-00390-y}
}

@InProceedings{Blochwitz2012,
  author     = {Blockwitz, Torsten and Otter, Martin and Akesson, Johan and Arnold, Martin and Clauss, Christoph and Elmqvist, Hilding and Friedrich, Markus and Junghanns, Andreas and Mauss, Jakob and Neumerkel, Dietmar and Olsson, Hans and Viel, Antoine},
  booktitle  = {Proceedings of the 9th International MODELICA Conference, September 3-5, 2012, Munich, Germany},
  title      = {Functional Mockup Interface 2.0: The Standard for Tool-independent Exchange of Simulation Models},
  year       = {2012},
  month      = nov,
  publisher  = {Linköping University Electronic Press},
  address    = {Linköping, Sweden},
  series     = {MODELLICA},
  collection = {MODELLICA},
  doi        = {10.3384/ecp12076173},
  issn       = {1650-3686},
  pages      = {173--184},
}

@InProceedings{Gomes2021a,
  author     = {Cláudio Gomes and Torsten Blochwitz and Christian Bertsch and Karl Wernersson and Klaus Schuch and Pierre R. and Oliver Kotte and Irina Zacharias and Matthias Blesken and Torsten Sommer and Masoud Najafi and Andreas Junghanns},
  booktitle  = {Proceedings of 14th Modelica Conference 2021, Linköping, Sweden, September 20-24, 2021},
  title      = {The FMI 3.0 Standard Interface for Clocked and Scheduled Simulations},
  year       = {2021},
  month      = sep,
  pages      = {27--36},
  publisher  = {Linköping University Electronic Press},
  address    = {Linköping, Sweden},
  series     = {Modelica 2021},
  volume     = {181},
  collection = {Modelica 2021},
  doi        = {10.3384/ecp2118127},
  issn       = {1650-3686},
}

@article{blas2022devs,
  title={DEVS-based formalism for the modeling of routing processes},
  author={Blas, Mar{\'\i}a Julia and Leone, Horacio and Gonnet, Silvio},
  journal={Software and Systems Modeling},
  volume={21},
  number={3},
  pages={1179--1208},
  year={2022},
  publisher={Springer},
  doi={10.1007/s10270-021-00928-4}
}

@incollection{kortenkamp2016robotic,
  title={Robotic systems architectures and programming},
  author={Kortenkamp, David and Simmons, Reid and Brugali, Davide},
  booktitle={Springer handbook of robotics},
  pages={283--306},
  year={2016},
  publisher={Springer},
  address={Cham},
  doi={10.1007/978-3-319-32552-1_12}
}

@article{Tao2017,
  author = {F. Tao and M. Zhang},
  title = {Digital twin shop-floor: a new shop-floor paradigm towards smart manufacturing},
  journal = {IEEE Access},
  volume = {5},
  pages = {20418--20427},
  year = {2017}
}

@article{Bhatti2021,
  author = {G. Bhatti and H. Mohan and R. R. Singh},
  title = {Towards the future of smart electric vehicles: Digital twin technology},
  journal = {Renewable and Sustainable Energy Reviews},
  volume = {141},
  pages = {110801},
  year = {2021}
}

@article{Xiong2022,
  author={Xiong, Minglan and Wang, Huawei},
  title = {Digital twin applications in aviation industry: A review},
  journal = {The International Journal of Advanced Manufacturing Technology},
  volume = {121},
  number = {9-10},
  pages = {5677--5692},
  year = {2022},
  doi={10.1007/s00170-022-09717-9},
  publisher={Springer}
}

@article{Zhang2022,
  author={Zhang, Houxiang and Li, Guoyuan and Hatledal, Lars Ivar and Chu, Yingguang and Ellefsen, André and Han, Peihua and Major, Pierre and Skulstad, Robert and Wang, Tongtong and Hildre, Hans Petter},
  title={A Digital Twin of the Research Vessel Gunnerus for Lifecycle Services: Outlining Key Technologies}, 
  journal = {IEEE Robotics \& Automation Magazine},
  year={2023},
  volume={30},
  number={3},
  pages={6-19},
  doi={10.1109/MRA.2022.3217745}
}

@article{Xie2023,
  author = {H. Xie and M. Xin and C. Lu and J. Xu},
  title = {Knowledge map and forecast of digital twin in the construction industry: State-of-the-art review using scientometric analysis},
  journal = {Journal of Cleaner Production},
  volume = {383},
  pages = {135231},
  year = {2023}
}

@article{Sun2023,
  author = {T. Sun and X. He and Z. Li},
  title = {Digital twin in healthcare: Recent updates and challenges},
  journal = {Digital Health},
  volume = {9},
  pages = {20552076221149651},
  year = {2023},
  publisher={Sage Publications Sage UK: London, England},
  doi={10.1177/20552076221149651}
}

@article{Mazumder2023,
  author = {Mazumder, A. and Sahed, M. F. and Tasneem, Z. and Das, P. and Badal, F. R. and Ali, M. F. and Islam, M. R.},
  title = {Towards next generation digital twin in robotics: Trends, scopes, challenges, and future},
  journal = {Heliyon},
  volume = {9},
  number = {2},
  year = {2023},
  pages = {e12982},
  doi = {10.1016/j.heliyon.2023.e12982}
}

@incollection{Hoebert2019,
  author = {Hoebert, T. and Lepuschitz, W. and List, E. and Merdan, M.},
  title = {Cloud-based digital twin for industrial robotics},
  booktitle = {Industrial Applications of Holonic and Multi-Agent Systems: 9th International Conference, HoloMAS 2019, Linz, Austria, August 26–29, 2019, Proceedings 9},
  editor = {},
  publisher = {Springer International Publishing},
  address={Switzerland},
  year = {2019},
  pages = {105--116},
  doi = {10.1007/978-3-030-25894-3_9}
}

@article{Li2021,
  author = {Li, X. and He, B. and Wang, Z. and Zhou, Y. and Li, G. and Jiang, R.},
  title = {Semantic-enhanced digital twin system for robot–environment interaction monitoring},
  journal = {IEEE Transactions on Instrumentation and Measurement},
  volume = {70},
  pages = {1--13},
  year = {2021},
  doi = {10.1109/TIM.2021.3077125}
}

@article{Yang2024,
  author = {Yang, H. and Qin, Z. and Xia, Y. and Cheng, F.},
  title = {Digital Twin-Based Autonomous Navigation and Control of Omnidirectional Mobile Robots},
  journal = {IEEE Transactions on Vehicular Technology},
  year={2025},
  volume={74},
  number={4},
  pages={5687-5697},
  publisher={IEEE},
  address = {New York, NY, USA},
  doi = {10.1109/TVT.2024.3254501}
}

@article{Li2022,
  author = {Li, C. and Zheng, P. and Li, S. and Pang, Y. and Lee, C. K.},
  title = {AR-assisted digital twin-enabled robot collaborative manufacturing system with human-in-the-loop},
  journal = {Robotics and Computer-Integrated Manufacturing},
  volume = {76},
  pages = {102321},
  year = {2022},
  doi = {10.1016/j.rcim.2021.102321}
}

@article{Kousi2021,
  author = {Kousi, N. and Gkournelos, C. and Aivaliotis, S. and Lotsaris, K. and Bavelos, A. C. and Baris, P. and Makris, S.},
  title = {Digital twin for designing and reconfiguring human–robot collaborative assembly lines},
  journal = {Applied Sciences},
  volume = {11},
  number = {10},
  pages = {4620},
  year = {2021},
  doi = {10.3390/app11104620}
}

@article{Song2023,
  author = {Song, Z. and Shi, H. and Bai, X. and Li, G.},
  title = {Digital twin‐assisted fault diagnosis system for robot joints with insufficient data},
  journal = {Journal of Field Robotics},
  volume = {40},
  number = {2},
  pages = {258--271},
  year = {2023},
  doi = {10.1002/rob.22177}
}

@book{zhang2024digital,
  author    = {Zhang, H. and Li, G. and Hildre, H. P.},
  title     = {Digital Twins for Vessel Life Cycle Service: Innovation in Maritime Industry},
  publisher = {Springer Nature},
  year      = {2024},
  address = {Singapore}
}

@article{major2021use,
  author    = {Major, P. and Li, G. and Hildre, H. P. and Zhang, H.},
  title     = {The use of a data-driven digital twin of a smart city: A case study of Ålesund, Norway},
  journal   = {IEEE Instrumentation \& Measurement Magazine},
  volume    = {24},
  number    = {7},
  pages     = {39--49},
  year      = {2021},
  doi={10.1109/MIM.2021.9549127}
}

@article{mao2025survey,
  author    = {Mao, R. and Li, Y. and Li, G. and Hildre, H. P. and Zhang, H.},
  title     = {A systematic survey of digital twin applications: Transferring knowledge from automotive and aviation to maritime industry},
  journal   = {IEEE Transactions on Intelligent Transportation Systems},
  volume    = {26},
  number    = {4},
  pages     = {Early Access},
  year      = {2025},
  doi       = {10.1109/TITS.2025.3535593}
}

@article{han2021fault,
  author    = {Han, P. and Ellefsen, A. L. and Li, G. and Æsøy, V. and Zhang, H.},
  title     = {Fault prognostics using LSTM networks: application to marine diesel engine},
  journal   = {IEEE Sensors Journal},
  volume    = {21},
  number    = {22},
  pages     = {25986--25994},
  year      = {2021}
}

@inproceedings{redmon2016you,
  title={You only look once: Unified, real-time object detection},
  author={Redmon, Joseph and Divvala, Santosh and Girshick, Ross and Farhadi, Ali},
  booktitle={2016 IEEE Conference on Computer Vision and Pattern Recognition (CVPR)}, 
  pages={779--788},
  year={2016},
  publisher={IEEE},
  address = {New York, NY, USA},
  doi={10.1109/CVPR.2016.91}
}

@article{he2017mask,
  author={He, Kaiming and Gkioxari, Georgia and Dollár, Piotr and Girshick, Ross},
  journal={IEEE Transactions on Pattern Analysis and Machine Intelligence}, 
  title={Mask R-CNN}, 
  year={2020},
  volume={42},
  number={2},
  pages={386-397},
  doi={10.1109/TPAMI.2018.2844175}
}

@inproceedings{roddick2020predicting,
  title={Predicting semantic map representations from images using pyramid occupancy networks},
  author={Roddick, Thomas and Cipolla, Roberto},
  booktitle={Proceedings of the IEEE/CVF Conference on Computer Vision and Pattern Recognition},
  pages={11138--11147},
  year={2020},
  publisher={IEEE},
  address = {New York, NY, USA},
  doi={10.1109/CVPR42600.2020.01115}
}

@inproceedings{agro2023implicit,
  title={Implicit occupancy flow fields for perception and prediction in self-driving},
  author={Agro, Ben and Sykora, Quinlan and Casas, Sergio and Urtasun, Raquel},
  booktitle={Proceedings of the IEEE/CVF Conference on Computer Vision and Pattern Recognition},
  pages={1379--1388},
  year={2023},
  publisher={IEEE},
  address = {New York, NY, USA},
  doi={10.1109/CVPR52729.2023.00139}
}

@inproceedings{nguyen2019review,
  title={Review of deep reinforcement learning for robot manipulation},
  author={Nguyen, Hai and La, Hung},
  booktitle={2019 Third IEEE International Conference on Robotic Computing (IRC)},
  pages={590--595},
  year={2019},
  publisher={IEEE},
  address = {New York, NY, USA},
}

@article{wang2019autonomous,
  title={Autonomous navigation of UAVs in large-scale complex environments: A deep reinforcement learning approach},
  author={Wang, Chao and Wang, Jian and Shen, Yuan and Zhang, Xudong},
  journal={IEEE Transactions on Vehicular Technology},
  volume={68},
  number={3},
  pages={2124--2136},
  year={2019},
  publisher={IEEE},
  address = {New York, NY, USA},
}

@article{pateria2021hierarchical,
  title={Hierarchical reinforcement learning: A comprehensive survey},
  author={Pateria, Shubham and Subagdja, Budhitama and Tan, Ah-hwee and Quek, Chai},
  journal={ACM Computing Surveys (CSUR)},
  volume={54},
  number={5},
  pages={1--35},
  year={2021},
  publisher={ACM},
  address = {New York, NY, USA},
  url = {https://doi.org/10.1145/3453160},
  doi = {10.1145/3453160}
}

@inproceedings{lykov2024llm,
  title={{LLM-BRAIn}: {AI}-driven Fast Generation of Robot Behaviour Tree based on Large Language Model},
  author={Lykov, Artem and Tsetserukou, Dzmitry},
  booktitle={2024 2nd International Conference on Foundation and Large Language Models (FLLM)},
  pages={392--397},
  year={2024},
  publisher={IEEE},
  address = {New York, NY, USA},
}

@article{spielberg2019neural,
  title={Neural network vehicle models for high-performance automated driving},
  author={Spielberg, Nathan A and Brown, Matthew and Kapania, Nitin R and Kegelman, John C and Gerdes, J Christian},
  journal={Science robotics},
  volume={4},
  number={28},
  pages={eaaw1975},
  year={2019},
  publisher={American Association for the Advancement of Science},
  doi={10.1126/scirobotics.aaw1975}
}

@article{wang2021incorporating,
  title={Incorporating approximate dynamics into data-driven calibrator: A representative model for ship maneuvering prediction},
  author={Wang, Tongtong and Li, Guoyuan and Hatledal, Lars Ivar and Skulstad, Robert and {\AE}s{\o}y, Vilmar and Zhang, Houxiang},
  journal={IEEE Transactions on Industrial Informatics},
  volume={18},
  number={3},
  pages={1781--1789},
  year={2021},
  publisher={IEEE},
  address = {New York, NY, USA},
  doi={10.1109/TII.2021.3088404}
}

@article{Oncay2023softrobotics,
   author = "Yasa, Oncay and Toshimitsu, Yasunori and Michelis, Mike Y. and Jones, Lewis S. and Filippi, Miriam and Buchner, Thomas and Katzschmann, Robert K.",
   title = "An Overview of Soft Robotics", 
   journal= "Annual Review of Control, Robotics, and Autonomous Systems",
   year = "2023",
   volume = "6",
   number = "Volume 6, 2023",
   pages = "1-29",
   doi = "10.1146/annurev-control-062322-100607",
   _url = "https://www.annualreviews.org/content/journals/10.1146/annurev-control-062322-100607",
   publisher = "Annual Reviews",
   issn = "2573-5144",
   _type = "Journal Article",
  }

@Article{Rafsanjani2019programmingSoftRobots,
  author   = {Ahmad Rafsanjani and Katia Bertoldi and André R. Studart},
  journal  = {Science Robotics},
  title    = {Programming soft robots with flexible mechanical metamaterials},
  year     = {2019},
  number   = {29},
  pages    = {eaav7874},
  volume   = {4},
  _eprint  = {https://www.science.org/doi/pdf/10.1126/scirobotics.aav7874},
  _url     = {https://www.science.org/doi/abs/10.1126/scirobotics.aav7874},
  abstract = {The complex behavior of highly deformable mechanical metamaterials can substantially enhance the performance of soft robots. The complex behavior of highly deformable mechanical metamaterials can substantially enhance the performance of soft robots.},
  doi      = {10.1126/scirobotics.aav7874},
}

@ARTICLE{Yim2002modularRobots,
  author={Yim, M. and Ying Zhang and Duff, D.},
  journal={IEEE Spectrum}, 
  title={Modular robots}, 
  year={2002},
  volume={39},
  number={2},
  pages={30-34},
  doi={10.1109/6.981854}}

@ARTICLE{Gilpin2010modularRobots,
  author={Gilpin, Kyle and Rus, Daniela},
  journal={IEEE Robotics and Automation Magazine}, 
  title={Modular Robot Systems}, 
  year={2010},
  volume={17},
  number={3},
  pages={38-55},
  doi={10.1109/MRA.2010.937859}}

@INPROCEEDINGS{Esterle2021verification,
  author={Esterle, Lukas and Porter, Barry and Woodcock, Jim},
  booktitle={2021 IEEE International Conference on Autonomic Computing and Self-Organizing Systems Companion (ACSOS-C)}, 
  title={Verification and Uncertainties in Self-integrating System}, 
  year={2021},
  volume={},
  number={},
  publisher={IEEE},
  address = {New York, NY, USA},
  pages={220-225},
  doi={10.1109/ACSOS-C52956.2021.00050}}

@INPROCEEDINGS{Barnes2019uncertainty,
  author={Barnes, Chloe M. and Esterle, Lukas and Brown, John N. A.},
  booktitle={2019 IEEE 4th International Workshops on Foundations and Applications of Self* Systems (FAS*W)}, 
  title={"When you Believe in Things that you don't Understand": the Effect of Cross-Generational Habits on Self-Improving System Integration}, 
  year={2019},
  volume={},
  number={},
  pages={28-31},
  publisher={IEEE},
  address = {New York, NY, USA},
  doi={10.1109/FAS-W.2019.00020}}

@article{Brambilla2013swarmRobotics,
	author = {Brambilla, Manuele and Ferrante, Eliseo and Birattari, Mauro and Dorigo, Marco},
	doi = {10.1007/s11721-012-0075-2},
	isbn = {1935-3820},
	journal = {Swarm Intelligence},
	number = {1},
	pages = {1--41},
	title = {Swarm robotics: a review from the swarm engineering perspective},
	_url = {https://doi.org/10.1007/s11721-012-0075-2},
	volume = {7},
	year = {2013},
  }

@article{esterle2016cyber,
  title={Cyber-physical systems: challenge of the 21st century},
  author={Esterle, Lukas and Grosu, Radu},
  journal={e \& i Elektrotechnik und Informationstechnik},
  volume={133},
  number={7},
  pages={299--303},
  year={2016},
  publisher={Springer},
  doi={10.1007/s00502-016-0426-6}
}

@article{Pianini2022coordinatemultirobot,
author = {Pianini, Danilo and Pettinari, Federico and Casadei, Roberto and Esterle, Lukas},
title = {A Collective Adaptive Approach to Decentralised k-Coverage in Multi-robot Systems},
year = {2022},
issue_date = {June 2022},
publisher = {Association for Computing Machinery},
address = {New York, NY, USA},
volume = {17},
number = {1–2},
issn = {1556-4665},
url = {https://doi.org/10.1145/3547145},
doi = {10.1145/3547145},
journal = {ACM Trans. Auton. Adapt. Syst.},
month = sep,
articleno = {4},
numpages = {39},
}

@article{Casadei2025collectives,
author = {Casadei, Roberto and Aguzzi, Gianluca and Audrito, Giorgio and Damiani, Ferruccio and Pianini, Danilo and Scarso, Giordano and Torta, Gianluca and Viroli, Mirko},
title = {Software Engineering for Collective Cyber-Physical Ecosystems},
year = {2025},
issue_date = {June 2025},
publisher = {Association for Computing Machinery},
address = {New York, NY, USA},
volume = {34},
number = {5},
issn = {1049-331X},
_url = {https://doi.org/10.1145/3712004},
doi = {10.1145/3712004},
journal = {ACM Trans. Softw. Eng. Methodol.},
month = may,
articleno = {153},
numpages = {40},
}

@Inbook{Kephart2017collectiveSelfAware,
author="Kephart, Jeffrey O.
and Diaconescu, Ada
and Giese, Holger
and Robertsson, Anders
and Abdelzaher, Tarek
and Lewis, Peter
and Filieri, Antonio
and Esterle, Lukas
and Frey, Sylvain",
title="Self-adaptation in Collective Self-aware Computing Systems",
bookTitle="Self-Aware Computing Systems",
year="2017",
publisher="Springer International Publishing",
address="Cham",
pages="401--435",
isbn="978-3-319-47474-8",
doi="10.1007/978-3-319-47474-8_13",
_url="https://doi.org/10.1007/978-3-319-47474-8_13"
}

@article{ahmed2020dynamic,
  title={Dynamic software updating: a systematic mapping study},
  author={Ahmed, Babiker Hussien and Lee, Sai Peck and Su, Moon Ting and Zakari, Abubakar},
  journal={IET Software},
  volume={14},
  number={5},
  pages={468--481},
  year={2020},
  publisher={Wiley Online Library},
  doi={10.1049/iet-sen.2019.0201}
}

@inproceedings{hosseini2023safety,
  title={A safety and security requirements management methodology in reconfigurable collaborative human-robot application},
  author={Hosseini, Ali M and Fischer, Clara and Bhole, Mukund and Kastner, Wolfgang and Sauter, Thilo and Schlund, Sebastian},
  booktitle={2023 IEEE 19th International Conference on Factory Communication Systems (WFCS)},
  pages={1--8},
  year={2023},
  publisher={IEEE},
  address = {New York, NY, USA},
  doi={10.1109/WFCS57264.2023.10144233}
}

@article{miyazawa2025diagrammatic,
  title={Diagrammatic physical robot models: A. Miyazawa et al.},
  author={Miyazawa, Alvaro and Ahmadi, Sharar and Cavalcanti, Ana and Baxter, James and Post, Mark and Ribeiro, Pedro and Timmis, Jon and Wright, Thomas},
  journal={Software and Systems Modeling},
  volume={24},
  number={5},
  pages={1549--1593},
  year={2025},
  publisher={Springer},
  doi={10.1007/s10270-025-01270-9}
}

@inproceedings{alberts2023development,
  title={Development and integration of self-adaptation strategies for robotics software},
  author={Alberts, Elvin},
  booktitle={2023 IEEE 20th International Conference on Software Architecture Companion (ICSA-C)},
  pages={131--136},
  year={2023},
  publisher={IEEE},
  address = {New York, NY, USA},
  doi={10.1109/ICSA-C57050.2023.00038}
}

@inproceedings{ingles2011towards,
  title={Towards the automatic generation of self-adaptive robotics software: An experience report},
  author={Ingl{\'e}s-Romero, Juan F and Vicente-Chicote, Cristina and Morin, Brice and Barais, Olivier},
  booktitle={2011 IEEE 20th International Workshops on Enabling Technologies: Infrastructure for Collaborative Enterprises},
  pages={79--86},
  year={2011},
  publisher={IEEE},
  address = {New York, NY, USA},
  doi={10.1109/WETICE.2011.54}
}

@inproceedings{ingles2010using,
  title={Using Models@ Runtime for designing adaptive robotics software: an experience report},
  author={Ingl{\'e}s Romero, Juan Francisco and Vicente Chicote, Cristina and Morin, Brice and Barais, Olivier},
  booktitle={International Workshop on Model Based Engineering for Robotics},
  year={2010},
  pages={1--11},
  publisher={Laurent Riox},
  address={Oslo, Norway}
}

@inproceedings{edwards2009architecture,
  title={Architecture-driven self-adaptation and self-management in robotics systems},
  author={Edwards, George and Garcia, Joshua and Tajalli, Hossein and Popescu, Daniel and Medvidovic, Nenad and Sukhatme, Gaurav and Petrus, Brad},
  booktitle={2009 ICSE Workshop on Software Engineering for Adaptive and Self-Managing Systems},
  pages={142--151},
  year={2009},
  publisher={IEEE},
  address = {New York, NY, USA},
  doi={10.1109/SEAMS.2009.5069083}
}

@article{medvidovic2010engineering,
  title={Engineering heterogeneous robotics systems: A software architecture-based approach},
  author={Medvidovic, Nenad and Tajalli, Hossein and Garcia, Joshua and Krka, Ivo and Brun, Yuriy and Edwards, George},
  journal={Computer},
  volume={44},
  number={5},
  pages={62--71},
  year={2010},
  publisher={IEEE},
  address = {New York, NY, USA},
  doi={10.1109/MC.2010.368}
}

@article{bao2015architecture,
  title={Architecture Model to Improve the Development of Robotics Online Reconfiguration},
  author={Bao, Xiaoan and Sun, Xiance and Gui, Ning and Zhang, Na and Lin, Hui and Liu, Shuhan},
  journal={International Journal of Control and Automation},
  volume={8},
  number={1},
  pages={69--82},
  year={2015},
  doi={10.14257/ijca.2015.8.1.06}
}

@inproceedings{baxter2025formal,
  title={Formal Architectural Patterns for Adaptive Robotic Software},
  author={Baxter, James and Van Acker, Bert and Kristensen, Morten and Wright, Thomas and Cavalcanti, Ana and Gomes, Cl{\'a}udio},
  booktitle={International Conference on Fundamental Approaches to Software Engineering},
  pages={145--165},
  year={2025},
  publisher={Springer Nature},
  address={Switzerland Cham},
  doi={10.1007/978-3-031-90900-9_8}
}

@article{alberts2025software,
  title={Software architecture-based self-adaptation in robotics},
  author={Alberts, Elvin and Gerostathopoulos, Ilias and Malavolta, Ivano and Corbato, Carlos Hern{\'a}ndez and Lago, Patricia},
  journal={Journal of Systems and Software},
  volume={219},
  pages={112258},
  year={2025},
  publisher={Elsevier},
  doi={10.1016/j.jss.2024.112258}
}

@inproceedings{silva2023suave,
  title={{SUAVE}: an exemplar for self-adaptive underwater vehicles},
  author={Silva, Gustavo Rezende and P{\"a}{\ss}ler, Juliane and Zwanepol, Jeroen and Alberts, Elvin and Tarifa, S Lizeth Tapia and Gerostathopoulos, Ilias and Johnsen, Einar Broch and Corbato, Carlos Hern{\'a}ndez},
  booktitle={2023 IEEE/ACM 18th Symposium on Software Engineering for Adaptive and Self-Managing Systems (SEAMS)},
  pages={181--187},
  year={2023},
  publisher={IEEE},
  address = {New York, NY, USA},
  doi={10.1109/SEAMS59076.2023.00031}
}

@inproceedings{rivera2021software,
  title={A Software Based Self-Recovering Robotic System Architecture Using ROS},
  author={Rivera, Luis J Figueroa and Chandrasekaran, Balasubramaniyan},
  booktitle={2021 7th International Conference on Mechatronics and Robotics Engineering (ICMRE)},
  pages={29--34},
  year={2021},
  publisher={IEEE},
  address = {New York, NY, USA},
  doi={10.1109/ICMRE51691.2021.9384816}
}

@inproceedings{nezhad2025towards,
  title={Towards A Standardized Framework For Developing Trustworthy Self-Adaptive Robotic Systems},
  author={Nezhad, Sahar Nasimi and Van Acker, Bert and De Meulenaere, Paul},
  booktitle={2025 Annual Modeling and Simulation Conference (ANNSIM)},
  pages={1--13},
  year={2025},
  publisher={IEEE},
  address = {New York, NY, USA},
}

@Article{Bencomo2019,
  author    = {Bencomo, Nelly and G{\"o}tz, Sebastian and Song, Hui},
  journal   = {Software \& Systems Modeling},
  title     = {{Models@run.time}: a guided tour of the state of the art and research challenges},
  year      = {2019},
  number    = {5},
  pages     = {3049--3082},
  volume    = {18},
  publisher = {Springer},
  doi={10.1007/s10270-018-00712-x}
}

@Article{Aldrich2019,
  author    = {Aldrich, Jonathan and Garlan, David and K{\"a}stner, Christian and Le Goues, Claire and Mohseni-Kabir, Anahita and Ruchkin, Ivan and Samuel, Selva and Schmerl, Bradley and Timperley, Christopher Steven and Veloso, Manuela and others},
  journal   = {IEEE software},
  title     = {Model-based adaptation for robotics software},
  year      = {2019},
  number    = {2},
  pages     = {83--90},
  volume    = {36},
  publisher={IEEE},
  address = {New York, NY, USA},
  doi={10.1109/MS.2018.2885058}
}

@InProceedings{Lotz2013,
  author       = {Lotz, Alex and Ingl{\'e}s-Romero, Juan F and Vicente-Chicote, Cristina and Schlegel, Christian},
  booktitle    = {International Workshop on Business Process Modeling, Development and Support},
  title        = {Managing run-time variability in robotics software by modeling functional and non-functional behavior},
  year         = {2013},
  publisher = {Springer},
  address = {Berlin, Heidelberg},
  pages        = {441--455},
  doi={10.1007/978-3-642-38484-4_31}
}

@InProceedings{Brugali2018,
  author       = {Brugali, Davide and Capilla, Rafael and Mirandola, Raffaela and Trubiani, Catia},
  booktitle    = {2018 Second IEEE International Conference on Robotic Computing (IRC)},
  title        = {Model-based development of {QoS}-aware reconfigurable autonomous robotic systems},
  year         = {2018},
  publisher={IEEE},
  address = {New York, NY, USA},
  pages        = {129--136},
  doi={10.1109/IRC.2018.00027}
}

@InProceedings{Cheng2020,
  author    = {Cheng, Betty HC and Clark, Robert Jared and Fleck, Jonathon Emil and Langford, Michael Austin and McKinley, Philip K},
  booktitle = {Proceedings of the 23rd ACM/IEEE International Conference on Model-Driven Engineering Languages and Systems},
  title     = {{AC-ROS}: Assurance case driven adaptation for the robot operating system},
  year      = {2020},
  pages     = {102--113},
  doi = {10.1145/3365438.3410952},
  publisher = {Association for Computing Machinery},
  address = {New York, NY, USA}
}

@article{cheng2024survey,
  title={A Survey on Testbench-Based Vehicle-in-the-Loop Simulation Testing for Autonomous Vehicles: Architecture, Principle, and Equipment},
  author={Cheng, Jingjun and Wang, Zhen and Zhao, Xiangmo and Xu, Zhigang and Ding, Ming and Takeda, Kazuya},
  journal={Advanced Intelligent Systems},
  volume={6},
  number={6},
  pages={2300778},
  year={2024},
  publisher={Wiley Online Library},
  doi={10.1002/aisy.202300778}
}

@inproceedings{tobin2017domain,
  title={Domain randomization for transferring deep neural networks from simulation to the real world},
  author={Tobin, Josh and Fong, Rachel and Ray, Alex and Schneider, Jonas and Zaremba, Wojciech and Abbeel, Pieter},
  booktitle={2017 IEEE/RSJ international conference on intelligent robots and systems (IROS)},
  pages={23--30},
  year={2017},
  publisher={IEEE},
  address = {New York, NY, USA},
  doi={10.1109/IROS.2017.8202133}
}

@article{cai2021modular,
  title={Modular deep reinforcement learning for continuous motion planning with temporal logic},
  author={Cai, Mingyu and Hasanbeig, Mohammadhosein and Xiao, Shaoping and Abate, Alessandro and Kan, Zhen},
  journal={IEEE robotics and automation letters},
  volume={6},
  number={4},
  pages={7973--7980},
  year={2021},
  publisher={IEEE},
  address = {New York, NY, USA},
  doi={10.1109/LRA.2021.3101544}
}

@book{international2011systems,
  title={Systems and Software Engineering: Systems and Software Quality Requirements and Evaluation (SQuaRE): System and Software Quality Models},
  author={International Organization for Standardization and Technical Committee ISO/IEC JTC 1, Information technology. Subcommittee SC 7, Software and systems engineering},
  year={2011},
  publisher={ISO},
  address = {Switzerland}
}

@inproceedings{isaku2025oodisar,
  title={Out of Distribution Detection in Self-adaptive Robots with {AI}-powered Digital Twins}, 
  author={Isaku, Erblin and Sartaj, Hassan and Ali, Shaukat and Sanguino, Beatriz and Wang, Tongtong and Li, Guoyuan and Zhang, Houxiang and Peyrucain, Thomas},
  booktitle={2025 40th IEEE/ACM International Conference on Automated Software Engineering (ASE)}, 
  year={2025},
  pages={3403-3414},
  publisher={IEEE},
  address = {New York, NY, USA},
  doi={10.1109/ASE63991.2025.00281}
}

@article{Rizk2019heteroRobots,
author = {Rizk, Yara and Awad, Mariette and Tunstel, Edward W.},
title = {Cooperative Heterogeneous Multi-Robot Systems: A Survey},
year = {2019},
issue_date = {March 2020},
publisher = {Association for Computing Machinery},
volume = {52},
number = {2},
issn = {0360-0300},
doi = {10.1145/3303848},
journal = {ACM Computing Surveys},
month = apr,
articleno = {29},
numpages = {31},
}

@article{Esterle2022heterogeneity,
author = {Esterle, Lukas and King, David W.},
title = {Loosening Control—A Hybrid Approach to Controlling Heterogeneous Swarms},
year = {2022},
issue_date = {June 2021},
publisher = {Association for Computing Machinery},
address = {New York, NY, USA},
volume = {16},
number = {2},
issn = {1556-4665},
doi = {10.1145/3502725},
journal = {ACM Transactions on Autonomous and Adaptive Systems},
month = mar,
articleno = {5},
numpages = {26},
}

@ARTICLE{Xing2023DistRobot,
  author={An, Xing and Wu, Celimuge and Lin, Yangfei and Lin, Min and Yoshinaga, Tsutomu and Ji, Yusheng},
  journal={IEEE Open Journal of the Computer Society}, 
  title={Multi-Robot Systems and Cooperative Object Transport: Communications, Platforms, and Challenges}, 
  year={2023},
  volume={4},
  number={},
  pages={23-36},
  doi={10.1109/OJCS.2023.3238324}}

@article{Maurelli2022localisation,
	author = {Maurelli, Francesco and Krupi{\'n}ski, Szymon and Xiang, Xianbo and Petillot, Yvan},
	date = {2022/06/01},
	doi = {10.1007/s41315-021-00215-x},
	id = {Maurelli2022},
	isbn = {2366-598X},
	journal = {International Journal of Intelligent Robotics and Applications},
	number = {2},
	pages = {246--269},
	title = {AUV localisation: a review of passive and active techniques},
	volume = {6},
	year = {2022}}

@article{billard2025roadmap,
  title={A roadmap for AI in robotics},
  author={Billard, Aude and Albu-Schaeffer, Alin and Beetz, Michael and Burgard, Wolfram and Corke, Peter and Ciocarlie, Matei and Dahiya, Ravinder and Kragic, Danica and Goldberg, Ken and Nagai, Yukie and others},
  journal={Nature Machine Intelligence},
  volume={7},
  number={6},
  pages={818--824},
  year={2025},
  publisher={Nature Publishing Group UK London},
  doi={10.1038/s42256-025-01050-6}
}

@MANUAL{SPECTRA,
	title = {Spectra Language and Spectra Tools User Guide},
	author = {S. Maoz and J. O. Ringert},
	url = {smlab.cs.tau.ac.il/syntech/spectra/},
	year = {2018}
}

@InProceedings{SHT2012,
	author="Maria Spichkova and Florian Hölzl and David Trachtenherz",
	title={{Verified System Development with the AutoFocus Tool Chain}},
	booktitle="Proceedings 2nd Workshop on Formal Methods in the Development of Software",
	year="2012",
	doi="10.4204/EPTCS.86.3",
    pages="17-24",
    publisher="EPTCS",
    address="Paris, France"
}

@InProceedings{Bastian2011,
  author    = {J.~Bastian and C.~Clau\ss and S.~Wolf and P.~Schneider},
  title     = {Master for Co-Simulation using {FMI}},
  booktitle = {Modelica Conf.},
  year      = {2011},
  pages     = {1-6},
  publisher = {Fraunhofer IIS / EAS},
  address   = {Dresden, Germany}
}

@InProceedings{Dhouib2012,
  author    = {Dhouib, S. and Kchir, S. and Stinckwich, S. and Ziadi, T. and Ziane, M.},
  title     = {{RobotML, a Domain-Specific Language to Design, Simulate and Deploy Robotic Applications}},
  booktitle = {SIMPAR 2012},
  year      = {2012},
  pages     = {149--160},
  publisher = {Springer},
  address = {Berlin, Heidelberg}
}

@InProceedings{Foughali2016,
  author    = {M.~Foughali and B.~Berthomieu and S.~Dal Zilio and F.~Ingrand and A.~Mallet},
  title     = {{Model Checking Real-Time Properties on the Functional Layer of Autonomous Robots}},
  booktitle = {Formal Methods and Soft. Eng.},
  year      = {2016},
  pages     = {383-399},
  publisher = {Springer},
  address = {Cham}
}

\end{document}